\documentclass[11pt,a4paper]{article}
\pdfoutput=1
\usepackage{jheppub}
\usepackage{lscape}
\usepackage{array}
\usepackage{hyperref}
\usepackage{amsfonts}
\usepackage{amssymb}
\usepackage{bbm} 
\usepackage{graphicx}
\usepackage{caption}
\usepackage{subcaption}
\usepackage{tabularx}

\usepackage[normalem]{ulem}

\newcommand{\T}{\widetilde{\tau}}
\newcommand{\LTotalz}{\,\text{i}\sum_{j=1}^{\infty} \frac{1}{j \sin \frac{\pi j}{\T}}\sum_{\ell=0}^{m-1} \,\cos\Bigl(\pi j\,\frac{2\xi_{\ell}+1}{\T} \Bigr)}

\newcommand{\Lz}{\text{i}\sum_{j=1}^{\infty} \frac{1}{j \sin \frac{\pi j}{\T}} \,\cos\Bigl(\pi j\,\frac{2\xi_{\ell}(z)+1}{\T} \Bigr)}

\newcommand{\diff}{\text{d}}
\renewcommand{\=}{\,= \,}

\newcommand{\s}{\sigma}
\renewcommand{\t}{\tau}

\newcommand\G{\Gamma}

\newcommand{\Br}[1]{\Bigl\{#1\Bigr\}}

\newcommand{\liMN}{\underset{\tau\to-\frac{n}{m}}{\simeq}}

\renewcommand{\i}{{\rm i}}

\newcommand{\Ge}{\Gamma_\text{e}}
\newcommand{\e}{{\bf e}}

\renewcommand{\th}{\theta}

\newcommand{\z}{\zeta}

\newcommand\bi{\begin{itemize}}
\newcommand\ei{\end{itemize}}

\newcommand\bspl{\begin{split}}
\newcommand\espl{\end{split}}

\newcommand{\susyL}[1]{\underset{\text{SUSY Locus}}{\longrightarrow}}
\newcommand{\be}{\begin{equation}}
\newcommand{\ee}{\end{equation}}
\newcommand{\bea}{\begin{eqnarray}}
\newcommand{\eea}{\end{eqnarray}}

\renewcommand{\=}{\,= \,}

\newcommand{\signo}{\text{sign}(-k)}
\newcommand{\ZGSUN}{Z_{\Gamma}}
\newcommand{\ZGUN}{Z^{U(N)}_{\Gamma}}

\title{On the 4d superconformal index near roots of unity:\\ Bulk and Localized contributions}

\author{Alejandro Cabo-Bizet}
\emailAdd{alejandro.cabo\_bizet@kcl.ac.uk}
\affiliation{Department of Mathematics, King's College London,\\
The Strand, London WC2R 2LS, U.K.}

\abstract{ We study the expansion near roots of unity of the superconformal index of 4d~$SU(N)$~$\mathcal{N}=4$ SYM. In such an expansion, middle-dimensional walls of non-analyticity are shown to emerge in the complex analytic extension of the integrand. These walls intersect the integration contour at infinitesimal vicinities and come from both, the vector and chiral multiplet contributions, and combinations thereof. We will call these intersections \emph{vector} and~\emph{chiral} \emph{bits}, and the complementary region~\emph{bulk}, and show that, in the corresponding limit, the integrals along the infinitesimal bits include, among other contributions, factorized products of either Chern-Simons and 3d topologically twisted partition functions.

In particular, we find that the leading asymptotic contribution to the index, which comes from  collecting all contributions coming from vector bits, reduces to an average over a set of~$N$ copies of three-dimensional $SU(N)$ Chern-Simons partition functions in Lens spaces $L(m,1)$ with~$m>1\,$, in the presence of background~$\mathbb{Z}^{N-1}_m$ flat connections. The average is taken over the background connections, which are the positions of individual vector bits along the contour.  We also find there are other subleading contributions, a finite number of them at finite $N$, which include averages over products of Chern-Simons and/or topologically~$A$-twisted Chern-Simons-matter partition functions in three-dimensional manifolds. This shows how in certain limits the index of 4d~$SU(N)$~$\mathcal{N}=4$ SYM organizes, \emph{via} an unambiguously defined coarse graining procedure, into~\emph{averages} over a finite number of lower dimensional theories. 

}

\usepackage{amsmath}

\begin{document}
 
\maketitle


\section{Introduction and summary of results}
\label{sec1}
In  recent years, superconformal and topologically twisted indices~\cite{Romelsberger:2005eg,Kinney:2005ej,Dolan:2008qi}\cite{Benini:2015noa} have been useful tools to understand the statistical meaning of the Bekenstein-Hawking entropy of BPS black holes in AdS$_4$~\cite{Benini:2015eyy} and~AdS$_5$~\cite{Cabo-Bizet:2018ehj,Choi:2018hmj,Benini:2018ywd}. In AdS$_5/$CFT$_4\,$, an unrefined superconformal index, denoted as~$\mathcal{I}(q)\,$, has been useful to uncover many zero-temperature phases of~$SU(N)$ $\mathcal{N}=4$ SYM~\cite{Cabo-Bizet:2019eaf,PaperPhases}. These phases are detected in the so-called generalized Cardy or Cardy-like limits of~$\mathcal{I}(q)$~\cite{Cabo-Bizet:2019eaf,ArabiArdehali:2021nsx,Jejjala:2021hlt}.


The index $\mathcal{I}(q)$  is a~$(-1)^F$ graded (and protected) trace over a Hilbert space~\cite{Romelsberger:2005eg,Kinney:2005ej} only receiving contributions from states in the cohomology of two complex conjugated supercharges.~$\mathcal{I}(q)$ can be also recast as a multi-dimensional integral depending on a single parameter~$q= e^{2\pi \text{i}\tau}\,$~\cite{Dolan:2008qi}\cite{Romelsberger:2005eg,Kinney:2005ej}
\be\label{IndexInitial}
\mathcal{I}(q)\= \int^{1}_{0}dv_1\,\ldots\,\int^{1}_{0}\, \diff v_{N-1} \,\prod_{i< j=1}^N I(v_{ij},q)
\ee
where $v_{ij} :=v_i-v_j$, $i,j\,=\,1,\,\ldots\,,\, N$ and~$v_N:=-\sum_{i=1}^{N_1} v_i$.~\footnote{It is also a zero-temperature limit of a refined thermal partition function~\cite{PaperPhases} which means that any phase detected with it is expected to be detected as well in a specific zero-temperature limit of a refined thermal partition function on which, simultaneously, the chemical~$\tau\,$ is taken to a rational number. This has been explained in details in~\cite{PaperPhases}. }
There are two ways to implement the generalized Cardy limits on~$\mathcal{I}(q)$
\begin{itemize}
    \item[1)] expand the integrand and integrate the result~\cite{Choi:2018hmj,Honda2019, ArabiArdehali:2019tdm,Kim:2019yrz,Cabo-Bizet:2019osg}\cite{GonzalezLezcano:2020yeb,Amariti:2021ubd,ArabiArdehali:2021nsx,Cassani:2021fyv}.
    \item[2)] integrate and expand the result~\cite{Benini:2018mlo,Cabo-Bizet:2019eaf,Goldstein:2020yvj,Jejjala:2021hlt,Lezcano:2021qbj}.~\footnote{Many of these references focus on a single limit~$\tau\to0$. The generic limits~$\tau\to -\frac{n}{m}$ have been studied in~\cite{Cabo-Bizet:2019eaf,Cabo-Bizet:2020nkr,Cabo-Bizet:2020ewf,ArabiArdehali:2021nsx,Jejjala:2021hlt}. Related discussions and results can be found in~\cite{Murthy:2020rbd,Agarwal:2020zwm,Gaiotto:2021xce}\cite{ArabiArdehali:2019orz}\cite{Copetti:2020dil}\cite{Lezcano:2019pae,Amariti:2019mgp,Lanir:2019abx,Benini:2020gjh}~\cite{David:2021qaa}\cite{Colombo:2021kbb}.}
\end{itemize}
Both approaches agree at very leading order in the expansions near roots of unity i.e. both~$1)$~\cite{ArabiArdehali:2021nsx} and $2)$~\cite{Cabo-Bizet:2020ewf}~\footnote{See also the analysis presented in~\cite{PaperPhases}.} predict a universal leading asymptotic expansion of the form
\be\label{LeadingBehaviour}
e^{-(N^2-1)\,\Bigl( \frac{\pi\i}{27 m\widetilde{\tau}^2}+\frac{2 \pi\i }{9 m\widetilde{\tau}}\,+\,\pi\i c_0(m,n)\,+\,\frac{8\pi\i}{27 m}\T\Bigr)}\,\mathcal{K}\,,\qquad \T\,\equiv\,m\tau+n\,,
\ee
as~$\tau \,\to\,-\frac{n}{m}\,$.~\footnote{\label{MWingLimit}This paper focuses on the limits~$\tau\to -\frac{n}{m}$ for which~$m>1\,$,~$\tau_2>0\,$, and~$m\tau_1+n>0\,$. In these limits the absolute value of~$e^{-\,\frac{\pi\i}{27 m\widetilde{\tau}^2}-\frac{2 \pi\i }{9 m\widetilde{\tau}}}$ grows. This choice of limits corresponds to the selection of~$M$-wings, in the language of~\cite{Lezcano:2021qbj}. For the limits in the~$W$-wings our conclusions would need to be modified. For example, in such cases there could be logarithmic corrections of the form~$\log (m\tau+n)$ to the effective action as shown in~\cite{Lezcano:2021qbj}, and recently argued in~\cite{Ardehali:2021irq}. In the Cardy-like limits we are studying such corrections are not present in the leading exponential order~\eqref{LeadingBehaviour} (They could be present in subleading exponential contributions, see the following subsection).  We thank A. A. Ardehali for a useful conversation regarding this point.} \footnote{The~$c_0$ is a real parameter depending on~$N\,$,~$m\,$, and~$n\,$.} The c-number~$\mathcal{K}$ is a $\tau$-independent contribution that in some cases, e.g.~$m=1$ $n=0$, is known to equal~$N$.


We expect that there exists a relation between~$1)$ and~$2)$ beyond their matching at leading order~\eqref{LeadingBehaviour}. Our expectation is that exponentially subleading corrections to~\eqref{LeadingBehaviour} from approach 1) could turn out to be helpful in identifying previously unnoticed Bethe roots of~\eqref{LeadingBehaviour} at finite~$N$, with a possible relevance in the holographically dual perspective. As a necessary intermediate and less ambitious step, this project improves our understanding on exponentially subleading contributions to the superconformal index from the perspective of 1).

\subsection{Summary of the main results} \label{SectionIntroResults}

Our main result is the definition of a coarse-graining procedure that reduces the superconformal index to an expansion in averages over lower dimensional theories. We represent such a reduction as
\be\label{AsymptoticTales}
\mathcal{I}(q)\, \underset{\tau\to -\frac{n}{m}}{\longmapsto}\,\sum_{\alpha}e^{-\pi \i\frac{P^{(3)}_{\alpha}(m,n;\tau)}{m(m \tau+n)^2}}\,\times \,N \,\times\,(\delta |m\tau+n|)^{d_{1\alpha}}\,\times\,\overline{Z}^{(\alpha)}\,,
\ee
where~$\alpha$ labels families of 3d partition functions obtained by gluing 3d blocks. These families include well known examples, such as Chern-Simons theories over Lens spaces~$L(p,1)$ with~$p\,\geq\,1\,$, and topologically twisted theories in oriented circle bundles of degree~$p$ over closed Riemann surfaces~$\Sigma_g\,$ of genus~$g$~\cite{Closset:2017zgf,Closset:2018ghr}. These families also include other more exotic partition functions that correspond to~\emph{coupled} and \emph{decoupled products} of 3d Chern-Simons and topologically twisted partition functions. The study of these more general cases is left for future work. Formula~\eqref{AsymptoticTales} together with~\eqref{AveragesPartitionF} and~\eqref{IntegralBlocks} below, will be called the~\emph{master formula(s)}. 

These sectors~$\alpha$ should have a holographic or string-theory dual interpretation. For instance one could say that in the limit~$\widetilde{\tau} \to 0$ the $SU(N)$~$\mathcal{N}=4$ SYM flows or \emph{coarse grains} into an effective theory associated to the leading sector~$\alpha$. If holography predicts the existence of a semiclassical gravitational dual realization of
the corresponding limit~$\widetilde{\tau}\to 0$ at large~$N$, then it is natural to expect that at large~$N$ the effective theory of subleading contributions~$\overline{Z}^{\alpha}$ could correspond to the effective theory of gravitational fluctuations around the dual gravitational background. In the future, it would be very interesting to interpret the topological degrees of freedom associated to the effective theory~$\alpha$ as gravitational or string-theory excitations around the large black holes that are known to dominate the gravitational picture in the large-$N$ expansion around~$\tau\sim 0\,$. Doing so, lies beyond the scope of this work. 

The asymptotic map~\eqref{AsymptoticTales}, which we will call \emph{coarse graining procedure} from now on, follows from the existence of an infinite set of identities of the form
\be\label{IntegrandSoloInitial}
I(v_{ij})\,=\, \prod_{ \Delta\, \in\,\bigl\{2\tau,\frac{2\tau-n_0}{3},\frac{2\tau-n_0}{3},\frac{2\tau-n_0}{3}\bigr\}\atop\Delta \sim \Delta +1\,\,\text{and}\,\, n_0=-1, 0, 1\text{ mod } 3}e^{-\pi \text{i}\frac{ R^{(3)(m,n)}(v_{ij}+\Delta)}{m (m\tau+n)^2}+L^{(m,n)}_{\Gamma_e}(v_{ij}+\Delta)}
\ee 
that can be applied to each one of the factors in the integrand of the index~\eqref{IndexInitial}. These identities are labelled by two integer co-primes~$m\geq1$ and~$n$. 

The function~$R^{(3)(m,n)}(z+\Delta)$ is a piecewise third-order polynomial function of~$z$, and the~$L^{(m,n)}_{\Gamma_e}(z+\Delta)$ is a transcendental function that vanishes exponentially fast in the expansion $\tau\to-\frac{n}{m}$ for almost every complex~$z$ except for at a discrete set of middle-dimensional sections in the~$z$-complex plane. Although at such sections~$L_{\Gamma_e}$ develops different limits from the left and right when~$\tau\,\to\,-\frac{n}{m}$, the limit of~$L_{\Gamma_e}$ at the sections is well-defined and non-vanishing. The sections across which the discontinuities in the limit $\tau\to-\frac{n}{m}$ of $L_{\Gamma_e}$ emerge will be called~\emph{walls}.~ \footnote{Their existence generates Chern-Simons classical corrections to the effective action associated to a sector~$\alpha$.}

~{It should be noted that these discontinuities are cancelled by those coming from the piecewise polynomial part~$R^{(3)}\,$ as it follows from the fact that~$I(z)$ is not discontinuous in the~$z$-complex plane.}~\footnote{That will be proven in section~\ref{CancellationBranchCuts}.}

The non-vanishing value of~$L_{\Gamma_e}$ at the walls (when~$\tau\to-\frac{n}{m}$) is recovered by the semi-sums of its lateral limits, which we denote as~$L_{\Gamma_e}^{\pm}$ below, i.e.,
\be\label{ValueLGammaWall}
L_{\Gamma_e}(\text{wall}+\Delta) \,=\,\frac{L_{\Gamma_e}^{+}(\text{wall}+\Delta)+L^-_{\Gamma_e}(\text{wall}+\Delta)}{2}\,.
\ee
We remark that these~$\pm$ limits are double limits in the sense that they include~$\tau\to-\frac{n}{m}$ and one of the two limits~$z\to wall^{\pm}$. We will show that these lateral limits are well-defined and non-vanishing. In the particular cases~$\Delta=2\tau\text{mod}1$, which corresponds to contributions to the integrand of the index coming from the~$\mathcal{N}=1$ vector multiplet, a drastic simplification happens. So, given this simplification we will branch the definition of walls in two: Those walls for which~$\Delta=2\tau\text{mod}1$, will be called~\emph{vector walls}, and those for which~$\Delta\neq 2\tau\text{mod}1s$, will be called~\emph{chiral walls}. The intersection of the vector and chiral walls with the contour of integration~$[0,1)^{\text{rk}(G)}$ will be called~\emph{vector} ($v$) and~\emph{chiral bits} ($c$) or simply \emph{bits}, respectively, and the complementary region~\footnote{...except for another set of infinitesimally small subdomains that will be called \emph{auxiliary bits}~$(b^\prime)$...} will be called~\emph{bulk}. 

 \subsubsection{A contour decomposition and definitions}

The coarse-graining procedure denoted with the symbol~$\underset{\tau\to -\frac{n}{m}}{\longmapsto}$ in~\eqref{AsymptoticTales} can be understood in various steps.

 The fist step is to define convenient $m$ and $n$-dependent contour decompositions of the superconformal index
\be\label{Decomposition}
\begin{split}
\mathcal{I}\,=\,\sum_{N_b\,=\,0}^{N\,-\,1}\sum_{ \widetilde{\lambda} \,\in\, \text{Par}(N-N_b-1)\atop \lambda \,\in\, \text{Par}(N_b)}\,\text{symm}(\widetilde{\lambda},\lambda) \,&\int_{\mathcal{M}_\epsilon(\widetilde{\lambda})} d\underline{x} \,\Bigl(\prod_{{v_{ij}}\text{'s} \,\text{in bits}}I(v_{ij})\Bigr)\\&\qquad\times\Bigl(\int_{\mathcal{M}_{b}(\lambda)} d\underline{y}  \prod_{{v_{ij}}\text{'s} \,\text{in bulk}}I(v_{ij})
\Bigr)\,\, .
\end{split}
\ee
For~$N=3\,$, this contour decomposition~\eqref{Decomposition} is derived from scratch in appendix~\ref{app:TheCountourDecN3}.

The~$\lambda$, $\widetilde{\lambda}$ are ordered partitions of~$N_b\,<\,N$ and $N\,-\,N_b\,-\,1\,$, respectively. The integer symmetry factor
\be
\text{symm}(\lambda, \widetilde{\lambda})\,\neq 0\,\qquad \text{if}\,\qquad \,|l(\lambda)-l(\widetilde{\lambda})|\,\leq\,1\,,
\ee
will be computed only in a case by case basis.~$l(\lambda)$ is the length of the partition~$\lambda$.~\footnote{Choices of partitions~$(\lambda^{\prime}, \widetilde{\lambda}^{\prime})$ for which~$|l(\lambda^{\prime})-l(\widetilde{\lambda}^{\prime})|\,>\,1\,$
turn out to be equivalent to a given choice~$(\lambda, \widetilde{\lambda})$ with~$|l(\lambda)-l(\widetilde{\lambda})|\,\leq\,1\,$. Thus, the contribution of the former subcontour integrals can be counted in the symmetry factor~$\text{symm}(\lambda, \widetilde{\lambda})$ of the latter without loss of generality.} 

The variables~$v_{ij}$'s in~\eqref{Decomposition} are sums of subsets of the~$N-1$ integration variables~$\underline{x}=\{x_k\}$ and~$\underline{y}=\{y_k\}$, more precisely
\be\label{eq:CombinationV}
v_{ij}\,=\,v_{ij}(\underline{x},\underline{y})\,=\, \,\sum_{k\,=\,i}^{j}v_{k,k+1}\,=\,\sum_{k\,=\,a_{\text{x}}}^{b_\text{x}} x_k +\sum^{b_y}_{k\,=\, a_{y}}
y_k\,,
\ee
where~$1\leq a_x\leq b_x \leq \sum_i \widetilde{\lambda}_i$ and~$1\leq a_y\leq b_y \leq \sum_i {\lambda}_i$ are positive integers that depend on~$i$ and~$j\,$.

~The~$v_{ij}$'s that do not depend on the variables~$\underline{y}$ are said to be located \emph{in bits}.~The~$v_{ij}$'s that are not in bits are said to be \emph{in the bulk}.

\paragraph{The bulk domains} The bulk domains~$\mathcal{M}_b(\lambda)$ are products of subregions that we call simple
\be\label{def:MbLambda}
\mathcal{M}_b(\lambda)=\underset{i}{\otimes} s\mathcal{M}^{(\lambda_{i})}_{b}\,,\qquad \text{with}\qquad  \lambda_{i}\,=\,i-\text{th element of~$\lambda\,$}.
\ee
The {simple} bulk region~$s\mathcal{M}^{(\lambda_i)}_{b}$ is defined as a~$\lambda_i$-dimensional region with coordinates~
\be\nonumber
\begin{split}
y_a\,:=\, v_{a_0+a} \,-\,v_{a_0+a+1}\,\in\,[-1,1],\,&\ \qquad a\,=\,1,\ldots, \lambda_i,\\&\qquad\text{ for some } 0 \,\leq \,a_0\, <\, N\,-\,\lambda_i\,,
\end{split}
\ee
such that the following~$\lambda_i(\lambda_i+1)/2$ linear combinations among the~$y_a$'s
\be\label{eq:BulkInfinitesimal}
\sum_{k\,\geq\, a}^{b}y_{k}\qquad, \qquad a\,=\,1\,,\,\ldots\,,\lambda_i\,, \quad b\,=\,a\,,\,\ldots\,,\,\lambda_i\,,
\ee
do not equal the position of any bit, this is,
\be\label{eq:BulkInterval}
\Biggl|\sum_{k\,\geq\, a}^{b}y_{k}-\text{choice of bit position}\Biggr| \,>\, (b-a)\,\varepsilon\,:=\,(b-a)\,\delta |m \tau +n| \,\liMN\, 0^+\,,
\ee
for every possible choice of bit position in the table~\ref{PositionsBitsIntro} below.~$\delta$ is a positive real number, independent of~$\tau\,$. This quantity~$\delta$ will be called \emph{cut-off} and it will play a relevant role in the following sections.

The coordinates of simple bulk spaces are in one-to-one relation with Cartan generators of simple subgroups of~$SU(N)$. The corresponding subgroup being included in the gauge symmetry transformations broken by the theory associated to the sector~$\alpha$ in~\eqref{AsymptoticTales}. There is a related symmetry-breaking classification of the sectors~$\alpha$ that will be briefly introduced in subsection~\ref{SymmetryBreakingBits}.


\paragraph{The bit domains} The bit domains~$\mathcal{M}_\varepsilon(\lambda)$ are unions of products of subregions that we call simple
\be
\mathcal{M}_\varepsilon(\widetilde{\lambda})=\sum_{\text{choices of bits}}\,\underset{i}{\otimes} s\mathcal{M}^{(\widetilde{\lambda}_{i})}_{\varepsilon}\,,\qquad \text{with}\qquad  \widetilde{\lambda}_{i}\,=\,i-\text{th element of~$\widetilde{\lambda}\,$}.
\ee
The simple bit region~$s\mathcal{M}^{(\widetilde{\lambda}_i)}_{\varepsilon}$ is defined as a~$\widetilde{\lambda}_i$-dimensional $\varepsilon$-infinitesimal region with coordinates~
\be\nonumber
\begin{split}
x_a\,:=\, v_{a_0+a} \,-\,v_{a_0+a+1}\,\in\,[-1,1],\,&\ \qquad a\,=\,1,\ldots, \widetilde{\lambda}_i,\\&\qquad\text{ for some } 0 \,\leq\,a_0\, <\, N\,-\,\widetilde{\lambda}_i\,,
\end{split}
\ee
such that the following~$\widetilde{\lambda}_i(\widetilde{\lambda}_i+1)/2$ linear combinations among the~$x_a$'s
\be\label{eq:BitsInfinitesimal}
\sum_{k\,\geq\, a}^{b}x_{k}\qquad, \qquad a\,=\,1\,,\,\ldots\,,\widetilde{\lambda}_i\,, \quad b\,=\,a\,,\,\ldots\,,\,\widetilde{\lambda}_i\,
\ee
belong to an infinitesimal~$\varepsilon$-vicinity of bits.
This latter set of conditions on~\eqref{eq:BitsInfinitesimal} follow from the~$\lambda_i$ conditions
\be\label{eq:BitsInterval}
\Biggl|x_{a}-\text{choice of bit position}\Biggr| \,<\, \varepsilon\,\liMN\,0^+.
\ee
That is because the total set of bit positions, -- as summarized in Table~\ref{PositionsBitsIntro} below --, is closed under the operation of addition. Consistently, this also implies, together with~\eqref{eq:BulkInterval}, that the~$v_{ij}(\underline{x},\underline{y})$'s ($j>i$) that depend non-trivially on~$\underline{y}$, which we have advanced to be 
located in the bulk, are such that 
\be\label{eq:BulkIntervalMixedVariables}
\Biggl|v_{ij}(\underline{x},\underline{y})-\text{choice of bit position}\Biggr| \,>\, (j-i)\,\varepsilon\,\liMN\,0^+,
\ee
for every possible choice of bit position, as it should be.

The simple vector bit domains are in one-to-one relation with Cartan generators of simple subgroups of~$SU(N)$. The corresponding total subgroup contains the gauge symmetry preserved by a theory associated to the sector~$\alpha$ in~\eqref{AsymptoticTales} as a subgroup.

\subsubsection{Integrating out bulk variables in Cardy-like limit}

Recall that if a~$v_{ij}$ is in the bulk, the transcendental contribution~$L_{\Gamma_e}$ vanishes exponentially fast in Cardy-like limit.
Then, given a partition $\lambda$, the integral over the~$\mathcal{M}_b(\lambda)$\,
\be
\int_{\mathcal{M}_b(\lambda)} d\underline{y}  \prod_{v_{ij}\text{'s}\text{ in the bulk}}I(v_{ij})
\ee   
reduces, in Cardy-like limit~$\tau\to -\frac{n}{m}\,$, to the integral of a deformed integrand obtained after dropping the transcendental contribution~$L_{\Gamma_e}$ in each of the relevant blocks~\eqref{IntegrandSoloInitial}. This integral is Gaussian and it can be solved exactly~\footnote{At least in the limit~$\delta\gg1$, which is enough.}

 Indeed, independently of the choice of bits and~$\widetilde{\lambda}$
\be\label{IntegralBulk}
\int_{\mathcal{M}_b(\lambda)} d\underline{y}  \prod_{v_{ij}\text{'s} \text{ in the bulk}}I(v_{ij}) \,\liMN\,  e^{-\pi \text{i}\frac{ P^{(3)(m,n)}(\tau)}{m (m\tau+n)^2}}C (\delta |m\tau+n|)^{d_{1}}
\ee
where~$C$ is a constant (to be absorbed in the Casimir pre-factor) that does not depend on the unintegrated variables~$\underline{x}$. The~$d_{1}$ is an integer number that we only know how to determine in a case by case  basis: it comes from counting contributions of~$|m\tau+n|$ out of constant pre-factors and Jacobian contributions (in integrations over the bulk variables) (Examples are given in section~\ref{subsec:35})~\footnote{In this expression the symbol~$\liMN$ means equal up to exponentially suppressed contributions in the limit~$\tau\to -\frac{n}{m}\,$.}

\subsubsection{The integral over bit variables in Cardy-like limit: Subleading~$3d$ ensembles} After evaluating~\eqref{IntegralBulk}, it remains to study the integral over the infinitesimally small bits
\be\label{BitIntegral}
\int_{\mathcal{M}_\varepsilon(\widetilde{\lambda})} d\underline{x}\,\Bigl(\prod_{v_{ij}\text{'s in bits
}}I(v_{ij})\Bigr) \,.
\ee
For~$\varepsilon:=\delta |m\tau+n|=0^+$ this integral would vanish, if the integrand would have an analytic limit within~$\mathcal{M}_{\varepsilon}(\widetilde{\lambda})$. However, as it will be shown in section~\ref{sec3.1}, the integrand develops non-analyticites in these infinitesimally smalls domains~$\mathcal{M}_{\varepsilon}(\widetilde{\lambda})\,$, and it does not vanish trivially at~$\varepsilon:=\delta |m\tau+n|=0^+$.

Indeed, in the limit~$\tau\to-\frac{n}{m}$ integral~\eqref{BitIntegral} reduces to an integration along walls (this will be explained in section~\ref{sec3.1}). Let~$u_{ij}$ be the coordinates along the wall coming from a bit~$v_{ij}(x)$. Walls sharing the same Casimir pre-factor form a sector~$\alpha$ in~\eqref{AsymptoticTales}. Different~$\alpha$'s are fixed by a choice of partitions $(\widetilde{\lambda};\lambda)$ and by a partial fixing of the ambiguity in the \emph{choice of bit positions} in the definition~\eqref{eq:BitsInterval} of~$\mathcal{M}_\varepsilon(\widetilde{\lambda})\,$. In the following subsection~\ref{SymmetryBreakingBits} we will expand on this classification of sectors~$\alpha\,$. In this subsection we focus on explaining how they are in one-to-one relation with ensembles over 3d gauge theory-like partition functions. 


The evaluation of $R^{(3)}$'s at the corresponding bits $v_{ij}=v_{ij}(x,y)$, and the contributions~$C$ coming from the integration over bulk regions, define the third order polynomial term~$P^{(3)}_{\alpha}$ in~\eqref{AsymptoticTales}.~\footnote{Other contributions could come from integrating out trivial~$U(1)$ modes, as we will explain next.} The non-vanishing value of the $L_{\Gamma_e}$'s at each of the walls in the set of bits denoted as~$\alpha\,$,~\eqref{ValueLGammaWall}, and quadratic terms in the variables~$v_{bits}$ coming from the~$R^{(3)}$'s in the factors
\be
\prod_{v_{ij}\text{ 's in the bulk}  }I(v_{ij})
\ee
define the 3d theories to average over in order to obtain the~$ \,N \,\times\,\overline{Z}^{(\alpha)}$ in~\eqref{AsymptoticTales}.

We find that~$\overline{Z}^{\alpha}$ equates to
\be\label{AveragesPartitionF}
\overline{Z}^{(\alpha)} \,:=\, \sum_{\underline{\ell}\,,\, \underline{\widetilde{\ell}}\,,\,\underline{\widetilde{\ell}^{\prime}}\,=\,0}^{m\,-\,1} \, e^{\pi\text{i}\Phi_{\alpha}[\ell,\widetilde{\ell},\widetilde{\ell}^{\prime}]} \,z^\alpha(\ell,\widetilde{\ell}\,,\, \widetilde{\ell}^{\prime})\,,
\ee
~\footnote{We know how to compute the real phase~$\Phi_\alpha(\ell,\widetilde{\ell},\widetilde{\ell}^\prime)$ in a case-by-case basis. For our present goal the important observation to keep in mind is that~$\Phi_\alpha(\ell,\widetilde{\ell},\widetilde{\ell}^\prime)$ only depends on~$\alpha$ and the averaging~$\mathbb{Z}_m$-variables. We leave for the future the derivation of a closed analytic expression for  this phase.}
which, in many cases, corresponds to an average over products of 3d Chern Simons or Chern-Simons-matter partition functions. The three sets of~$N-1$ indices $\underline{\ell}:=\{\ell_{i,i+1}\}_{i=1,\ldots, N-1}$, $\underline{\widetilde{\ell}}:=\{\widetilde{\ell}_{i,i+1}\}_{i=1,\ldots, N-1}$, and~$\underline{\widetilde{\ell}}^\prime:=\{\widetilde{\ell}^{\prime}_{i,i+1}\}_{i=1,\ldots, N-1}$ parameterize the positions of a set of~$v$,~$c$, and~$b^\prime$ bits denoted as~$\alpha$.


In summary, one obtains
\be\label{IntegralBlocks}
z^\alpha(\ell,\widetilde{\ell},\widetilde{\ell}^\prime)\,:=\, C^\alpha\,\int_{\Gamma_\alpha} d\underline{u} \,\Bigl(\prod_{(i,j)_{v}}\,\mathcal{V}_{m}[u_{ij},{\ell}_{ij}] \Bigr)\,\Bigl(\prod_{(i,j)_c}\,\mathcal{C}^{}_{m}[{u}_{ij},\widetilde{\ell}_{ij}]\Bigr)\,\Bigl(\prod_{(i,j)_{b^{\prime}}}\,\mathcal{B}^{\prime}_{m}[{u}_{ij},\widetilde{\ell}^{\prime}_{ij}]\Bigr)\,.
\ee
 ~\footnote{The undetermined constant~$C^\alpha$, which could depend on~$\tau$ can be absorbed in the Casimir prefactor~$e^{-\pi \text{i}\frac{P^{(3)}_{\alpha}(m,n;\tau)}{m(m \tau+n)^2}}\,$, but for later convenience we have chosen to leave it in this expression. The $C^\alpha$'s come from the collection of the~$C$'s in~\eqref{IntegralBulk}.}
 where
 \be
 \ell_{ij}\,=\,\sum_{k\,=\,i}^{j}\ell_{k,k+1}
 \quad,\quad \widetilde{\ell}_{ij}\,=\,\sum_{k\,=\,i}^{j}\widetilde{\ell}_{k,k+1} \quad,\quad \widetilde{\ell}^\prime_{ij}\,=\,\sum_{k\,=\,i}^{j}\widetilde{\ell}^{\prime}_{k,k+1}\,.
 \ee

 The factors~$\mathcal{V}_m$, $\mathcal{C}_m$, and $\mathcal{B}^{\prime}$ come from the~$v$,~$c$ and~$b^\prime$ types of bit in the group~$\alpha$, respectively.~$(i,j)_v$,~$(i,j)_c$,~$(i,j)_{b^\prime}$ denote the set of indices~$i$ and~$j$ for which the original variables~$v_{ij}(\underline{x})$ are~$\varepsilon$-infinitesimally close to the set of~$v$,~$c$ and~$b^\prime$ bits definning the sector~$\alpha$, respectively. The definition of the contour~$\Gamma_\alpha$ and the new integration variables~$\underline{u}=\{u_i\}$ will be given towards the end of this subsection. The integrand is defined in terms of the building blocks~\footnote{... assuming a parameter that will be defined below as~$n_0$ equals to~$-1$ and~$m+n=1\,\text{mod}\,3\,$...}
\be\label{HolomorphicBlocks}
\begin{split}
\mathcal{V}_{m}[x,\ell]&\,=\,e^{{\pi \text{i} m} x^2}\mathcal{V}_{m}^{+}[x,\ell]\,\mathcal{V}_{m}^{-}[x,\ell]\,, \\  \mathcal{C}_{m}[x,\ell]&\,=\,e^{-\frac{\pi \text{i} m}{2} x^2}\mathcal{C}_{m}^{+}[x,\ell]\,\mathcal{C}_{m}^{-}[x,\ell],\,   \\ \mathcal{B}^{\prime}_{m}[{x},\widetilde{\ell}^{\prime}]&\,=\,e^{-\frac{\pi \text{i} m}{2} x^2}\,,
\end{split}
\ee
where the quantities with superindices~$\pm$ are defined as
\be
\begin{split}
\mathcal{V}^{\pm}_m[u,\ell]&\,:=\, \,e^{\,\mp\,\pi \text{i} (x+\frac{\ell}{m}) - \Omega^{\pm}(x+\frac{\ell}{m})}\,, \\
\mathcal{C}^{\pm}_m[u,\ell]&\,:=\,  \,e^{3\pi\i(g-1) \Omega^{\pm}(x+\frac{r+{\ell}}{m})} \,\Bigl(\mathcal{G}^{\pm}_{\frac{m}{2}}(x+\frac{r+{\ell}}{m})\Bigr)^3 \,,
\end{split}
\ee
and
\be\label{LeftAndRightBlocks}
\begin{split}
e^{2\pi\i \Omega^{+}(x)}&:=\,e^{ \text{Li}_1(e^{+2\pi\i (x)})}\,,\\ 
e^{2\pi\i {\Omega}^{-}(x)}&=\,e^{ \text{Li}_1(e^{-2\pi\i (x)})}=e^{-2\pi \text{i}(x-\frac{1}{2})}e^{2\pi\i {\Omega}^{+}(x)}\,,\\
\mathcal{G}^{+}_{{m}}(x) &:= e^{\frac{m}{2\pi\i} \,\text{Li}_2(e^{+2\pi\i (x)}) \,-\,  m x\text{Li}_1(e^{+2\pi\i (x)})}\,,\\ {\mathcal{G}}^{-}_{{m}}(x)&:= e^{-\frac{m}{2\pi\i} \,\text{Li}_2(e^{-2\pi\i (x)}) \,-\,  m x\text{Li}_1(e^{-2\pi\i (x)})}\,=\, e^{\frac{\pi \text{i}}{6}\,-\,\pi \text{i} m x^2 }\mathcal{G}_{{m}}^{+}(x)\,.
\end{split}
\ee
These expressions follow from the results obtained in section~\ref{sec:ChiralBits}. The variable~$r=\frac{2}{3}$ is the superconformal R-charge of the three~$\mathcal{N}=1$ chiral multiplets within 4d~$\mathcal{N}=4$ SYM.

At this point it should be noted that~$\mathcal{G}^{+}_{m}(x)$ happens to be the fibering operator~$\mathcal{G}^{\Phi}_{1,m}$ of a 3d chiral multiplet as defined by Closset, Kim, and Willet (Please refer to the definition given in equation (4.64) of~\cite{Closset:2018ghr}). The quantities $\Omega^{\pm}$ can be understood as contributions to a dilaton profile in the framework of~\cite{Closset:2018ghr}, and the polynomial contributions in the exponents are essentially classical Chern-Simons contributions to both, the dilaton and twisted superpotentials~\cite{Closset:2018ghr}.~\footnote{Although we have only checked this is some particular examples, we expect it to be true for generic~$\alpha$'s.} A more detailed comparison with the framework and conventions of~\cite{Closset:2018ghr} will be given elsewhere. 

Some of the sectors~$\alpha$ in~\eqref{AsymptoticTales} correspond to a coupled set of theories at different genus~$1\,<\,g=g(\ell)=\ell+2\,\leq\, m+1$~\footnote{Or at~$g\,=\,0$ as it is the case of the contributions coming from vector bits.}. In some other cases the component theories \emph{decouple}. In that case the theory~$\alpha$ is interpreted as a product of independent theories. By decoupling we mean that the integral~$\Gamma_{\alpha}$ factorizes in products of integrals that are partition functions of theories living on a 3d spacetime characterized by a single value for the genus parameter~$g$. 
 
Interestingly, the functions with~$\pm$ superindex in~\eqref{HolomorphicBlocks} emerge from the exponentials of the left and right limits in~\eqref{ValueLGammaWall}. Namely, from the contributions
\be
e^{\,\frac{L_{\Gamma_e}^{+}(\text{wall}+\Delta)}{2}} \,,\qquad e^{\frac{L^-_{\Gamma_e}(\text{wall}+\Delta)}{2}\,}
\ee
that can arise in the lateral limits~$\tau\to-\frac{n}{m}$ of any of the $N(N-1)/2$ factors in the integrand of the index~
\be\label{IntegrandSolo}
I(v_{ij})\,,
\ee 
when the~$v_{ij}$ hits a~\emph{wall}. The discontinuities in the limit~$\tau\to-\frac{n}{m}$ of the~$L_{\Gamma_e}$'s, which can be related to exponentials of quadratic and linear functions of the coordinate along the wall, translate into mixed and unmixed classical Chern-Simons contributions to the effective action that underlies the integral~\eqref{IntegralBlocks}.

The integration variables~$u$ in~\eqref{IntegralBlocks}, which can be fewer than or equal to the rank of the gauge group, are a deformation of the bit variables~$\underline{x}\,$. The latter lie along the original contour of integration, the former are supported over an integration contour~$\Gamma_\alpha$ which is an unbounded multi-dimensional domain: an infinitesimal deformation of the real contour $(-\infty,\infty)^{\text{dim}\Gamma_\alpha \,\leq\,\text{rk}(G)}\,$. The infinitesimal deformation is determined by requiring the integral~\eqref{IntegralBlocks} to be convergent. As the asymptotic of the integrand of~\eqref{IntegralBlocks} is dominated by the Gaussian term coming from classical Chern-Simons contributions, the infinitesimal deformation must be such that the deformed contour extends up to the infinitely far regions on which the leading Gaussian term vanishes. We have only constructed~$\Gamma_\alpha$ for very simple examples of~$\alpha\,$. For generic~$\alpha$ we expect the prescription that defines~$\Gamma_\alpha$ to be similar to the Jeffrey-Kirwan recipe, or more precisely, to the variations of the latter that have been previously proposed and studied by Closset, Kim, and Willet, for 3d topological twisted theories over a large variety of three dimensional manifolds~\cite{Closset:2018ghr,Closset:2017zgf}. Section~\ref{sec:JKContour} formulates the initial steps that hint at a general prescription. The completion of that analysis is left for the future.

\paragraph{A couple of examples of sectors~$\alpha$}  A sector~$\alpha$ is defined by a selection of two partitions~$\lambda$ and $\widetilde{\lambda}$ and a distribution of \emph{vector} ($v$), \emph{chiral} ($c$)\,, and \emph{auxiliary} ($b^{\prime}$) bit positions in the definitions of bit domains,~\eqref{eq:BitsInterval}. For instance, one can take the trivial partition~$\lambda=0$, which imply~$\widetilde{\lambda}=N-1\,$, and then take only vector bits~$v$. Such sector~$\alpha$ is called the \emph{maximally-symmetric sector}. This is the sector with the leading exponential Casimir pre-factor, the one corresponding to the exponential of the entropy function of the dual black hole,~\eqref{LeadingBehaviour}. Subsection~\ref{MaximallySymmetricSector} will study this example.

Another possibility is to assign a vector bit position~$v$ to a single seed, say to the~$v_{1,2}$ and let every other seed~$v_{ij}$ to be in the bulk, which corresponds to assuming two partitions of unit length,~$\lambda=N-2\,$ and~$\widetilde{\lambda}=1$. This last sector~$\alpha$ corresponds to an ensemble over~$SU(2)$ Chern-Simons partition functions at quantum corrected level~$k=2\,$, as follows from the master formula~\eqref{IntegralBlocks}.~\emph{Non maximally symmetric}~$\alpha$'s like the latter will be called~\emph{symmetry-breaking} sectors.

\subsection{Symmetry-breaking classification of sectors~$\alpha$}\label{SymmetryBreakingBits}
This subsection introduces a classification of contributions to sectors~$\alpha$ in terms of the amount of gauge symmetry they preserve and the type of bits or walls that contribute to~$z^\alpha(\ell,\widetilde{\ell},\widetilde{\ell}^\prime)\,$.
The~$SU(N)$ index~\eqref{IndexInitial} can be written as
\be\label{IndexInitial2}
 \int^{1}_{0}dv_1\,\ldots\,\int^{1}_{0}\, \diff v_{N} \,\prod_{i< j=1}^N I(v_{ij},q)\,\delta\Bigl(\frac{1}{N}\sum_{i}v_i\Bigr)\,.
\ee
\label{Seeds}

There are~$N-1$ linearly independent positions among the~$N (N-1)$ $v_{ij}$'s in the integrand of~\eqref{IndexInitial2}  and without loss of generality they can be defined to be
 \be
 v_{i, i+1}\,:=\,v_{i}-v_{i+1}\,\in\, [-1,1]\,, \qquad~i=1\,,\,\ldots\,, N-1\,.
 \ee
These variables will be called~\emph{seeds} from now on. This is,  linear combinations of the seeds generate all the other~$v_{ij}$'s
\be\label{eq:VariablesConvenient}
v_{ij}= \sum_{k\,=\,i}^{j\,-\,1} v_{k,k+1}\,.
\ee
We find convenient to change variables in the integral~\eqref{IndexInitial2} from
 \be
\{v_i\} \to \,\{v_{i,i+1}\}\,\cup\, \{\frac{1}{N}\sum_i v_i\}\,.
 \ee
This transformation has trivial Jacobian. The only dependence of the integrand on the center of mass variable~$\frac{1}{N}\sum_{i}v_i$ comes from the Dirac delta. Thus, integrating the center of mass variable one obtains
\be\label{eq:CleverContour}
\eqref{IndexInitial2}\,=\,
 \int^{1}_{-1}dv_{12}\,\ldots\,\int^{1}_{-1}\, \diff v_{N-1, N} \,\prod_{i< j=1}^N I(v_{ij},q)\,.
\ee
\begin{figure}[h]\centering
\includegraphics[width=8cm]{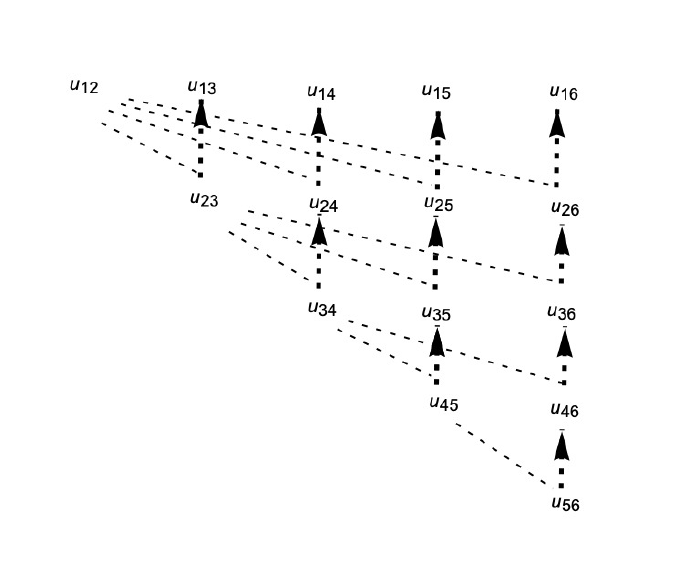} 
\caption{The element indicated by the head of an arrow can be obtained from the sum of the two elements in the initial and middle points of the corresponding broken arrows. That relation,~\eqref{eq:VariablesConvenient}, implies that any element outside of the main diagonal can be written as a linear combination of the ones in the main diagonal: the~\emph{seeds}.
}
\label{fig:IndepVars}
\end{figure}

In virtue of~\eqref{eq:VariablesConvenient}, the positions of the bits and bulk along the contour of integration in the right-hand side of~\eqref{eq:CleverContour} are defined by~$N-1$ conditions
\be\label{BitsConditionIntroduction}
 v_{i,i+1}\= v^{(0)}_{i}\,,
\ee
where the possible values of~$v^{(0)}_i$ are classified in table~\ref{PositionsBitsIntro}
\begin{center}\label{tab:Tabla}
\begin{tabular}{||c c||} 
 \hline
 \qquad & $v_i^{(0)}$  \\ [0.5ex] 
 \hline
vector bits ($v$) & $\frac{n\ell_{ i,i+1}}{m}\,\text{mod}\, 1$ \\ 
 \hline
 chiral bits ($c$) & $-\frac{(m n_0 -n)}{m}\, \frac{1}{3}\,+\, \frac{n\widetilde{\ell}_{i,i+1}}{m}\,\text{mod}\, 1$  \\ & where $(m n_0-n)$ can not be a multiple of~$3$\\
 \hline
bulk ($b$) & 
$\,\in\, \mathcal{M}_{b}(\lambda)$ for some~$\lambda$\\ \hline
auxiliary bits~$(b^\prime)$ &  $-\frac{(m n_0 -n)}{m}\, \frac{2}{3}\,+\, \frac{n\widetilde{\ell}^{\prime}_{i,i+1}}{m}\,\text{mod}\, 1$  
\\ [1ex] 
\hline
\end{tabular}
\captionof{table}{The positions of vector, chiral, auxiliary bits and the bulk. The seeds in positions~$b$ are to be integrated over the bulk of the original integration contour. They define the integration variables entering, for instance, in the integrals~\eqref{IntegralBulk}. The factors of $I(v_{ij})$ that enter in~\eqref{IntegralBulk} are defined by the relations depicted in figure~\ref{fig:IndepVars}, the sum-rules~\eqref{assignationRules} and~\eqref{LastRule}. They are the ones labelled by the~$(i,j)$ that are assigned to the letter~$b$. These factors~$I(v_{ij})$ do not contribute to the bit integral~\eqref{IntegralBlocks}.}
\label{PositionsBitsIntro}
\end{center}
where~$\ell_{i,i+1}, \widetilde{\ell}_{i,i+1}\,,\,\widetilde{\ell}^{\prime}_{i,i+1}={0,\ldots\,,\, m-1}\,$. The positions~$b$ are integrated over the bulk of the contour, namely they are integrated out in the initial step~\eqref{IntegralBulk}. 

Once the position of each of the~$N-1$ \emph{seeds}~\eqref{BitsConditionIntroduction} is fixed to either $``v"= $ vector bits, $``c"= $chiral bits,~$``b"= $bulk, or~$``b^{\prime}"=$ auxiliary bits, the positions of the other~$v_{i,j}$'s with~$j\neq i+1$ follow from the relations depicted in figure~\ref{fig:IndepVars}, or equivalently~\eqref{eq:VariablesConvenient}, and the commutative (closed set of) sum rules
\be\label{assignationRules}
\begin{split}
v+v\, \to\, v\,,
v + c\, \to\, c\,, v+b^\prime \,\to\, b^\prime\,,  \\
c+c \to\,b^\prime \,,\,b^{\prime}+b^{\prime}\,\to\, c\,\,,\,c+ b^\prime \,\to\, v\,,
\end{split}
\ee
which follow from taking linear combinations of elements in table~\ref{PositionsBitsIntro}, and at last, 
\be\label{LastRule}
v+b\to b\,,\, c+b\to b\, , \, b^{\prime}+b\to b \,, \, b\,+\,b\,\to\, b.
\ee
The rules~\eqref{LastRule} follow from the definition of~$\mathcal{M}_{b}(\lambda)$ given in~\eqref{def:MbLambda}, and the previous rules~\eqref{assignationRules}. The contributions coming from the factors corresponding to bits, can then be collected in the form given in the integrand of~\eqref{IntegralBlocks}.

\paragraph{On the simple bulk~$\mathcal{M}_b^{(\lambda)}$ and bit~$\mathcal{M}_{\varepsilon}^{(\widetilde{\lambda})}$ regions}

Placing seeds at the bulk~$b$ breaks the~$SU(N)$ gauge symmetry into smaller subgroups. Fixing one seed in the middle of the diagonal to~$b$ and the remaining ones as vector bits~$v$, generates a number of off-diagonal~$b$'s that splits the net of~$v_{i,j}$'s into two disconnected subsets of vector bits, isomorphic to the set of weights of~$SU(\widetilde{\lambda}_1)$ and~$SU(\widetilde{\lambda}_2
)$, respectively. We say then that this sector~$\alpha$ preserves $SU(\widetilde{\lambda}_1)\times SU(\widetilde{\lambda}_2)\,$. The seed that generates the bulk contributions~$b$ belong to a simple bit domain~$\mathcal{M}_{b}(\lambda=1)$. The seeds that generate the two disconnected subsets of~$v_{ij}$'s belong to simple bit domains~$s\mathcal{M}_\varepsilon^{(\widetilde{\lambda}_i)}$ with $i=1,2\,$. The coordinates in the Cartan torus of~$SU(\widetilde{\lambda}_1)\,\times\,SU(\widetilde{\lambda}_2)$ are in one-to-one relation with the coordinates of the simple bit domains~$s\mathcal{M}_\varepsilon^{(\widetilde{\lambda}_i)}\,$. 




    More general symmetry breaking patterns are
possible. For example, if we locate the pink variable~$v_{3,4}$ in the diagram~\ref{fig:IndepVars} at the bulk, i.e. $v_{3,4}\in \mathcal{M}_{b}(\lambda=1)$ , then the possible symmetry-breaking patterns are~$SU(6)\,\to\, SU(3)\times SU(3\,)$ or~$SU(6)\,\to\, SU(3)\times SU(2)\times SU(2)$ or~$SU(2)^4$. If both, green and yellow elements are located at vector bits, then integrating out the seed variable in the bulk $v_{3,4}$ leaves a factorized partition function~\eqref{IntegralBlocks} with two decoupled~$SU(3)$ Chern-Simons partition functions on Lens spaces out of the integral~\eqref{IntegralBlocks}. 

On the other hand, if the green and yellow factors in the diagonal correspond to vector~$v$ and chiral bits~$c$, respectively, still the integral~\eqref{IntegralBlocks} factorizes in two pieces, but now the corresponding~$\alpha$ is slightly less gauge symmetric as it preserves an~$SU(3)\times SU(2)\times SU(2)$ symmetry. This further symmetry-suppression is due to the fact that the rule~$c+c= b^\prime \neq c$ breaks gauge invariance, i.e., to the fact that some of the gauge transformations within the~$SU(3)$ ones that map the yellow elements among themselves map the two chiral seeds into an auxiliary one in the off-diagonal position; only an~$SU(2)\times SU(2)\subset SU(3)$ does so. At last, if both, yellow and green seeds are located at chiral bits then the corresponding integral~\eqref{IntegralBlocks} factorizes in two factors, both of them being the partition function of A-twisted theories with gauge group~$SU(2)^2$ and matter charged with respect to the two~$SU(2)$ factors.
\begin{figure}[h]\centering
\includegraphics[width=7cm]{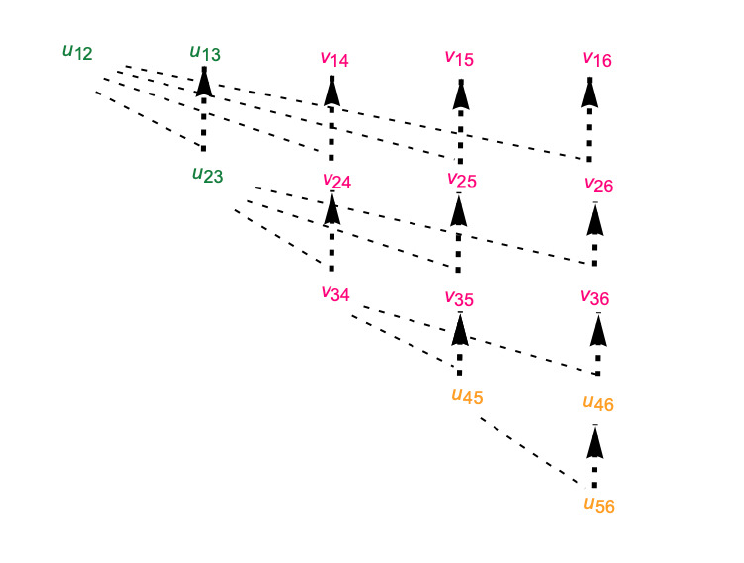} \includegraphics[width=7cm]{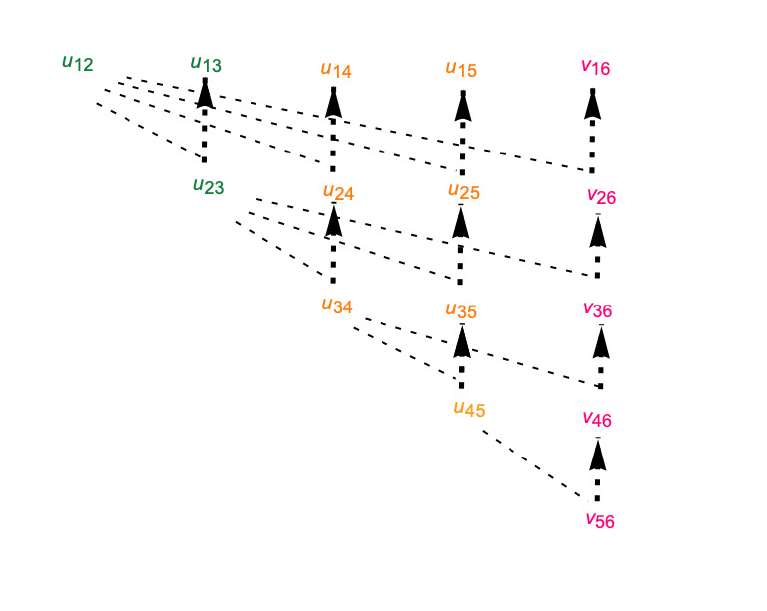}  
\caption{Two symmetry-breaking patterns. The~$v's$ denote variables in the bulk~$b$ which are integrated out trivially in the limits to root of unity. The~$u$'s denote variables along either vector, chiral, or auxiliary bits. Once a distribution of colours is given to the seed elements in the diagonal the colours of the elements in the off-diagonal are implied by the relations~\eqref{assignationRules}. Chiral and auxiliary walls, which are inverse elements to each other are represented with the same colour.}
\label{fig:IndepVars}
\end{figure}

If the seed variable~$v_{56}$ is the one assigned to the bulk and integrated out, then if green and yellow seeds are assigned to vector bits, the integral~\eqref{IntegralBlocks}
reduces to the partition function of~$SU(5)_{k=5}$ Chern-Simons theory over Lens spaces up to spurious factors that contribute to the Casimir pre-factor. On the other hand if the green and yellow seeds correspond to  vector and chiral bits, respectively, then the integral~\eqref{IntegralBlocks} does not factorize. In that case, we say that the corresponding theory is a \emph{coupled product} of an~$SU(3)$ Chern-Simons theory and an A-twisted model. 

 Summarizing, there are various possibilities for~$\alpha$, but only a finite number of them at a fixed~$N$. They can include, for instance, the partition function of multiple decoupled copies of three-dimensional Chern-Simons, and topologically twisted theories. More generally, there are also sectors in which the multiple copies of Chern-Simons and the A-twisted theories do not decouple. Meaning, that the integral over~$\Gamma_\alpha$ does not factorize into products of partition functions of Chern-Simons (the~$\mathcal{V}_m$ blocks) and/or Chern-Simons-matter theories (the~$\mathcal{C}_m$ blocks). We leave for the future a more detailed study and   of such~$\alpha$'s.~\footnote{It would be also interesting to explore whether this symmetry-breaking classification relates to the classification of vacua of~$\mathcal{N}=1^*$ on~$\mathbb{R}^{3,1}$ of~\cite{Donagi:1995cf}\cite{Dorey:1999sj}, which has been conjectured to correspond to Bethe roots of the~$SU(N)$~$\mathcal{N}=4$ index~\cite{Benini:2021ano,ArabiArdehali:2019tdm}.}

\subsection{An example: The maximally symmetric sector~$\alpha$}\label{MaximallySymmetricSector}

As mentioned before, the maximally symmetric sector~$\alpha$ corresponds to selecting elements only in the first row of~\eqref{PositionsBitsIntro}. In that case the conditions~\eqref{BitsConditionIntroduction} take the form
\be
v_{i\, j} \=( \frac{n}{m}\,{\ell}_{ij}) \,\text{mod} \,1\,, \qquad
\ee
where
\be
\ell_{i j}\,=\,\sum_{k=i}^{j}\ell_{i,i+1}\,.
\ee
In the expansion near-roots of unity these bits contribute to the index as follows
 \be
e^{-(N^2-1)\,\Bigl( \frac{\pi\i}{27 m\widetilde{\tau}^2}+\frac{2 \pi\i }{9 m\widetilde{\tau}}\,+\,\pi\i c_0(m,n)\,+\,\frac{8\pi\i}{27 m}\T\Bigr)}\,\times \,N\,\times\,\overline{Z}^{\alpha}\,,
\ee
where from the master formulas~\eqref{AveragesPartitionF} and~\eqref{IntegralBlocks}, it follows that~$\overline{Z}^{(\alpha)}$ equals the average
 \be\label{ZMaximalBits}
\sum_{{\ell}_1\,,\,\ldots \,,\,{\ell}_{N-1}=0}^{m\,-\,1}\,e^{\pi \text{i} k \sum_{i=1}^{N-1} \Phi(\ell_{i,i+1
})}\,Z^{L(m,1)}_{SU(N)}(k,{{\ell}})\,.
 \ee
with~\eqref{ZMaximalBits}
\be\label{PartionFunctionVectorBitsIntro}
Z^{L(m,1)}_{SU(N)}(k,{{\ell}})\,:=\,\int \prod_{i=1}^{N-1} du_i\, e^{\, \pi\i k \sum_{i=1}^{N-1} \,m u_{i}^2\,+\,2{{\ell}_{i} u_{i}}} \prod_{i \,>\, j\, =\,1}^{N} \Bigl( 2 \sin{\pi(u_{ij})}\, 2\sin{\pi(u_{ij})} \Bigr)\,. \ee
where~$u_{ij}=\sum_{k=i}^{j}u_{k,k+1}$ and~$u_i$ and~$\ell_i$ variables are defined by the relations
\be
\begin{split}
 \text{For } 1\leq\,i \,<\,N-1\,:& \qquad u_{i,i+1}=:u_i-u_{i+1}\,, \qquad \ell_{i,i+1}=:\ell_i-\ell_{i+1}\\
 \text{For } \qquad i \,=\,N-1\,:  & \qquad u_{i,i+1}=:2 u_{N-1}\,+\,\sum_{k=1}^{N-2}u_k\,,\,\qquad  \ell_{i,i+1}=:2 \ell_{N-1}\,+\,\sum_{k=1}^{N-2}\ell_k\,.
\end{split}
\ee
\eqref{PartionFunctionVectorBitsIntro} is the Chern-Simons partition function over Lens spaces~$L(m,1)$  (in some normalization), in the presence of background $\mathbb{Z}^{N-1}_m$ flat connections ${\ell}_i\sim {\ell}_i+m$, and quantum corrected level~$k$ such that~$|k|= N\,$~\cite{Brini:2008ik,2002math......9403H}.

\subsection{Outline of the remaining part of the paper} 
The remaining part of the paper deepens into the details of the conclusions just presented.

Subsection~\ref{subsec:Reps} introduces the representations~\eqref{IntegrandSoloInitial}. Subsection~\ref{subsec:VectorWalls} introduces the concept of vector walls. Subsection~\eqref{sec:ChiralBits} introduces and develops the concept of chiral walls (which we recall include the vector walls as limiting cases). Section~\ref{sec:ChiralBits} computes all the elements summarized in equation~\eqref{LeftAndRightBlocks}, which are, essentially, the main technical insight in this paper, the one that leads to the master formula~\eqref{AsymptoticTales}.

Subsection~\ref{Subsec:SUIndex} studies the maximally symmetric sector $\alpha$ for~$SU(2)$ gauge group. Subsection~\ref{UNVectorBits} computes the maximally-symmetric bit contributions ($\ell_i=0$) for gauge group~$SU(N)$: This is a particular case of the general result presented in subsection~\eqref{PartionFunctionVectorBitsIntro}.

Subsection~\ref{subsec:35} studies particular examples of symmetry-breaking sectors~$\alpha$ with only vector bit contributions. In particular, it details how the integration over bulk variables does not backreact the integration over bits. Section~\ref{ChiralWalls} studies examples of chiral bit contributions and explains how they reduce to sum over vacua of an underlying A-model. 

In section~\ref{Sec:Future} some prospective open questions and observations are presented. Further supporting material can be found in the appendices.

\section{The index near roots of unity}
\label{sec3.1}

In this section we show how the non-analytic walls emerge in the limits to roots of unity of the index~$\mathcal{I}\,$. The goal is to introduce from scratch the concept of bit and bulk. 

\subsection{The index and useful representations}\label{subsec:Reps}

The~$SU(N)$ superconformal index can be represented as the following integral
\be\label{N4SYM}
\begin{split}
\mathcal{I}&\,\equiv\,\kappa\,\oint_{|\zeta|=1} \prod_{i=1}^{\text{rk}(G)}{\frac{d\zeta_i}{\zeta_i}}\, e^{-S_\text{eff}}\,\equiv\, \kappa \,\oint_{|\zeta|=1} \prod_{i=1}^{\text{rk}(G)}{\frac{d\zeta_i}{\zeta_i}}\,  I_{\text{v}}\,I_c\,.\, 
\end{split}
\ee
The pre-factor~$\kappa$ is defined as
\be\label{Kappa0}
\begin{split}
\kappa\quad&\,\equiv\, \frac{(p;p)^{\text{rk}(G)}(q;q)^{\text{rk}(G)}}{N!}\,
\Bigl(\,{\prod_{I=1}^{3} \G_{\text{ell}}(t_{I};\,p,\,q)}\Bigr)^{\text{rk}(G)}\,.
\end{split}
\ee
Moreover
\be\label{IntegrandSCI}
\begin{split}
{I}_v(v)&\,\equiv\, \prod_{i\,<\,j=1}^{N} \, \theta_{\text{ell}}(\frac{\zeta_{i}}{\zeta_j};p)\, {\theta_{\text{ell}}(\frac{\zeta_{j}}{\zeta_i};q)}\,,\qquad
{I}_c(v)\,\equiv\, \prod_{i,j=1\atop i \neq j}^{N} \,\,{\prod_{I=1}^{3} \G_{\text{ell}}(\frac{\zeta_{i}}{\zeta_j}\,t_{I};\,p,\,q)}\, .
\end{split}
\ee
At some points we will use~$G=U(N)$ but our scope is~$SU(N)$. For~$G=SU(N)\,$
\be
\zeta^{-1}_N=\prod_{i=1}^{N-1}\zeta_{i}\,,\qquad\,\zeta_i\,=\,\e(v_i)\,.
\ee
To recover the~$SU(N)$ index from the~$U(N)$ one we use the identity
\be\label{SUNVSUN}
\mathcal{I}_{SU(N)}=\mathcal{I}_{U(N)}/\mathcal{I}_{U(1)}\,,
\ee
where
\be\label{Kappa}
\mathcal{I}_{U(1)}\,\equiv\,(p;p)\,(q;q)\,
\Bigl(\,\prod_{I=\textbf{1}}^{\textbf{3}} \G_{\text{ell}}(t_{I};\,p,\,q)\Bigr)\,.
\ee
At some points we will assume
$p=q=\e(\tau)\,$,~$t_{\textbf{1}}=t_{\textbf{2}}=t_{\textbf{3}}=q^{\frac{2}{3}}\,\e{\bigl(-\frac{n_0}{3}\bigr)}\,$,~\footnote{In this paper~$\textbf{e}(x)\,:=e\,^{2\pi\text{i}x}\,$.} and~$n_0=-1,0,1\,$.~\footnote{But many more general cases $p\neq q$ can be recovered from our discussion.}
In that case the index is a function of~$q$ and~$n_0\,$. Eventually we will fix~$n_0=-1\,$. Let us also define
\be
\theta_0(z;\tau)\,\equiv\,\theta_{\text{ell}}(\zeta;q) \,,\qquad \Ge(z;\tau)\,\equiv\,\Gamma_{\text{ell}}(\zeta;q,q)\,, \qquad \zeta\,=\,\e(z)\,.
\ee
The usual representations of~$\theta_{\text{ell}}$ and~$\Gamma_e$ are given in~\eqref{ThetaDef} and~\eqref{GammaeDef}. Here we will rely on the following set of representations that were originally put forward in~\cite{Cabo-Bizet:2019eaf}
\begin{align}\label{IdentityTheta00}
\log \theta_0(z)&\,\equiv\, \pi\, \text{i}\,\sum_{\ell=0}^{m-1} B_{2,2}(\xi_{\ell}|\T,-1)\,+\,L_{\theta_0}(z)\,,
\end{align}
and
\be \label{repmnG00}
\begin{split}
\log \Gamma_e(z;\, \tau) & \,=\, \sum_{\ell=0}^{2 (m-1)}  \frac{\pi i}{3}\,(m-|\ell-m+1|)\, B_{3,3}(\xi_{\ell}|\T,\T,-1)\,+\, L_{\Gamma_e}(z)\,,
\end{split}
\ee\footnote{The first term in the right-hand side of this equation equals the term~$-\pi \text{i}\frac{ R^{(3)(m,n)}(z)}{m (m\tau+n)^2}$ in the exponent of~\eqref{IntegrandSoloInitial}.}
where
\be
\T\,\equiv\, m \,\tau\,+\,n\,,
\ee
and
\begin{equation}\label{DefinitionZetaLTau0}
\xi_{\ell}\,=\, \xi_{\ell}(z;\tau)\,=\, \xi_{\ell}(z)\,\equiv\,z\,-\,\text{Integer}\,+\,\ell \tau\,.
\end{equation}
The definition of the Integer will be given below. These representations are convenient to study the expansion around roots of unity.

The objects~$L_{\theta_0}$ and~$L_{\Gamma_e}$, which will play an important role, are defined as
\be
\begin{split}
  L(z)&\,\equiv\,L_{\theta_0}(z)\,=\,\LTotalz\,,  \\  L_{\Gamma_e}(z)&\,\equiv\,  \sum_{\ell=0}^{2 (m-1)} \i\,(m-|\ell-m+1|)\,\Bigl(\,\sum_{j=1}^{\infty}{  \frac{-\,2\, \T +2 \xi_{\ell}+1 }{2 \,j\, \T\, \sin \left(\frac{\pi  j}{\T }\right) }}\,\times\\&\times\,
	 \cos{  \frac{\pi j (2\xi_{\ell}+1)}{\T }}\,- \, \frac{\pi  j \cot \left(\frac{\pi  j}{\T }\right)+\T }{2 \pi\,  j^2\, \T \, \sin \left(\frac{\pi  j}{\T }\right)} \,
	\sin{ \frac{\pi  j (2\xi_{\ell} +1)}{\T}}\,\Bigr) \,.
\end{split}
\ee
We will assume~$m>0$ and~gcd$(m,n)=1\,$. In particular, that implies that if~$n=0\,$, there is a single choice~$m>0\,$, which is~$m=1\,$. For other values of~$n$ there are other choices of~$m\,$. 

The integer in~\eqref{DefinitionZetaLTau0} is selected by the condition
\begin{equation}\label{ConditionNoLog0}
-1<\xi_{\ell \perp \T}<0\,.
\end{equation}
The real numbers~$\xi_{\ell ||}\,\equiv\,\xi_{\ell || \T}\,$, and~$\xi_{\ell \perp}\,\equiv\,\xi_{\ell \perp \T}\,$, are the components of the complex number~$\xi_{\ell}$ in the basis of the complex plane defined by~$1$ and~$\T\,$, i.e.~$\xi_{\ell}=\xi_{\ell\perp}\,+\,\T\, \xi_{\ell||}\,$.

Representations~\eqref{IdentityTheta00} and~\eqref{repmnG00} are absolutely convergent, and thus uniformly convergent, for generic~$z\in\mathbb{C}\,$, except for:
\begin{itemize}
\item[1)] At the~$z$'s for which~$\xi_{\ell\perp}$ hits the boundaries of the region~\eqref{ConditionNoLog0}. These regions are lines in the complex~$z$-plane that we will call~\emph{walls}. As we will see below, the lateral limits of both~$L(z)$ and~$L_{\Gamma_e}(z)\,$, to the walls~\eqref{ConditionNoLog0}, converge, except for at isolated points.  For~$L(z)$, the lateral limits, when they exist, coincide. That is not the case for~$L_{\Gamma_e}(z)$, where the lateral limits, when they exist, are different.
 
\item[2)] At the~$z$'s that correspond to the zeros of~$\theta_0\,$, and the zeroes and poles of~$\Gamma_{e}\,$, where the series diverges.
\end{itemize}
The divergences are always at positions~$z$'s for which~$\xi_{\ell\perp}$ hits the boundaries of~\eqref{ConditionNoLog0}. They are the points in the walls where both lateral limits diverge. Thus, in a sense,~$2)$ is included in~$1)$. 

\begin{figure}[h]\centering
\includegraphics[width=7cm]{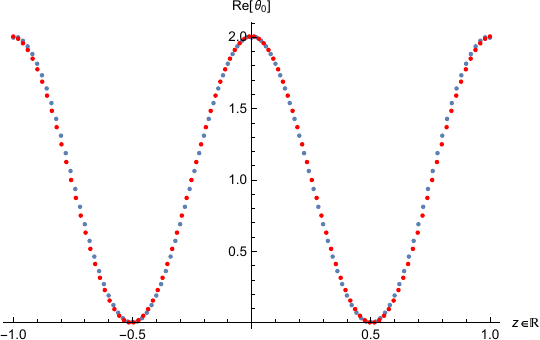} 
\includegraphics[width=7cm]{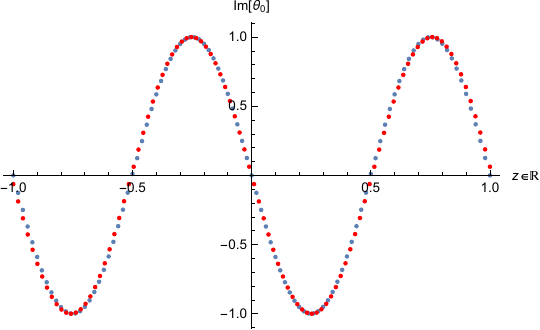} 
\caption{The plot of Real and Imaginary parts of~$\theta_0(\frac{1}{2}(2\tau+1)+z;\tau)$ for~$z\in [-1,1]$ and~$\tau=\text{i}\,$. The blue dots come from the representation~\eqref{IdentityTheta00} for~$m=2$ and~$n=1$, truncated at the element~$j=100$.  The red dots come from the product representation~\eqref{ThetaDef} truncated at the element~$j=20\,$.}
\label{fig:RepMNTheta}
\end{figure}

For generic complex~$\zeta$ away from the walls, both~$L$ and~$L_{\Gamma_e}$ vanish exponentially fast in the limit~$\tau\to-\frac{n}{m}\,$. In such a limit, only the piece-wise polynomial part coming from the~$B_{2,2}$'s and~$B_{3,3}$'s is relevant. In contradistinction, for complex~$z$ infinitesimally close to the walls, the series contributions do not vanish as~$\tau\to-\frac{n}{m}\,$, producing a finite non-piece-wise polynomial and non-analytic~$\mathcal{O}(1)$ correction to the effective action as function of~$z\,$. Such a correction diverges only at the zeroes of~$\theta_0\,$, and at the zeroes and the poles of~$\Gamma_e\,$. In the case of~$\Gamma_e$, besides the zeroes and poles, a branch cut opens up for~$L_{\Gamma_e}$ along the walls; however, as expected, the jump is cancelled by an opposite branch cut contribution coming from the piece-wise polynomial pre-factor. We will prove this last statement below.

\subsection{$\theta_0(z)$ for~$q$ near roots of unity: (Vector) walls of non-analyticity} \label{subsec:VectorWalls}
Let us study~$\theta_0(z)$ at a generic point in the complex~$z$-plane, when~$q$ approaches a root of unity. For this it is convenient to use the representation~\eqref{IdentityTheta00}. Specifically
\be\label{SubtlePiece}
L(z)\,\equiv\,\text{i}\sum_{j=1}^{\infty} \frac{1}{j \sin \frac{\pi j}{\T}}\sum_{\ell=0}^{m-1} \,\cos\Bigl(\pi j\,\frac{2\xi_{\ell}+1}{\T} \Bigr)\,,
\ee
with
\be
\xi_{\ell}(z) = z+\ell \tau-\lfloor z_{\perp}- \ell \frac{n}{m}\rfloor -1\,.
\ee
For generic complex~$z$ we define the real components~$z_{\perp,\,||}$, as the two real numbers defined by the condition
\be
z \=  z_{||}\, \T\,+\,z_{\perp}\,.
\ee
Using these definitions, one finds that
\be
\xi_{\ell||}(z)\= z_{||}\,+\,\frac{\ell}{m}\,,\,\quad \xi_{\perp}(z)\= z_{\perp}\,-\,n\frac{\ell}{m}\,-\,k_0\,,
\ee
where for~$z_{\perp}\,-\,n\frac{\ell}{m}\,\notin\,\mathbb{Z}\,$
\be
k_0\,\equiv\,\lfloor z_{\perp}\,-\,n\frac{\ell}{m} \rfloor\,-\,1\,.
\ee
A computation shows that for generic complex~$z$
\be\label{GenericLimit}
\exp(L(z))\,\underset{\tau\to -\frac{n}{m}}{\longrightarrow}\, 1\,.
\ee
However, for~$z_{\perp}\,=\,0\,$, for example,~\eqref{GenericLimit} does not hold. The technical reason being the presence of the~$\ell=0$ term in the exponent of~\eqref{SubtlePiece}.

In that case, the contribution of~$e^{L(z)}$ to the even product~$\theta_{\text{ell}}(\zeta)\,\theta_{\text{ell}}(\frac{1}{\zeta})$ reduces to
\be\label{ContributionVector}
\log \Bigl((-1)\,\Bigl(2\sin{ \pi (z_{||})}\Bigr)\,  \Bigl(2\sin{ \pi (-z_{||})}\Bigr)\Bigr)\,,
\ee
where~$z_{||}$ is the distance from the origin to a point whose radial vector is parallel to~$\T\,$. For~$\tau\approx -\frac{n}{m}$ the long-range potential~\eqref{ContributionVector} only arises in a very narrow region of the direction defined by the radial vector associated to~$\T\,$. The width of the region vanishes in the strict limit~$\tau\to-\frac{n}{m}\,$. This is saying that the continuity (and thus analyticity) of
\be\label{LogL}
L_{+}(z)\,\equiv\,L(z)\,+\,L(-z)
\ee
in the complex~$z$-plane is broken in the Cardy-like expansion at order~$\mathcal{O}(1)\,$. The breaking occurs at the lines where a one-dimensional logarithmic profile~\eqref{ContributionVector} arises (See plot~\ref{fig:Bits}). 

In appendix~\ref{OtherBits} we demonstrate that there are many other such lines of non-analyticity emerging in Cardy-like limit. Their positions are fixed by the condition
\be
\chi_\ell(z)\,\equiv\,2\xi_{\ell\perp}(z) + 1\,=\,\pm\,1
\ee
with~$\ell\,=\,0\,,\,1\,,\,\ldots\,,\,m-1\, $. We call them (vector)~walls, where the~$p$ labels the position of the point at which they intersect the real axis. 
That point of intersection will be called~bit. The classification of bits for generic co-primes~$m$ and~$n$ can be found in appendix~\ref{OtherBits}. The contributions of bits to the original integral~\eqref{N4SYM} includes the ones argued in~\cite{ArabiArdehali:2021nsx} for the cases with~$n=1$. 
\begin{figure}[h]\centering
\includegraphics[width=10cm]{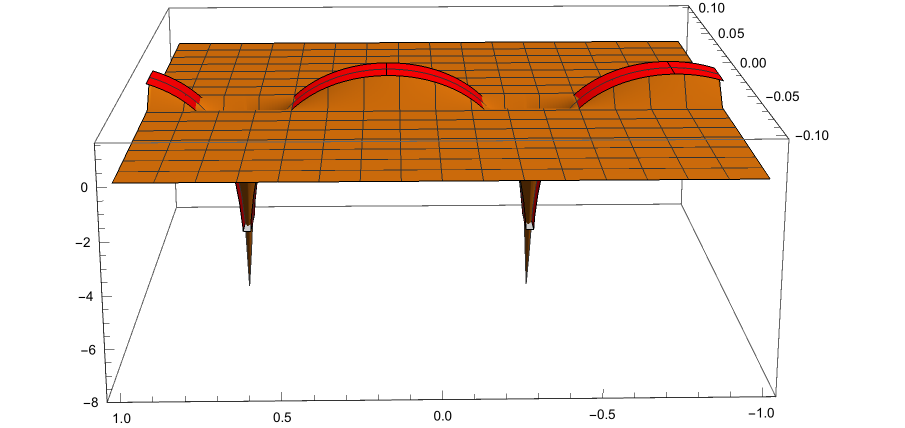}\vspace{1cm}
\includegraphics[width=10cm]{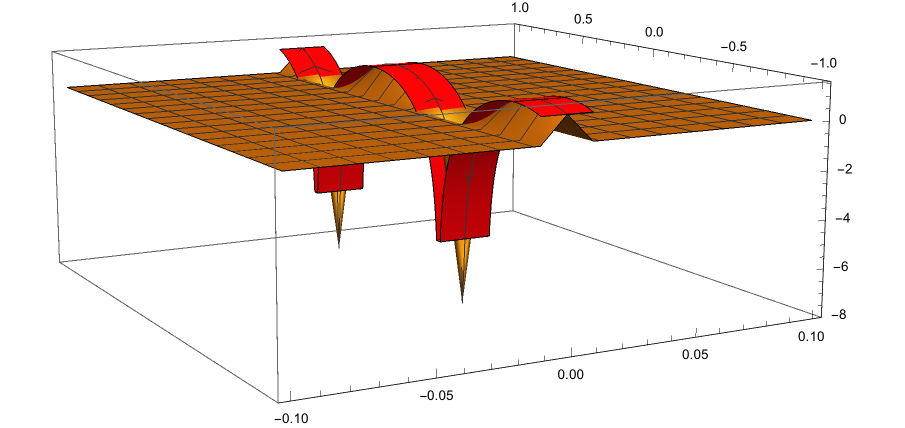}
\caption{An example of a wall of non-analyticity in Cardy-like limit~$\tau\to-\frac{1}{3}\,$.}
\label{fig:Bits}
\end{figure}
Figure~\ref{fig:Bits} plots~$\text{Re}(L(z)+L(-z))$ for~$(m,n)=(3,1)\,$, and~$\tau\approx-\frac{1}{3}\,$. The horizontal axes correspond to the real coordinates~$0\,<\,x_1\,<\,1$ and~$-0.1\,<\,x_2\,<\,0.1$ in~$z$-plane, with~$z\,=\,(x_1+\frac{1}{3})(3\tau+1)+x_2\,$.  We fixed~$\tau\,=\, 0.0005 \,(\i \,-\, 1)\, -\, 1/3\,$, which is close to the Cardy-like limit~$\tau\to-\frac{1}{3}\,$. To produce the orange plot we truncated the series in the right-hand side of~\eqref{IdentityTheta00} at $j=1000\,$, and evaluated it at square lattice with~$100\times 100$ points within the domain~$(x_1,x_2)$ covered by the plot. 
There are three points to remark. First, notice that the function vanishes almost everywhere, except for, at a thin strip located at~$x_2\,=\,0\,$. The width of the strip is of order~$|3\tau+1|$ for~$\tau$ close enough to~$-\frac{1}{3}\,$. Second, the profile of the limiting function along~$x_2\,=\,0$ i.e. along the ray~$z\,=\,(x_1+\frac{1}{3})(3\tau+1)\,$, is the one of the function~$\text{Re}\log{\Bigl(2 \sin{\pi (x_1 + 1/3)}\,2 \sin{\bigl(-\pi (x_1 + 1/3)}\bigr)\Bigr)}\,$. The profile of the red strip is the plot of the latter function. Third, note that the original contour of integration, which extends along the ray~$x_1\,=\,-\frac{1}{3}\,$, only intersects the non-vanishing region at a small segment around~$z=x_2=0$ that becomes a point in Cardy-like limit. At the intersection point the function diverges~$\text{Re}\Bigl(L(z)\,+\,L(-z)\bigr)\Bigr)$ to~$-\infty\,$. That was known a priori because~$\theta_0(z)\,\theta_0(-z)$ has a double zero at~$z=0\,$. One could naively expect that as the latter region becomes infinitesimally small in Cardy-like limit it is possible to approximate the theta functions with the exponential of the first piecewise-polynomial factor in~\eqref{IdentityTheta00}, and ignore the series contribution above plotted. This naive expectation turns out to be incorrect, because the integral along the infinitesimal region where that series contribution cannot be discarded, turns out  to give a finite and non-vanishing contribution in Cardy-like limit. Such contributions, i.e. contributions coming from the intersection points between non-analyticities and the contour of integration (the bits) will be called localized contributions.

\subsection{$\Gamma_{e}(z)$ for~$q$ near roots of unity: (Chiral) walls of non-analyticity}
\label{sec:ChiralBits}

There is another sort of localized contribution. These emerge in the Cardy-like limit of the elliptic Gamma functions~$\Gamma_e$ and include the residues associated with the poles of the analytic continuation of the integrand of~\eqref{N4SYM}. 

Let us use the following decomposition of~$L_{\Gamma_e}$
\be\label{DefLTilded}
\begin{split}
L_{\Gamma_e}&\=  \sum_{\ell=0}^{2 (m-1)} \i\,(m-|\ell-m+1|)\,L^{\ell}_{\Gamma_e} \,,  \\
&\=\sum_{\ell=0}^{m-1} \i\,\Bigl((\ell+1)\,L^{\ell}_{\Gamma_e} \,+\,(m\,-\,\ell\,-\,1)\,L^{\ell+m}_{\Gamma_e} \Bigr)\, \\
&\,\equiv\, \sum_{\ell\=0}^{m\,-\,1} \widetilde{L}^{\ell}_{\Gamma_e}\,,
\end{split}
\ee
where
\be
L^{\ell}_{\Gamma_e}\,\equiv\,L^{\ell(1)}_{\Gamma_e}+L^{\ell(2)}_{\Gamma_e}+L^{\ell(3)}_{\Gamma_e}+L^{\ell(4)}_{\Gamma_e}\,,
\ee
and
\be\label{Lell}
\begin{split}
L^{\ell(1)}_{\Gamma_e}&\=    \, \sum_{j=1}^{\infty}{  \frac{-\,1 }{ \,j\,  \sin \left(\frac{\pi  j}{\T }\right) }}\,
	 \cos{  \frac{\pi j (2\xi_{\ell}+1)}{\T }}     \\
L^{\ell(2)}_{\Gamma_e}& \=  \sum_{j=1}^{\infty}{  \frac{2 \xi_{\ell}\,+\,1 }{2 \,j\, \T\, \sin \left(\frac{\pi  j}{\T }\right) }}
	 \cos{  \frac{\pi j (2\xi_{\ell}+1)}{\T }}      \\
L^{\ell(3)}_{\Gamma_e}&\=      \sum_{j=1}^{\infty} \, \frac{-\,\pi  j \cot \left(\frac{\pi  j}{\T }\right) }{2 \pi\,  j^2\, \T \, \sin \left(\frac{\pi  j}{\T }\right)} \,
	\sin{ \frac{\pi  j (2\xi_{\ell} +1)}{\T}}\,     \\
L^{\ell(4)}_{\Gamma_e}&\= \sum_{j=1}^{\infty} \, \frac{-1 }{2 \pi\,  j^2\, \, \sin \left(\frac{\pi  j}{\T }\right)} \,
	\sin{ \frac{\pi  j (2\xi_{\ell} +1)}{\T}}\,.
\end{split}
\ee
As for~$L\,$, in the limit of~$q$ to roots of unity, these functions vanish exponentially fast almost everywhere in the complex $\xi_{\ell}$-plane except for at infinitesimally thin walls centred at the positions fixed by the condition
\be\label{EllipticWalls}
\chi_\ell(z)\,\equiv\,2\xi_{\ell\perp}(z)\, +\, 1 \=\pm\,1\,.
\ee
\begin{figure}\centering
\includegraphics[width=10cm]{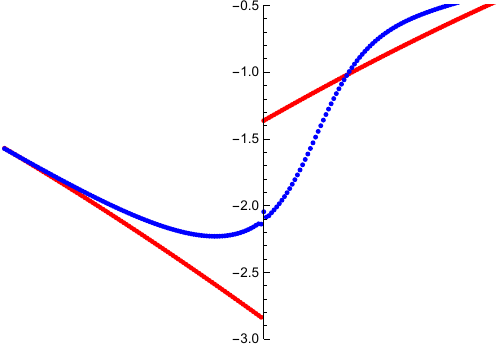}
\caption{ 
The blue points represent~$\log \Gamma_{e}(\xi_{\perp}+0.2\tau,\tau,\tau)$ for~$-0.1<\xi_{\perp}<0.1$ and~$\tau=\frac{\i-1}{10}\,$. We have used a truncation of~\eqref{repmnG00} at~$j=1000\,$. The red points represent the piece-wise polynomial part~$\frac{\pi i}{3}\, B_{3,3}(\xi_{\perp}-\lfloor\xi_{\perp}\rfloor+0.2 \tau |\tau,\tau,-1)\,$. The branch cut of the piece-wise polynomial part is cancelled by the corresponding series~$L_{\Gamma_e}$ in such a way that
~$\Gamma_{e}(\xi_{\perp}+0.2\tau,\tau,\tau)$ has no branch cut. 
}
\label{fig:BranchCuts}
\end{figure}
For later convenience, we note that assuming the integration variables~$v_i$'s to be real, i.e., $v_{2i}=0$, the condition~\eqref{EllipticWalls} translates, essentially, to requiring~$\rho(v)=\sum_i \rho^i v_i$~\footnote{...with~$\rho=\{\rho^i\}$ being a given gauge charge (a weight vector)} to approximate specific values, that we will quoted below in equation~\eqref{PositionChiralBits}.

To extract the contribution of~$L_{\Gamma_e}$ along the chiral walls, in Cardy-like limit, one needs to zoom into an infinitesimal region around the walls defined by~\eqref{EllipticWalls}. In particular, we need to understand the details of a double limit: the limit~$\tau\to-\frac{n}{m}$, together with each one of the lateral limits to the walls~\eqref{LimitTowalls}
\be\label{LimitTowalls}
\chi_\ell(z)\,\equiv\,2\xi_{\ell\perp}(z) \,+\, 1 \,\to\,\pm 1^{\mp}\,.
\ee
In contradistinction to the case of $L_{\theta_0}$, in the case of~$L_{\Gamma_e}$, the lateral limits to the walls do not match. 

A computation shows that the lateral double limits of~$L_{\Gamma_e}^{\ell(1)}\,$, $L^{\ell(2)}_{\Gamma_e}\,$, $L^{\ell(3)}_{\Gamma_e}\,$ are
\be\label{DoubleExpansions}
\begin{split}
L^{\ell(1)}_{\Gamma_e}&\, \underset{\tau\to -\frac{n}{m}}{\longrightarrow} \, -\,\i \,\sum_{j\,=\,1}^{\infty}\, \frac{\e{(\pm j \xi_{\ell||} )}}{j} \,,
\\
L^{\ell(2)}_{\Gamma_e}&\,\liMN \,\i\,\xi_{\ell||} \,\sum_{j\,=\,1}^{\infty}\, \frac{\e{(\pm j \xi_{\ell||} )}}{j}\,+\,L^{\ell(\T)}_{\Gamma_e}\,,
\\  L^{\ell(3)}_{\Gamma_e}\, &\liMN\,-\,L^{\ell(\T)}_{\Gamma_e} \,,\,\qquad\qquad\qquad\,,\\
L^{\ell(4)}_{\Gamma_e}&\, \underset{\tau\to -\frac{n}{m}}{\longrightarrow}\,\mp\,\sum_{j\,=\,1}^{\infty}\, \frac{\e{(\mp j \xi_{\ell||}})}{2 \pi j^2}
\end{split}
\ee
where
\be
L^{\ell(\T)}_{\Gamma_e}\=\pm\frac{\i}{\T} \,\sum_{j\,=\,1}^{\infty} \frac{\e{\Bigl(\pm\,j \xi_{\ell||}\Bigr)}}{2 j}\,.
\ee
Notice that for~$L^{\ell(2)}_{\Gamma_e}\,$ and~$L^{\ell(3)}_{\Gamma_e}\,$ the expressions in~\eqref{DoubleExpansions} are asymptotic expansions not really limits, however, for~$L^{\ell(2)}_{\Gamma_e}\,+\,L^{\ell(3)}_{\Gamma_e}\,$, which is what we need to move forward, the limit~$\tau\to-\frac{n}{m}$ is well-defined.

\paragraph{Cancellation of the branch cut}\label{CancellationBranchCuts}
Let us prove that the branch cuts of~$L^{\ell}_{\Gamma_e}$ are cancelled by branch cuts of the piece-wise polynomial contributions. Define
\be
W(\xi_{\ell||})\,\equiv\,W_1(\xi_{\ell||})\,+\,W_2(\xi_{\ell||})\,+\,W_3(\xi_{\ell||})\,,
\ee
where
\be
\begin{split}
-\i W_1(\xi_{\ell||})&\,\equiv\,L^{\ell(1)+}_{\Gamma_e}-L^{\ell(1)-}_{\Gamma_e}\=-\,2\pi\,B_{1per}(\xi_{\ell||}) \\
-\i W_2(\xi_{\ell||})&\,\equiv\,L^{\ell(2)+}_{\Gamma_e}+L^{\ell(3)+}_{\Gamma_e}-L^{\ell(2)-}_{\Gamma_e}-L^{\ell(3)-}_{\Gamma_e}\=  2\pi\,\xi_{\ell||} \,B_{1per}(\xi_{\ell||})\\ 
-\i W_3(\xi_{\ell||})&\,\equiv\,L^{\ell(4)+}_{\Gamma_e}-L^{\ell(4)-}_{\Gamma_e}\=\,-\,\, \pi \,B_{2per}(\xi_{\ell||})\,,
\end{split}
\ee
and for~$n\,\geq\,1$
\be
B_{n per}(\zeta)\,\equiv\,-\, \frac{n!}{(2\pi\i)^n }\,\sum_{j\,=\, -\infty\atop j\,\neq\,0}^{\infty}\, \frac{e^{2\pi\i j\, \zeta }}{j^n}\,.
\ee
The~$W(\xi_{\ell||})$ is the jump of~$L^{\ell}_{\Gamma_e}$ across the branch cut at~$\xi_{\ell\perp}\,=\,0\,\sim\,-1$, and it is a function of~$\xi_{\ell||}\,$.

For later convenience we note that for~$0\,<\,\xi_{\ell||}\,<\,1$
\be\label{Condi1}
-\i W(\xi_{\ell||})\=\pi\,\xi_{\ell||}^2\,-\,2\pi\xi_{\ell||}\,+\,\pi\frac{5}{6}\,.
\ee
To reconstruct the function~$W(\xi_{\ell||})$ outside the domain~$0\,<\,\xi_{\ell||}\,<\,1$ we just need to use the periodicity property
\be\label{Condi2}
e^{W(\xi_{\ell||}+1)-W(\xi_{\ell||})}\= e^{2\pi\i B_{1per}(\xi_{\ell||})}\=e^{2\pi \i B_{1}(\xi_{\ell||})}\,.
\ee
Notice that the periodic Bernoulli polynomial~$B_{1per}$ can be substituted by the ordinary Bernoulli polynomial~$B_1$ because the difference between the two exponentiates to the unity.

Next, define
\be
\widetilde{W}(\xi_{\ell||})\,\equiv\,\frac{\pi\i}{3}\, B_{3,3}\Bigl(\xi_{\ell||}\T+\xi_{\ell\perp}\,,\,\T,\,\T\,,\,-1\Bigr)\Bigl|_{\xi_{\ell\perp}=-1}^{\xi_{\ell\perp}=0}\,.
\ee
which is, by the definition, the jump of the piece-wise polynomial factor~$\frac{\pi\i}{3}\, B_{3,3}(\xi_{\ell}|\T, \T,-1)$ in~\eqref{repmnG00} across the branch cut located at~$\xi_{\ell\perp}\,=\,0\,\sim\,-1\,$ at the point at the cut fixed by a specific value of the coordinate~$\xi_{\ell||}\,$. Two computations show that
\be\label{Condi3}
\widetilde{W}(\xi_{\ell||})\=-\pi\i\,\xi_{\ell||}^2\,+\,2\pi\i\xi_{\ell||}\,-\,\pi\i\,\frac{5}{6}\,,
\ee
and that
\be\label{Condi4}
e^{\widetilde{W}(\xi_{\ell||}+1)\,-\,\widetilde{W}(\xi_{\ell||})}\=e^{-\,2\pi \i B_{1}(\xi_{\ell||})}\,.
\ee
Thus, after comparing~\eqref{Condi1} and~\eqref{Condi2} with~\eqref{Condi3} and~\eqref{Condi4}, respectively, one concludes that for every real~$\xi_{\ell||}$
\be
e^{\widetilde{W}(\xi_{\ell||})\,+\,{W}(\xi_{\ell||})}\=1\,.
\ee
This proves that in Cardy-like expansion, the branch cuts of~$L_{\Gamma_e}$ along the walls of non-analyticities are cancelled by the ones of the piece-wise polynomial contribution in~\eqref{repmnG00}, up to an irrelevant addition of an integer multiple of~$2\pi\i\,$. Note that the jump of the piece-wise polynomial part across the branch cut is independent of~$\tau$.
\footnote{Indeed, the jump of~$L_{\Gamma_e}$ across the branch cut is independent of the value of~$\tau$. That can be proven directly from definitions~\eqref{Lell}.}

In summary, effectively, along the walls we can use
\be
L^{\ell}_{\Gamma_e}(\chi_\ell\,=\,\pm 1)\, \underset{\tau\,\to\, -\,\frac{n}{m}}{\sim}\,\frac{L^{\ell}_{\Gamma_e}(\chi_\ell\,\to\, 1^-)\,+\,L^{\ell}_{\Gamma_e}(\chi_\ell\,\to\, -1^+)}{2}\,+\,\text{tbc}\,,
\ee
where~tbc (to be cancelled) stands for the branch-cut contributions that are cancelled by the piece-wise polynomial part of~\eqref{repmnG00}. We have substituted the limit symbol~$\rightarrow$ by~$\sim$ to make clear that the correspondence is not quite a limit of~$L^{\ell}_{\Gamma_e}(\chi_\ell\,=\,\pm 1)$.

Putting all the pieces together one gets
\be\label{IntermediateCHiralWall}
\begin{split}
L^{\ell}_{\Gamma_e}(\chi_\ell\,=\,\pm 1)&\, \underset{\tau\,\to\, -\,\frac{n}{m}}{\sim} \,-\,\Bigl(\xi_{\ell||}\,-\,1\Bigr) \,\sum_{j\,=\,1}^{\infty}\, \frac{\cos{\Bigl(2\pi j \xi_{\ell||} \Bigr)}}{j}\,+\,\sum_{j\,=\,1}^{\infty} \frac{\sin(2 \pi j \xi_{\ell||} )}{2 \pi j^2} \\&\qquad\qquad\qquad\qquad\,+\,\text{tbc}\\&\quad\,=\,\Bigl(\xi_{\ell||}\,-\,1\Bigr)\,\log \Bigl(2\,|\sin{\Bigl(\pi\, \xi_{\ell||} \Bigr)|}\Bigr)\,-\,\frac{\i}{4\pi}\Bigl( \text{Li}_2(e^{2\pi\i \xi_{\ell||}})-\text{Li}_2(e^{-2\pi\i \xi_{\ell||}})\Bigr)\\ &\qquad\qquad\qquad\qquad\,+\,\text{tbc}\,.
\end{split}
\ee
The first term carries the information about the poles and zeroes of the elliptic Gamma functions.

Using~$\xi_{\ell+m||}\= \xi_{\ell ||}\,+\,1\,$ and~\eqref{IntermediateCHiralWall} in the definition of $\widetilde{L}^{\ell}_{\Gamma_e}$ given in~\eqref{DefLTilded}, one obtains
\be
\begin{split}
\widetilde{L}^{\ell}_{\Gamma_e}(\chi_\ell\,=\,\pm 1)&\, \underset{\tau\,\to\, -\,\frac{n}{m}}{\sim}
\\\quad\,&\Bigl(m\,\Bigl(\xi_{\ell||}\,-\,1\Bigr)\,+\,(m\,-\,\ell\,-\,1)\Bigr)\,\log \Bigl(2|\sin{\Bigl(\pi\, \xi_{\ell||} \Bigr)}|\Bigr)\, \\ &\qquad \qquad\,-\,m\,\frac{\i}{4\pi}\Bigl( \text{Li}_2(e^{2\pi\i \xi_{\ell||}})-\text{Li}_2(e^{-2\pi\i \xi_{\ell||}})\Bigr)\,+\,\text{tbc}\,.
\end{split}
\ee
For a chiral multiplet with weights~$\rho=\{\rho^i\}$, the relation between the variable~$z$ and the Cartan components of the~\emph{hermitian eigenvalues}~$v_i$ is
\be\label{DefinitionZChiral}
z\,=\,\rho(v)\,+\,\Delta\,, \,\qquad\Delta\,=\Delta_2\,\tau\,+\,\Delta_1\,,
\ee
where
\be\label{IntroducingR}
\rho(v)\=\sum_{i} \rho^{i}\,v_i\,, \,\quad \Delta_1\,=\,-\,n_0\,\frac{r}{2}\,,\,\quad  \Delta_2\,=\,r\,.
\ee
Equation~\eqref{DefinitionZChiral} implies that along chiral walls
\be
\xi_{\ell||} = \rho(u)\,+\,\frac{\Delta_2+\ell}{m} \,, 
\ee
and~$\xi_{\ell\perp}$ is determined by condition~\eqref{LimitTowalls}. The classification of positions of chiral walls out the conditions~\eqref{LimitTowalls} is equivalent to the one of vector walls, which has been reported in appendix~\ref{OtherBits}.  The difference is an extra shift in the positions which is linear in the~$R$-charge~$r$. This is, the chiral walls intersect the real locus~$v_i\,\in\,\mathbb{R}$ at the positions defined by
\be\label{PositionChiralBits}
\begin{split}
\rho(v)&\=-\,n\,\frac{\Delta_2\,+\,\ell}{m}\,+\,\Delta_1 \,\text{mod}\, 1\,. \\
&\=\Bigl(-\,n\,\frac{\Delta_2}{m}\,+\,\Delta_1 \,+\,\frac{\ell^*}{m}\Bigr)\, \text{mod}\, 1\,,
\end{split}
\ee
~\footnote{ Whereas for the vector walls~$p=n\frac{\ell}{m}\text{mod}1$ and~$\ell=0,\ldots, m-1\,$.} where~$\rho$ can be any non-vanishing adjoint weight and
\be
\ell^* \,=\, ({n}\,\ell)\, \text{mod} \,m \= 0,1,\dots, m-1,
\ee
is a label that we can use, as an alternative to~$\ell$, in order to enumerate set of chiral walls or bits that intersect the original domain of integration (See more around~\eqref{DefLStar}).

\paragraph{On the solutions to~\eqref{PositionChiralBits}} \label{ChiralSymmBreaking} There is an observation regarding equation~\eqref{PositionChiralBits} that we discuss for later convenience. To fix a solution~$v^i_{0}$ with~$i=1,\ldots, N$ out of the set of equations~\eqref{PositionChiralBits} one needs to fix~$N-1$ roots~$\rho$'s.  The ansatz~$v^i_0$ does not necessarily solve~\eqref{PositionChiralBits} for the remaining non-vanishing roots~$\rho$.
For instance, there are chiral bit solutions for the following sets of~$U(N)$ roots
\be\label{SetsWeights}
\textbf{R} \=\bigl\{\rho:=e_{i}-e_{i-1}\bigr\}_{i=1,\ldots, N}\,,
\ee
where~$e_0:= e_N$\,. This set of weights defines the \emph{seeds} introduced in section~\ref{Seeds}.
In this definition~$e_i$ is the $N$-vector with a single non-vanishing -- and unit -- component, the~$i$-th one. For generic values of~$m$,~$n$,~$\Delta_1$ and~$\Delta_2$, the set of weights~$\textbf{R}$ and permutations thereof form the largest set of~$\rho$'s that can allow for~\eqref{PositionChiralBits} to be solvable.

Generically, for the values of~$v$ that solve~\eqref{PositionChiralBits} with~$\rho\in \textbf{R}\,$, the weights~$\rho^\prime \notin \textbf{R}$ are such that~$\rho^\prime(v)$ does not belong nor to chiral nor to vector walls.~{\label{ChiralSubtle}Note that this assumption does not hold if~
\be\label{ConditionMixing}
(- n\,\Delta_2\,+\, m \,\Delta_1) \,\times\,l \,\in\,\mathbb{Z}
\ee
for some integer~$0\,\leq\, l\,<\,N\,$. We highlight this point because \eqref{ConditionMixing} is satisfied by the usual choice for~$\Delta_{1}$ and~$\Delta_{2}$,~ $\Delta_1\,=\,-\,n_0\,\frac{r}{2}\,,\,\quad  \Delta_2\,=\,r\,=\,\frac{2}{3}$ for large enough values of~$N$, more precisely, for~$N>3\,$. The important thing to recall, is that in these cases there will be some of the roots~$\rho^\prime$ for which contributions from vector bits mix with chiral bits as summarized in the section~\ref{SymmetryBreakingBits}, and in virtue of the sum-rules~\eqref{assignationRules}.}

For example the chiral bit defined by the choices
\be 
m=n+1=1\,,\qquad \ell=\ell^\star=0\,\quad \text{and}\quad \rho \,\in\,\textbf{R}\,,
\ee
is located at the position dicated by the~$v_i\in\mathbb{R}$ that solve the equations
\be
v_{i j}\= \Delta_1\,\text{mod}\,1, \qquad j = i+1\, \qquad i=1,\ldots, N-1\,.
\ee
which can be parametrized in the form
\be
v_i \= i\, (\Delta_1) \,\text{mod}\,1 \,+\,\text{constant}.
\ee
The constant can be fixed by demanding the tracelessness constraint~$\sum_{j=1}^N v_i\,=\,0$~mod~$1$.

\paragraph{Partial summary} Along the chiral walls defined by the condition~\eqref{EllipticWalls}, in the limit~$\tau\to-\frac{n}{m}\,$, the profile of~$L_{\Gamma_e}$ takes the form
\be\label{ProfileText}
\begin{split}
\Bigl(m\,\Bigl(\xi_{\ell||}\,-\,1\Bigr)\,+\,(m\,-\,\ell\,-\,1)\Bigr)\,\log \Bigl(2|\sin{\Bigl(\pi\, \xi_{\ell||} \Bigr)|}\Bigr) \\
\,-\,m\,\frac{\i}{4\pi}\Bigl( \text{Li}_2(e^{2\pi\i \xi_{\ell||}})-\text{Li}_2(e^{-2\pi\i \xi_{\ell||}})\Bigr)\,\,+\,\text{tbc}
\end{split}
\ee
with
\be\label{ParallelVariable}
\xi_{\ell||}\=\rho(u)\,+\,\frac{\Delta_2\,+\,\ell}{m}\,.
\ee
The real variable~$\rho(u)\,=\, (\rho(v) -\rho(v^{(0)}) )/\T$ runs along the wall that intersects the contour of integration at the position~$\rho(v) = \rho(v^{(0)})$ that solves the linear condition~\eqref{PositionChiralBits}, not along the original integration contour. Equivalently, for~$\xi_{\ell\perp}$ not in a small enough vicinity of a wall, i.e. for $\rho(v)\not\approx \rho(v^{(0)})$ the series~$L_{\Gamma_e}$ vanishes exponentially fast in the limit~$\tau\to -\frac{n}{m}\,$. Note that~\eqref{ProfileText} contains information about the poles of the~$\Gamma_e$'s in the integrand of~\eqref{N4SYM}, that is, for negative enough~$\xi_{\ell||}$ the exponential of~\eqref{ProfileText} has poles coming from the term in the first line.

Although the chiral walls intersect the contour of integration at infinitesimal segments (the bits), the integral along such segments gives a finite and non-vanishing contribution in Cardy-like expansion. How this happens will be explained in the following section with an explicit example of vector bits. The analysis for chiral bits is analogous.

\paragraph{Chiral wall contributions are suppressed} The profile of the integrand~$\mathcal{I}(q)$ along the vector and chiral walls receives piece-wise polynomial contributions that must be added to~$L$, and $L_{\Gamma_e}$, respectively. The exponential of such contribution divided by the Casimir pre-factor of the maximally symmetric secto~$\alpha$, is, for~$G=SU(N)$
\be\label{RelWeight}
\begin{split}
&e^{-(\sum_{\rho>0}2\mathcal{F}^{(m,n)})\,+\,\frac{(\text{dim}G-\text{rk}G)}{2}\,2\mathcal{F}^{(m,n)}(0)}
\end{split}
\ee
where
\be
\mathcal{F}^{(m,n)}\=\mathcal{F}^{(m,n)}(x) \,\equiv\,\sum_{\text{a}\,\in\,\{\,V,\,I\,=\,\textbf{1},\,\textbf{2},\,\textbf{3}\}}\mathcal{F}_{\text{a}}^{(m,n)}(z_a(x))\,.
\ee
The object~$\mathcal{F}_V$ is defined in~\eqref{FcdVector} and the~$\mathcal{F}_{I}$ in~\eqref{Fcd}\,for~$r_I=\frac{2}{3}$. The computation of~\eqref{RelWeight} for vector walls will be reviewed in appendix~\ref{EffectiveActionComputation} and~\ref{PhasesLens}, here we note that the answer is of order~$\mathcal{O}(1)$ in the expansion near roots of unity (The part of the~$O(1)$ contributions that depend on~$x$ can be read from equation~\eqref{ExpansionF0}). On the contrary, generically, the contribution of chiral walls is exponentially suppressed in such limits. Along chiral walls
\be\label{RHOVChiral}
\rho(v)=:x\=\rho(u)\T \,+\,(-\,n\,\frac{\Delta_2\,+\,\ell}{m}\,+\,\Delta_1)\,,
\ee
where for the integrand of $\mathcal{I}(q)$, there is only one possible choice~
\be\label{Delta}
\Delta_1\,=\,-\,n_0\,\frac{r}{2}\,,\,\quad  \Delta_2\,=\,r\,=\,\frac{2}{3}\,,
\ee 
and a computation then shows that for a given gauge charge~$\rho$ 
\be\label{EquationPotentialInitial}
e^{\,2\Bigl(-\mathcal{F}^{(m,n)}(x)\,+\,\mathcal{F}^{(m,n)}(0)\Bigr)}\,\liMN\,e^{\,\,\frac{V_2\Bigl(-\,n\,\frac{\Delta_2\,+\,\ell}{m}\,+\,\Delta_1\Bigr)}{ \T^2}\,+\,\frac{V_1\Bigl(-\,n\,\frac{\Delta_2\,+\,\ell}{m}\,+\,\Delta_1\Bigr)}{ \T}\,}\,e^{\mathcal{O}(1)}\,,
\ee
where~$V_1$ and~$V_2$ are periodic pure imaginary even functions of period~$\frac{1}{m}$.~The function~$V_1$ and~$V_2$ are piecewise linear and quadratic, respectively, and their imaginary parts are either semipositive or seminegative-definite in dependence of the value of~$m$,~$n$ and~$n_0\,$. Both functions have zeroes at the $1/m$-periodic images of~$0\,$, and away from the zeroes, their imaginary part has a definite signature
\be
\text{Sign}(-\i V_{1,2})\= -n_0 \chi_1(m-n_0n)\,.
\ee
\footnote{$\chi_1$ is defined in~\eqref{BackgroundContribution}.} This signature is such that in the Cardy-like limits,~\footnote{... for which the absolute value of the leading background contribution grows exponentially fast...} the pre-factor~\eqref{RelWeight} vanishes exponentially fast, unless~
\be\label{ConditionDelta}
m \,\Delta_1\,-\,n\,\Delta_2\,\in\,\mathbb{Z}\,.
\ee
This is the condition that defines the positions of the zeroes in the exponent of the first exponential in the right-hand side of~\eqref{EquationPotentialInitial}. For values of~$m$ and~$n$ that solve~\eqref{ConditionDelta}, the factor~$|e^{-\mathcal{F}^{(m,n)}(0)}|$ does not grow as~$\tau\to-\frac{n}{m}\,$. Thus, the corresponding~$(m,n)$ limits do not define one of the Cardy-like limits studied in this paper.

In conclusion, in the expansions near roots of unity and for generic choices of~$\Delta_1$ and~$\Delta_2\,$, the contribution from chiral walls is exponentially suppressed with respect to the vector one.

\paragraph{A remark about chiral wall contributions at~$N=2$}  

In this paragraph we show that the Casimir pre-factor of chiral wall contributions to the~$SU(2)$ index~$\mathcal{I}(q)$, is finite in the limit~$\tau\to-\frac{n}{m}$. The absence of such exponential growth, which is also a defining property of one of the three eigenvalue distributions entering in the Bethe ansatz representation of the~$SU(2)$ index, indicates that chiral wall contributions could be related to such eigenvalue distribution. The conclusion for~$SU(N)$ is more involved.

For~$n_0=-1$ and the Cardy-like limit~$\tau\to-\frac{n}{m}\,$,\be\label{ParticularCasePotential}
\begin{split}
V_1\Bigl(-\,n\,\frac{\Delta_2\,+\,\ell}{m}\,+\,\Delta_1\Bigr)&\=V_1\Bigl(\Delta_1\Bigr)\= V_1(\frac{1}{3})\= \frac{2\pi\i} {3}\,,\\
V_2\Bigl(-\,n\,\frac{\Delta_2\,+\,\ell}{m}\,+\,\Delta_1\Bigr)&\=V_2\Bigl(\Delta_1\Bigr)\= V_2(\frac{1}{3})\= \frac{\pi\i} {9}\,.
\end{split}
\ee
To evaluate the last equality in each of the two lines, we used the definition of the potentials~$V_1$ and~$V_2$ which will be presented in the next section. For~$N=2$ the values~\eqref{ParticularCasePotential} imply that the leading behaviour of chiral non-analyticities in the expansion~$\tau\to -\frac{n}{m}$ is of order~$\mathcal{O}(1)$ i.e.
\be\label{Remark1}
\begin{split}
e^{-\sum_{\rho>0}2\mathcal{F}^{(m,n)}(\rho(v))\,-\,\frac{\text{rk}G}{2}\,2\mathcal{F}^{(m,n)}(0)}&\=e^{-2\mathcal{F}^{(m,n)}(2v)\,-\,\mathcal{F}^{(m,n)}(0)} \\
&\=e^{-3\mathcal{F}^{(m,n)}(0)\,-\,2 \bigl(\mathcal{F}^{(m,n)}(2v)- \mathcal{F}^{(m,n)}(0)\bigr)} \\
&\liMN\,e^{-3\mathcal{F}^{(m,n)}(0)\,+\, \frac{V_2(\Delta_1)}{\widetilde{\tau}^2} +\frac{V_1(\Delta_1)}{\widetilde{\tau}}\,+\,2\pi\i \mathcal{O}(1) } \\
&\liMN\, e^{-3\mathcal{F}^{(m,n)}(0)\,+\, \frac{\pi \i}{9\widetilde{\tau}^2} +\frac{2\pi\i}{3 \widetilde{\tau}}\,+\,2\pi\i \mathcal{O}(1)} \\
&\liMN\, e^{-3( \frac{\pi \i}{27\widetilde{\tau}^2}+\frac{2\pi\i}{9\widetilde{\tau}})\,+\, \frac{\pi \i}{9\widetilde{\tau}^2} +\frac{2\pi\i}{3 \widetilde{\tau}} \,+\,2\pi\i\mathcal{O}(1) }\,\liMN\, e^{2\pi\i \mathcal{O}(1)}\,.
\end{split}
\ee
In these steps we used equations~\eqref{EquationPotentialInitial} and~\eqref{ParticularCasePotential}. We also used the expression for~$\mathcal{F}^{(m,n)}(0)$ reported in equation~\eqref{ConstantPhase0} with~$n_0\,=\,-1\,$. 

An independent computation shows that~\eqref{Remark1} is also the leading behaviour of one of the three contributions in the~$SU(2)$ Bethe ansatz formula studied in~\cite{Lezcano:2021qbj,Benini:2021ano}. The one coming from the Bethe root~$(0,1)$
\be\label{Root01}
\rho(v)\=2v\=\frac{1}{2}\,.
\ee
 Indeed out of the three relevant roots there is only one that has order $\mathcal{O}(1)$ behaviour for every Cardy-like limit~$\tau\to -\frac{n}{m}\,$, which is~\eqref{Root01}. 
This matching indicates that chiral bits are related to the Bethe root~\eqref{Root01}. It would be interesting to explore this relation in more depth.

\section{Bulk and vector bit contributions: some examples }\label{sec:SU2}

This section starts by studying the~$SU(2)$ index from scratch and makes explicit contact with the main concepts enunciated in the introduction. 

In the case of~$SU(2)$ there are only two possible choices of partitions in the contour decomposition~\eqref{Decomposition},~$(\lambda=1,\widetilde{\lambda}=0)$ (pure bit regions, which can be vector, or chiral or auxiliary chiral~\footnote{In the case of~$SU(2)$ the contribution of auxiliary chiral bits vanishes for~$\varepsilon=0^+$. So, it will be ignored from now on. }) and~$(\lambda=0,\widetilde{\lambda}=1)$ (a pure bulk region).

The main focus is to show from scratch how the maximally symmetric contribution ($\widetilde{\lambda}\,=\,1$) dominates the total integral in Cardy-like index, even if the integration domain~$\mathcal{M}_\varepsilon(\lambda=1)$ is $\varepsilon$-infinitesimally small. Secondly, it is shown in detail how the process of integrating over the bulk domain~$\mathcal{M}_b(\lambda=1)$ reduces to the computation of a trivial Gaussian integral. Moreover, contributions from chiral bits to the integral~\eqref{Decomposition} are shown to be exponentially suppressed with respect to the leading maximally symmetric~$\alpha\,$.

Subsection~\ref{UNVectorBits} computes a specific maximally-symmetric bit ($\ell_i=0$) for gauge group $SU(N)$: a particular contribution to the general result summarized in~\eqref{PartionFunctionVectorBitsIntro}. Subsection~\ref{subsec:35} studies bulk contributions in the case~$N=3\,$. 

\subsection{Bulk and vector bit contributions}\label{Subsec:SUIndex}
Let us study the~$SU(2)$ index:
\be\label{SU2Index}
\mathcal{I}\= \kappa\,\int_{0}
^1 \, \diff v \,I_{\text{v}}\,I_{c}\,.
\ee
The two factors in the integrand are defined as
\be
\begin{split}
I_{\text{v}}&\=\theta_0(2v;\tau)\, \theta_{0}(-2v;\tau)\,, \\
I_{c}&\=\left(\Ge(2v+\frac{2\tau-n_0}{3};\tau)\,\Ge(-2v+\frac{2\tau-n_0}{3};\tau)\right)^3\,.
\end{split}
\ee
Let us define the zero-dimensional effective action as (with some convenient choice of branch for the logarithm) 
\be\label{effectiveAction}
S_{\text{eff}}\,\equiv\, -\,\log I_{\text{v}}\, I_c\,.
\ee
Thinking on taking a Cardy-like limit~$\tau\to -\frac{n}{m}\,$, it is
possible to split the integral~$\mathcal{I}$ into two pieces that we call bulk and localized
\be
\mathcal{I}\= \mathcal{I}_{bulk}\,+\,\mathcal{I}_{bit}\,.
\ee
This follows obviously from the particularization of the  decomposition~\eqref{Decomposition} to the case~$N=2$. 
The bulk contribution is
\be
\mathcal{I}_{bulk}\,\=\,\kappa \int^{\prime\, 1}_0 \diff v \, I_{\text{v}} I_c\,.
\ee
This variable~$v$ is $\frac{1}{2}$ the seed variable~$v_{12}$ defined in the Introduction, thus it runs in a period of length one instead of two as the latter.
The prime means integration over the segment~$(0,1)$ excluding the segments~$(p-\delta|\T|,p+\delta|\T|)$, with~$p$ being the location of a bit, and~$\delta |\T|$ being small enough. Notice that in Cardy-like expansion, these isolated regions have an infinitesimal width of order~$|\T|\,$ for any finite~$\delta\,$. By definition, the value of the integral is independent of~$\delta\,$.

As already mentioned, and contrary to the naive intuition, the contribution of infinitesimal bit regions,~$\mathcal{I}_{bit}\,$, is not subleading in the Cardy-like limit. Among such infinitesimal regions, one finds the neighbourhoods
\be
|2 v\,-\,p|<\varepsilon\,=\,\delta \,|\T| 
\underset{\tau\to-\frac{n}{m}}{\longrightarrow} 0\,, \qquad \,(\delta=\text{finite}\in\mathbb{R})\,
\ee
of the points
\be\label{Vicinities}
2v\= p= \frac{\ell^*}{m} +\text{integers}\,,\qquad \ell^*\=0,\,1,\,\ldots,\,m-1\,.
\ee
The localized contribution can be arranged as follows
\be\label{LocalizedSU2}
\mathcal{I}_{bit} \= \sum_{\ell^*=0}^{m-1} w\times \mathcal{I}_{\ell^*}\,,
\ee
where the positive integer~$w=2$ is the number of times that the integration cycle~$(0,1)$ wraps the smallest period of~$I_{\text{v}}\, I_c\,$, which is~$(0,\frac{1}{2})\,$. Let us explain this. The sum over~$\ell^*$, and the factor~$w=2$ come from the original sum over the bits intersected by the integration contour~$2v\in(0,2)\,$. The total number of such vicinities is~$w m\=2 m\,$, i.e. these are the~$w(=2)$ groups of~$m$ bits~\eqref{Vicinities}. Namely,~$w$ is the number of integer choices~$\mathbb{Z}$ in~\eqref{Vicinities} for which~$2v$ is intersected by the integration contour. In the present case, those choices are~$0$ and~$1$, and consequently,~$w=2$.

The object~$\mathcal{I}_{\ell^*}$ is defined as an integral over a bit intersected by~$(0,1)$ (this is one of the components in the integral over~$\mathcal{M}_{\varepsilon}(\widetilde{\lambda}=1)$ \eqref{BitIntegral}),
\be
\mathcal{I}_{\ell^*}\,\equiv\,\kappa\,\int^{p+\delta |\T|}_{p-\delta |\T| }\diff v \, I_{\text{v}} I_c\,.
\ee
Notice that in the limits~$\tau\to -\frac{n}{m}$ the factor of~$|\T|$ is infinitesimal. 
\begin{figure}\centering
\includegraphics[width=8cm]{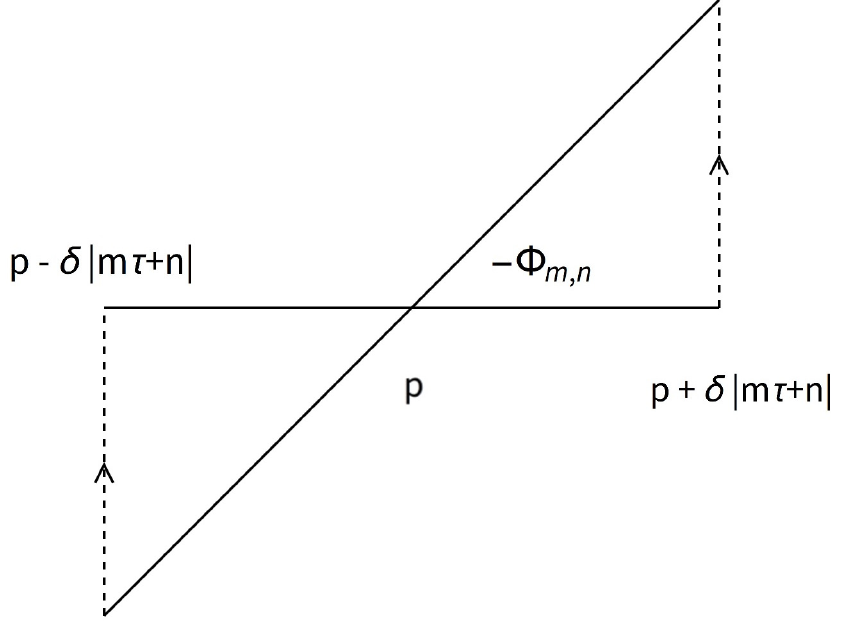} 
\caption{ The integral along the diagonal bit matches the one along the horizontal bit. Before assuming the Cardy-like expansion, we can slightly deform the real contour (avoiding crossing poles) so that in a complex vicinity of the~bit, the deformed contour matches the diagonal piece in the figure and not the horizontal one. Then to compute contributions of~bits, one can do the integral along the diagonal contour instead of the horizontal one. To further justify the previous argument, one would need to prove that contributions from dashed contours cancel in Cardy-like limit: this will be proven in the appendix~\ref{OnCourtourDiagonal}.   }
\label{fig:ContourGammaMN0}
\end{figure}
In appendix~\ref{OnCourtourDiagonal} (see also the preliminary argument given in the caption of figure~\ref{fig:ContourGammaMN0}) we will show that
\be
\mathcal{I}_{\ell^*} \,\liMN\, \int_{\mathcal{D}^p_{m,n}} \diff v \, I_{\text{v}} I_{c}\,,
\ee
where the new contour
\be\label{ContourComplexBit}
\mathcal{D}^p_{m,n}\=\Bigl(p-{\delta}\,\T\,,\,p\,+\,{\delta}\,\T\Bigr)\,,
\ee
is a segment lying (almost) along the~$\T$ direction in the complex~$v$-plane, and running between the points~$p\pm{\delta}\,\T\,$. In~\eqref{ContourComplexBit} we have substituted~$\delta\to \delta\,\cos{\Phi_{m,n}}$ where the angle~$\Phi_{m,n}$ is defined from
\be
e^{\i(\Phi_{m,n}\,+\, \epsilon)}\equiv \underset{\tau\to-\frac{n}{m}}{\lim} {\frac{|\T|}{\T}}\,,
\ee
where~$\epsilon= O(|\T|^{>0})\,$ is an infinitesimal tilt needed for convergence in the eventual limit~$\delta\to\infty\,$. The angle~$\Phi_{m,n}$ is part of the ambiguity one has in defining the Cardy-like limit. In particular, not every value of~$\Phi_{m,n}$ defines a~\emph{well-behaved} Cardy-like limit (See equation~\eqref{ConditionMB} below).

\subsubsection*{Bulk contributions}
\label{BulkCon}

Using representations~\eqref{IdentityTheta00} and~\eqref{repmnG00}, a computation shows that in the bulk of the contour of integration~$(0,1)\,$, the integrand takes the form
\be\label{CardyFormBulk0}
\kappa\, I_{\text{v}} I_c\, \liMN \,\kappa\,\times\,e^{- 2 \mathcal{S}_{(m,n)}+\pi \i\mathcal{O}(1)}\times e^{V(2v)}\,,
\ee
~\footnote{In the limit~$\tau\to-\frac{n}{m}$ in which the absolute value of~$e^{-\mathcal{S}_{(m,n)}}$ grows to~$+\infty\,$. See equation~\eqref{ConditionMB}.} 
where
\be\label{BackgroundContribution}
\begin{split}
\mathcal{S}_{(m,n)}\,\equiv\, \frac{\pi\i}{27 m}\frac{(2 \T-n_0 \chi_1(m-n_0n))^3}{\T^2},
\end{split}
\ee
and~$\chi_1(x)\=-1\,$,~$0\,$, or~$1$ if~$x=-1\,$,~$0$ or~$-1$~mod~$3\,$. This term and the imaginary constant~$\pi\i \mathcal{O}(1)$ in~\eqref{CardyFormBulk0} come from the~$B_{2,2}$ and~$B_{3,3}$ that arise from substituting~\eqref{Fcd} in the definition of effective action~\eqref{effectiveAction}, and evaluating the result at~$v=0$ (Further details are given around~\eqref{ConstantPhase0}).

Note that~$|e^{-\mathcal{S}_{(m,n)}}|$ grows (exponentially fast) iff~$\tau\to -\frac{n}{m}\,$ along the directions
\begin{equation}\label{ConditionMB}
 \left\{\begin{array}{cc}
\frac{\pi}{2}\,<\,-\,\Phi_{m,n}
\,<\,{\pi}&\quad \text{for} \quad (m-n_0n)\=-1\,\text{mod}\,3\, \\\\
0\,<\,-\,\Phi_{m,n}
\,<\,\frac{\pi}{2}&\quad \text{for}\quad (m-n_0n)\=\,\,\,\,1\,\text{mod}\,3\,
\end{array}\right..
\end{equation}
These are the limits we will focus on.
The potential~$V$ in~\eqref{CardyFormBulk0} has the form
\be\label{PotentialBulk}
V(v)\,\equiv\,\frac{V_2(2v)}{\T^2}\,+\,\frac{V_1(2v)}{\T}\,+\, V_{0}(2v)\,,
\ee
where~$V_{2}\,$,~$V_1$ and~$V_0$ are piece-wise polynomial functions of~$\nu=2v\,$. The~$V_{2}$,~$V_{1}$ and~$V_0$ are periodic in~$\nu$ with period~$1\,$, and independent of~$\tau\,$. To simplify the presentation, let us assume~$n_0=-1$ and~$\chi_1(m-n_0 n)=1\,$. The other cases are analogous.

\begin{figure}\centering
\includegraphics[width=7.3cm]{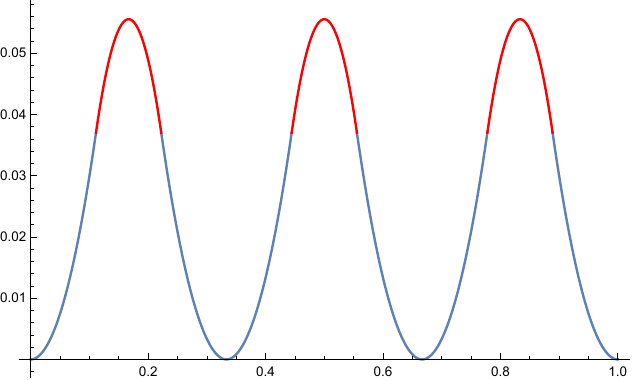}\quad
\includegraphics[width=7.3cm]{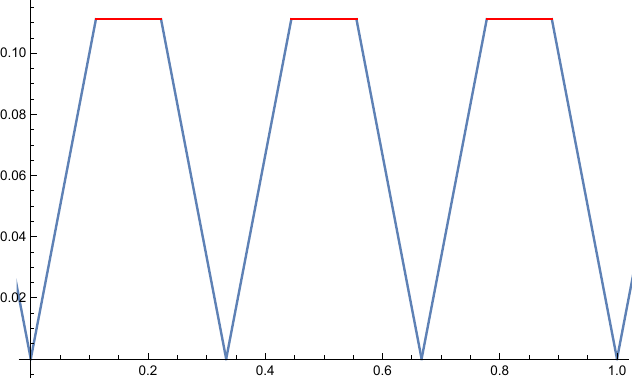}
\caption{ The portions of parabolas and cusps in the dominant and subdominant regions for~$m=3\,$. The figure to the left shows a plot of~$V_2(\nu)$ as a function of~$\nu\,$. The figure to the right shows a plot of~$V_1(\nu)$ as a function of~$\nu\,$. In both plots~$
\nu$ ranges from zero to one. The blue portions correspond to the dominant regions. The red portions correspond to the subdominant regions.    }
\label{fig:Cusps}
\end{figure}

The function~$V_2(\nu=2v)$ is made of a collection of two portions of parabolas, one of the two collections corresponds to
\be
 \frac{V_2(\nu)}{\T^2}\=\frac{(\pi \i m)}{\T^2}\, (\nu\,-\,\nu_0)^2
\ee
with domain in segments centered at the positions
\be
\nu_0\=\nu_0(\ell)\,\equiv\,\frac{\ell}{m}\,,\qquad \ell\,\in\,\mathbb{Z}
\ee
and of length~$\frac{2}{3m}$\,. The union of the latter segments will be called~\emph{dominant region}. The second collection corresponds to
\be
 \frac{V_2(\nu)}{\T^2}\=\frac{\pi \i }{6 m ^2 \T^2} \,-\,   \frac{2\pi \i }{\T^2}\, (\nu\,-\,\nu_0)^2\,
\ee
with domain segments centered at the positions
\be
\widetilde{\nu}_0\= \widetilde{\nu}_0({\ell})\,\equiv\,\frac{2{\ell}\,+\,1}{2m}\,,\qquad {\ell}\,\in\,\mathbb{Z}\,,
\ee
and having length~$\frac{1}{3m}$\,. The union of the latter segments will be called~\emph{subdominant region}. We have plotted one concrete case in figure~\ref{fig:Cusps}.

For later reference, we note that in the limits~$\tau\to -\frac{n}{m}\,$, for which~$|e^{-\mathcal{S}_{(m,n)}}|$ grows,
\be
e^{\frac{V_{2}(\nu)}{\T^2}}\, \longrightarrow\, 0\,.
\ee
Thus after extraction of the factor~$e^{-2\mathcal{S}_{(m,n)}}$, the remaining integrand vanishes in the strict Cardy-like limit. This is not the end of the story, though.

The function~$V_1(\nu=2v)$ is made of a collection of cusps and constant pieces. The cusps fill the dominant region. Precisely, they correspond to
\be
\frac{V_1(\nu)}{\T}\=\pm\,\frac{2\pi\i}{\T} \,(\nu\,-\,\nu_0)\,. 
\ee
where the~$+$ corresponds to the subregions
\be
\nu\,\in\, (\nu_0,\,\nu_0\,+\,\frac{1}{3m})
\ee
and the~$-$ corresponds to the subregions
\be
\nu\,\in\, (\nu_0\,-\,\frac{1}{3m}\,,\, \nu_0)\,.
\ee
The collection of constant values fills the subdominant region, and corresponds to
\be
\frac{V_1(\nu)}{\T}\= \frac{2\pi\i}{3 m \T}\,.
\ee
The~$V_0(\nu)$ is a piece-wise constant and imaginary function.~\footnote{ The jumps happen within the dominant region and also from the dominant to the subdominant region. However, jumps within the dominant region are always an integer multiple of~$2\pi\text{i}$, and thus are spurious.} The exponential of~$V_0(\nu)$
\be\label{PhaseConstantRegions}
e^{V_0(\nu)}
\ee
remains constant within both the dominant and the subdominant regions, although generically, it changes when one moves from the dominant to subdominant region or vice versa. Let~$e^{\i \Phi_{\emph{dom}}}$ (resp.~$e^{\i \Phi_{\emph{subdom}}}$ ) denote the value of~$e^{V_0(\nu)}$ in the dominant (resp. subdominant) region. The explicit expression for the phases~$e^{\i\Phi}$'s can be recovered by following the previous explanations.

Focus on the integral
\be
\begin{split}
\int_0^{\prime\,1}\diff u \,e^{\frac{V_2(2v)}{\T^2}\,+\,\frac{V_1(2v)}{\T}\,+\, V_{0}(2v)}&\=\int^{\prime 1}_0 d\nu \,e^{\frac{V_2(\nu)}{\T^2}\,+\,\frac{V_1(\nu)}{\T}\,+\, V_{0}(\nu)}\,\\
&\= e^{\i \Phi_{\emph{dom}}}\,\sum_{\ell=0}^{m-1} \mathcal{I}_{\ell}\,+\,e^{\i \Phi_{\emph{subdom}}}\,\sum_{\ell=0}^{m-1} \widetilde{\mathcal{I}}_{\ell}\,,
\end{split}
\ee
where
\be
\begin{split}
\mathcal{I}_{\ell}(\tau)&\,\equiv\,\int^{\prime \,\nu_0(\ell)+\frac{1}{3m}}_{\nu_0(\ell)-\frac{1}{3m}}\,\diff \nu \, e^{\pi\i m\,\frac{(\nu\,-\,\nu_0(\ell))^2}{\T^2}\,+\,\frac{2\pi\i}{\T} \,|\nu\,-\,\nu_0(\ell)|\,. }\,\\
\widetilde{\mathcal{I}_{\ell}}(\tau)&\,\equiv\,\int^{\prime\,\widetilde{\nu}_0(\ell)+\frac{1}{6m}}_{\widetilde{\nu}_0(\ell)-\frac{1}{6m}}\,\diff \nu \, e^{\frac{\pi \i }{6 m \T^2} \,-\,   \frac{2\pi \i }{\T^2}\, (\nu\,-\,\widetilde{\nu}_0(\ell))^2}\,,
\end{split}
\ee
and
\be
\int^{\prime x\,+\,y}_{x\, -\, y}\,\equiv\, \int^{\prime x\,-\,\delta |\T|}_{x\, -\, y}\,+\,\int^{\prime\, x\, +\, y}_{x\,+\,\delta |\T|}\,.
\ee
The integrals~$\mathcal{I}_{\ell}$ and~$\widetilde{\mathcal{I}}_{\ell}$ do not depend on~$\ell\,$. Thus,
\be
\sum_{\ell=0}^{m-1} \mathcal{I}_{\ell}\= m \, \mathcal{I}_{\ell=0}\,, \qquad \sum_{\ell=0}^{m-1} \widetilde{\mathcal{I}_{\ell}}\= m \, \widetilde{\mathcal{I}}_{\ell=0}\,.
\ee
A computation shows that in the limits~\eqref{ConditionMB}\,, which include those for which the quotient~$\frac{\tau_2}{m\tau_1+n}$ remains finite, with~$\tau_2\,>\,0$ and~$m\tau_1\,+\,{n}\,<\,0\,\,$,
\be
\mathcal{I}_{\ell}\sim O(\T^1)\,, \qquad \widetilde{\mathcal{I}}_{\ell}\sim O\bigl(e^{\frac{\pi (n+m\tau_1)\tau_2}{3|\T|^4}}\,\T^1\bigr)\,.
\ee
Thus, in such limits both~$\mathcal{I}_{\ell}$ and~$\widetilde{\mathcal{I}}_{\ell}$ vanish. The former vanishes as a linear function of~$\T\,$, and the latter vanishes exponentially fast. As it will be shown next, this does not mean that~$\mathcal{I}_{bulk}$ does not contribute to the index in the Cardy-like limit. So far, we have ignored the factor of~$\kappa$ in~\eqref{SU2Index}. In Cardy-like limit
\be
\kappa\,\longrightarrow \,\frac{1}{2}\,\frac{e^{-\,\mathcal{S}_{(m,n)}}}{ (\i\,\T)}\, e^{\pi\i\,\mathcal{O}(1) }\,.
\ee
Collecting results one concludes that
\be
\begin{split}
\mathcal{I}_{bulk}&\,\liMN\, \frac{m e^{\i \Phi_{\emph{dom}}}}{2 \i}\,\times\,e^{-3 \mathcal{S}_{(m,n)}\,+\, \pi\i \mathcal{O}(1)}\,\times\,\frac{\mathcal{I}_{\ell\=0}}{ \T}\,.
\end{split}
\ee
Solving the Gaussian integral~$\mathcal{I}_{\ell=0}$ and writing the result in terms of the error-function we obtain
\be\label{LimitIL0}
\begin{split}
\frac{\mathcal{I}_{\ell\=0}}{\T} &\= \frac{\sqrt[4]{-1} e^{-\frac{i \pi }{m}} \left(\text{erf}\left(\frac{(-1)^{3/4} \sqrt{\pi
   }\,(1+\delta \frac{|\T|}{\T})}{\sqrt{m}}\right)-\text{erf}\left(\frac{(-1)^{3/4} \sqrt{\pi } (3 \T+1)}{3 \sqrt{m}
   \T}\right)\right)}{\sqrt{m}} \\
   &\,\liMN\,\frac{\sqrt[4]{-1} e^{-\frac{i \pi }{m}} \left(\text{erf}\left(\frac{(-1)^{3/4} \sqrt{\pi
   }\,(1+\delta \frac{|\T|}{\T})}{\sqrt{m}}\right)\,-\,1\right)}{\sqrt{m}}\,,
\end{split}
\ee
where in the second line, we have taken the Cardy-like limit.~\footnote{To recover the result in the limit~$\tau\to\frac{n}{m}=0\,$, for instance, we recall that for~$n=0\,$, the only possible value of~$m>0$ is~$m=1\,$. That is an implicit assumption that we have used in the intermediate steps that lead to~\eqref{LimitIL0}. Analogously, for generic values of~$n\,$,~$m>0$ can run over the set of relative primes of~$n$ only.} This result shows that the bulk contributions~$\mathcal{I}_{bulk}$ are not necessarily subleading in Cardy-like expansion. Their contribution depends on the value of the parameter~$\delta$, which is independent of~$\tau\,$. 

Because~$\delta$ is a parametrization of the contour-decomposition~\eqref{Decomposition}, the total integral can not depend on~$\delta\,$.~\footnote{During the completion of the first version of this paper we became aware of the results of~\cite{Ardehali:2021irq}, which used a similar decomposition of the contour of integration. It would be interesting to compare and complement the approaches and tools here used with those of~\cite{Ardehali:2021irq}. We thank A.A.Ardehali for conversations on this point. } Consequently, it is safe to assume~$\delta\,\gg\,1\,$ and finite as $\widetilde{\tau},\varepsilon \to 0$. In this second limit~$\delta\to \infty$, after the Cardy-like one, the bulk contributions are exponentially suppressed. We will keep~$\delta$ finite in many equations, but in the very end it will be convenient to assume~$\delta\,\gg\, 1\,$.

\paragraph{Remark 2:} Before leaving behind the exponentially suppressed contributions~$\widetilde{\mathcal{I}}_{\ell=0}$  (in Cardy-like limit)  there is an observation to make. Focus on the Cardy-like limit~$\tau\to0\,$. A computation shows that in such limit
\be\label{Remark2}
\,e^{-3 \mathcal{S}_{(1,0)}}\,\times\,\frac{\widetilde{\mathcal{I}}_{\ell\=0}}{ \tau}\,\underset{\tau\to 0}{\simeq}\, e^{\frac{1}{18}\frac{\pi\i}{\tau^2}\,+\,\mathcal{O}(\frac{1}{\tau})} \,.
\ee
~\eqref{Remark2} is the same leading behaviour in the expansion around~$\tau= 0$ of the contribution to the~$SU(2)$ Bethe ansatz formula studied in~\cite{Lezcano:2021qbj,Benini:2021ano} coming from the Bethe root~$(1,1)$
\be\label{Root11}
\rho(v)\=2v\=\frac{1+\tau}{2}\,,
\ee
which is given by
\be
e^{-\sum_{\rho>0}2\mathcal{F}^{(m,n)}(\rho(v))\,-\,\mathcal{F}^{(m,n)}(0)}\,=\, e^{\frac{1}{18}\frac{\pi\i}{\tau^2}\,+\,\mathcal{O}(\frac{1}{\tau})} \,.
\ee
where~$\mathcal{F}^{(m,n)}(\rho(v))$ was defined in~\eqref{DefF}. This matching at very leading order suggests that in the expansion~$\tau
\,\to\,0\,$, the bulk integrals~$\widetilde{I}_{\ell}$ could be related to the Bethe root~\eqref{Root11}.~\footnote{One point that makes us cautious of drawing a conclusion is that the next subleading corrections do not match: This remains a puzzle that we leave for the future to explain. }

\subsubsection*{Bit contributions}

Let us now focus on the infinitesimal~bits.
Using representations~\eqref{IdentityTheta00} and~\eqref{repmnG00}, a computation shows that along the tilted~bit~$\mathcal{D}^{p}_{m,n}\,,$ the integrand takes the form
\be\label{CardyFormBulk}
\kappa\, I_{\text{v}} I_c\, \liMN \,\kappa\,\times\,e^{- 2 \mathcal{S}_{(m,n)}+\pi \i\mathcal{O}(1)}\times e^{V(2v)}\,e^{-2 \pi \i\, \varphi^{(m,n)}_p}\,,
\ee
where~$\varphi^{(m,n)}_p$ is a constant phase that depends on the discrete variable~$p\,$.~$\varphi^{(m,n)}_p$ has been defined in~\eqref{PhaseP}, for example~$\varphi^{(m,n)}_{p=0}=0\,$.
Along the~bit~$\mathcal{D}^p_{m,n}\,$, for which,
\be
\nu\=2v\= 2u\T\,+\,p,\qquad u\,\in\, \mathbb{R}\,.
\ee
the potential takes the form (which is different from~\eqref{PotentialBulk})
\be\label{PotentialBits}
\begin{split}
V&\,=\, 4\pi\i m u^2 \,+\, \log \Bigl( 2 \sin{\pi(2 u\,+\,\frac{\ell}{m})}\, 2\sin{\pi(-2 u\,-\,\frac{\ell}{m})} \Bigr)\,+\,\log(-1)\,.
\end{split}
\ee
The integer~$\ell$ ranges from~$0$ to~$m-1$ and it is determined in terms of~$\ell^*$ by the condition
\be
\ell^* = n \ell \mod{m}\,.
\ee
\footnote{Recall that we are assuming~gcd$(m,n)=1\,$. For example, if~$n=0$ then~$m=1\,$, and thus~$\ell=\ell^*=0\,$. If~$n=1$ then~$m$ can be any positive integer and~$\ell=\ell^*= 0,\ldots, m-1\,$. If~$n>1\,$, generically,~$\ell^*\neq \ell\,$.} Using~\eqref{PotentialBits} one obtains
\be
\begin{split}
w\mathcal{I}_{\ell^*}&\=\frac{w}{2 \i \T}\,\times\,e^{-3 \mathcal{S}_{(m,n)}\,+\, \pi\i \mathcal{O}(1)}\,e^{-2 \pi \i\, \varphi^{(m,n)}_p}\,\int_{\mathcal{D}_{m,n}} \diff v \,e^{V} \\
&\= e^{-3 \mathcal{S}_{(m,n)}\,+\, \pi\i \mathcal{O}(1)}\,\mathcal{J}_{\ell^*}\,,
\end{split}
\ee
where
\be\label{JLStar}
\begin{split}
\mathcal{J}_{\ell^*}&\,\equiv\,\frac{w}{2}\,\times\,e^{-2 \pi \i\, \varphi^{(m,n)}_p}\,\times\,\int_{\Gamma_{m,n}} \frac{\diff u}{i} \, e^{ \pi\i \,m (2 u)^2}\, \Bigl|2 \sin{\pi(2 u\,+\,\frac{\ell}{m})}\Bigr|^2\\&\= \frac{w}{2}\,\times\,e^{-2 \pi \i\, \varphi^{(m,n)}_p}\,\times\,\int^{\delta\, e^{\i \epsilon}}_{-\delta e^{\,i\epsilon}} \frac{\diff u}{i} \, e^{ \pi\i \,m (2 u)^2}\, \Bigl|2 \sin{\pi(2 u\,+\,\frac{\ell}{m})}\Bigr|^2\,.
\end{split}
\ee
The domain~$\Gamma_{m,n}$ is defined as
\be\label{ContourGMNFinal}
\Gamma_{m,n}\equiv\frac{\mathcal{D}^{p=0}_{m,n}}{\T}=(-\delta e^{\,i\epsilon},\delta e^{\,i\epsilon})\,.
\ee
After taking~$\tau\to-\frac{n}{m}$, we take the splitting parameter~$\delta$ to be very large. To obtain a finite result in such double limit, one must take the infinitesimal tilt
\be
\epsilon\= 0^{-}\,.
\ee
The tilt can be non-infinitesimal as long as the deformed integral remains convergent. The convergent result in the limit~$\delta\to\infty$ is independent of the magnitude of the tilting.

After changing variables to~$\sigma={2\pi\i u}$ and taking the double limit,~$\mathcal{J}^*_{\ell}$ becomes (up to a constant phase that does not depend on~$p$)
\be\label{ChernSimonsSU2}
\frac{w}{2}\,\times\, e^{-2 \pi \i\, \varphi^{(m,n)}_p}\,\times\int_{\Gamma} \frac{\diff\sigma}{2\pi} e^{-\frac{\i\,m\, (\sigma)^2}{\pi }}\, \,\Bigl(2 \sinh{\Bigl(\sigma\,+\,\frac{\pi \i \ell}{m}}\Bigr)\Bigr) \,\Bigl(2 \sinh{\Bigl(-\sigma\,-\,\frac{\pi \i \ell}{m}}\Bigr)\Bigr),
\ee
where~$\Gamma$ is a generic straight segment with its two extrema going to~$\infty$ in the second and fourth quadrants of the $\sigma$-complex plane (This will be explained around figure~\eqref{fig:ContoursConve} below). For generic~$m$ and~$\ell=\ell^*=0$ the integral~\eqref{ChernSimonsSU2} is, up to a phase,~$w(=2)$ times the~$SU(2)$ Chern-Simons path integral (For more details on this, see around equation~\eqref{CSMIntegral}). The total contribution from vector bits to the~$SU(2)$ index can be recovered from~\eqref{ChernSimonsSU2} and~\eqref{LocalizedSU2}.

\paragraph{Remark 3:} A computation shows that the leading exponential factor of the vector bit contributions, $e^{- 3 \mathcal{S}_{(1,0)}}$, matches, in the expansion~$\tau\to 0\,$, the contribution of the root~$(1,0)$ 
\be\label{Root10}
\rho(u)\=2u\=\frac{\tau}{2}\,.
\ee
to the~$SU(2)$ Bethe ansatz formula studied in~\cite{Lezcano:2021qbj,Benini:2021ano} up to corrections of order~$O(1)$.
This matching suggests that in the expansion~$\tau
\,\to\,0\,$, the vector bit integrals are related to one of the three contributions that compose the~$SU(2)$ Bethe ansatz formula, the one corresponding to the Bethe root~\eqref{Root10}.

\subsection{An example of maximally symmetric bit for~$SU(N)$}
\label{UNVectorBits}

Let us compute the~$0$-bit contribution to the~$SU(N)$ index, for any~$\tau\to-\frac{n}{m}$ limit along the directions~\eqref{ConditionMB} followed by the limit~$\delta\to \infty\,$.~\footnote{The final answer completes a formula of~\cite{ArabiArdehali:2021nsx} for the limits~$\tau\to -\frac{n}{m}\,$ with~$n=1\,$, and generalizes it to the case~$n\neq 1$.} The result we obtain after analogous computations for generic maximally symmetric bits has been already reported in equation~\eqref{PartionFunctionVectorBitsIntro}.

Focus on
\newcommand{\Seffmn}{S_{p=0}(u)}
\be\label{GammaMN}
\mathcal{I}_{U(N)\ell^*=0}\=\kappa\,\T^N\,\int_{\Gamma_{m,n}^N}\prod_{i=1}^N\diff u_i\, I_{\text{v}}\, I_c\,,
\ee
To recover the~$SU(N)$ bit integral~$\mathcal{I}_{\ell^*=0}\,$ from the~$U(N)$ one we use
\be\label{RelationUNSUN}
\mathcal{I}_{\ell^*=0}\=\frac{1}{2\delta}\, \frac{\mathcal{I}_{U(N)\ell^*=0}}{\mathcal{I}_{U(1)}}\,.
\ee
The factor of~$2\delta$ is the value of the integral of the center-of-mass mode, which equals the length of the segment~$\Gamma_{m,n}$.

Using the Cardy-like expansion of the integrand along the~$0$-th wall (See appendix~\eqref{EffectiveActionComputation} equation~\eqref{EffectiveActionSmn}), and after some algebraic manipulations, one obtains
\be\label{BitsATOrigin}
\begin{split}
\mathcal{I}_{U(N)\ell^*=0}&\, \liMN \,e^{-N^2 \mathcal{S}_{(m,n)}\,+\, \pi\i \mathcal{O}(1)}\, \mathcal{J}_{U(N)\ell^*=0}\,,
\end{split}
\ee
where
\be\label{IntegralOriginal}
\mathcal{J}_{U(N)\ell^*=0}\,\equiv\,\frac{1}{N!}\int_{\Gamma_{m,n}} \prod_{i=1}^N \,\frac{du_i}{i}\, e^{V}\,,
\ee
and~\cite{GonzalezLezcano:2020yeb}
\be\label{ExpansionT0}
V \, \equiv \, \,\pi\i\, k \sum_{i\,=\,1}^N \Bigl(u_i\,-\,\frac{1}{N}\sum_{j=1}^N u_j\Bigr)^2\, +\,\sum_{i\,\neq\, j}\log 2 |\sin{ \pi u_{ij}|}\,.
\ee
The integer~$k$ takes the values
\be\label{Level}
k\,\equiv\, n_0\,m \,N\chi_1(m-n_0n)\,.
\ee
In reaching~\eqref{BitsATOrigin} we have used
\be\label{LimitKappa}
\kappa \liMN  \frac{e^{-\,\text{rk}(G)\,\times\,\mathcal{S}_{(m,n)}}}{N!\, (\i\,\T)^{\text{rk}(G)}}\, e^{\pi\i\,\mathcal{O}(1) }\,,\qquad \mathcal{I}_{U(1)}\liMN  \frac{e^{-\,\mathcal{S}_{(m,n)}}}{ (\i\,\T)}\, e^{\pi\i\,\mathcal{O}(1) }\,,
\ee
to obtain
\be
\kappa\,\T^N\,e^{-(N^2-N)\, \mathcal{S}_{(m,n)}}\=\frac{1}{\i^N}\,e^{-\,N^2\, \mathcal{S}_{(m,n)}\,+\, \pi\i \mathcal{O}(1)}\,.
\ee
This last equation illustrates how the potential logarithmic contributions in~$\widetilde{\tau}$ to the effective action completely cancel out.
 
The expansions~\eqref{LimitKappa} were obtained from the definition~\eqref{Kappa} and identity~\eqref{QPochIdentity}.
To recover the~$SU(N)$\,$0$-th bit of integral~$\mathcal{J}_{\ell^*=0}\,$, we use the asymptotic identity
\be\label{IntegralGammaMNIdentity}
\int_{\Gamma_{m,n}} \prod_{i=1}^N \,\diff u_i e^{V}\,\underset{\delta\,\to\,\infty}{\simeq}\, 2\delta \,\times \,N\,\times \int_{\Gamma_{m,n}} \prod_{i=1}^{N-1} \,\diff u_i e^{V}\,,
\ee
where in the integrand of the right-hand side~$u_N\=-\sum_{i=1}^N u_i\,$. The proof of~\eqref{IntegralGammaMNIdentity} is reported in~\eqref{IntegralOriginalApp}. From~\eqref{IntegralGammaMNIdentity} and~\eqref{RelationUNSUN} one obtains
\be
\mathcal{J}_{\ell^*=0}\=  \,N\, \times\,\frac{1}{N!}\,\times \int_{\Gamma_{m,n}} \prod_{i=1}^{N-1}\, \frac{ \diff u_i}{\i} e^{V}\,.
\ee
Defining~$\sigma^i\,\equiv\, 2 \pi \i\, u_i$ and noting that iff~$\sigma^N\=-\sum_{i=1}^N \sigma^i$
\be\label{RelationCS}
\begin{split}
V&\=- S_{\text{CS}, k}(\sigma) \,+\,\pi
\i \mathcal{O}(1)\,, 
\end{split}
\ee
we conclude that
\be\label{JLStart}
\begin{split}
\mathcal{J}_{\ell^*=0}&\=  \,N\, \times\,{e^{\pi \i \mathcal{O}(1)}}\times Z_{\Gamma}\,, \qquad Z_{\Gamma}\=\int_{\Gamma}\prod_{i=1}^{N-1} \frac{\diff\sigma^i}{2\pi}\, e^{-S_{\text{CS},k}(\sigma)}\,.
\end{split}
\ee
~$S_{CS,k}\,$, as defined in~\eqref{CSMIntegral}, is the effective action of~$SU(N)_k$ Chern-Simons matrix integral on~$S_3\,$, and~$Z_\Gamma$ is, up to a normalization factor, the~$SU(N)_{k}$ Chern-Simons integral on~$S_3\,$. The relation between~$Z_\Gamma$ and the~$U(N)_k$ Chern-Simons integral (which was reviewed in appendix~\ref{app:CSW}) is
\be\label{ZSUNMain}
\ZGSUN\= \text{sign}(-k)\,\Bigl(\frac{\i k}{N}\Bigr)^{\frac{1}{2}}\,\ZGUN\,.
\ee\footnote{Notice that the original~$U(N)$ integral~\eqref{IntegralOriginal} is not equal to the~$U(N)_k$ Chern-Simons integral~$Z^{U(N)}_{\Gamma}$ that was defined in appendix~\ref{app:CSW}\,. In particular, the former is invariant under rigid center-of-mass translations~$u_i\to u_i+c$, and the latter is not. However, the latter can be interpreted as a Chern-Simons integral with non-diagonal Chern-Simons matrix-level. The matrix-level having integer eigenvalues.}

\begin{figure}\centering
\includegraphics[width=15cm]{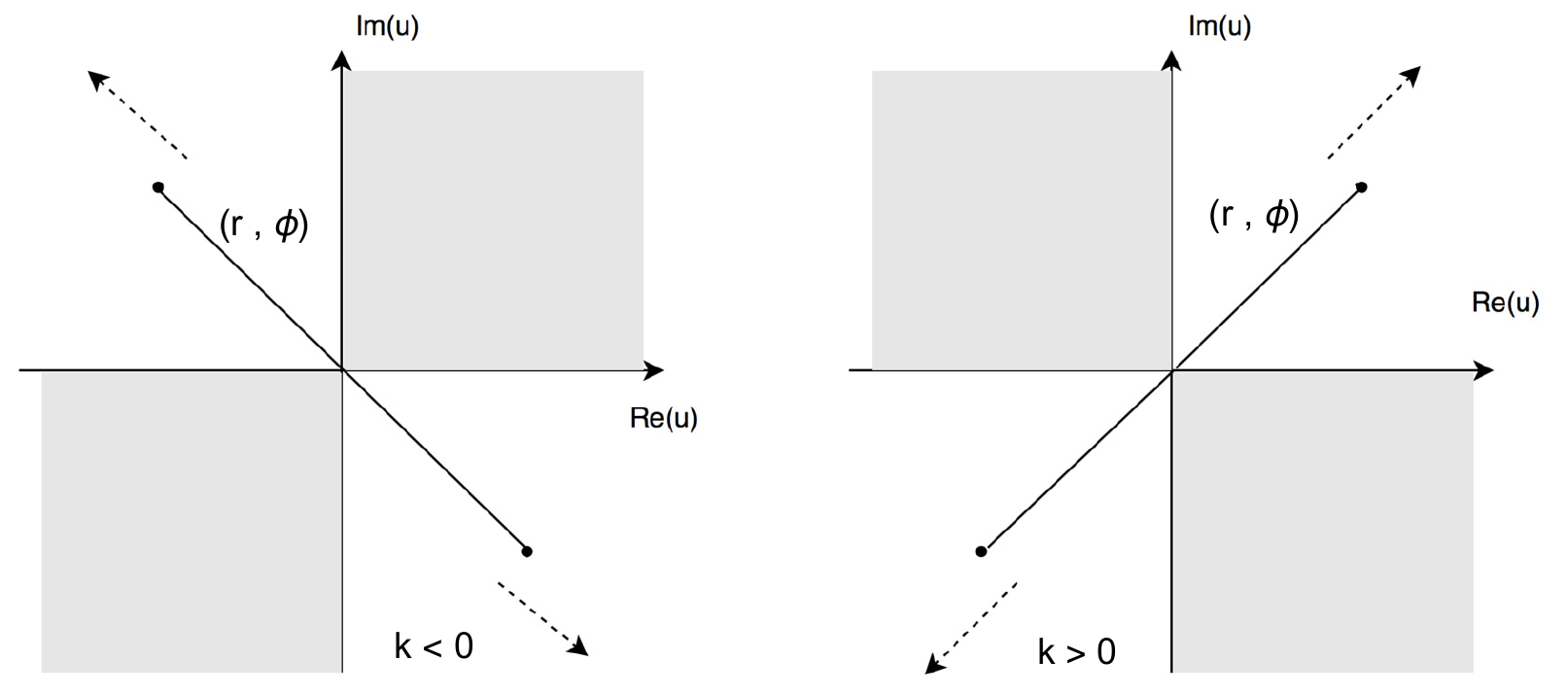} 
\caption{ Any connected contour with two endings going to the opposite asymptotic regions in white can be used as integration contour for the Chern-Simons integral~\eqref{CSMIntegral} in variables~$u_i\,=\,\frac{\sigma^i}{2\pi\text{i}}\,$:~$k$ is the Chern-Simons level.
}
\label{fig:ContoursConve}
\end{figure}

\subsection{Symmetry-breaking sectors: Integrating out bulk variables}
\label{subsec:35}

This section shows how the integration over seed variables in the bulk generates further suppressions of the Casimir pre-factor and it does not affect the effective action for the remaining seed variables, which we assume to be of vector type~$v$.

We will focus on a particular example, but the general analysis is the same. Let us start from the~$U(3)$ index in the particular limit~$\tau\to 0$ i.e. choose~$n=0$ and~$m=1$. To obtain the contributions to the~$SU(3)$ index we use formula~\eqref{SUNVSUN}. Let the three Cartan potentials of~$U(3)$ be
\be
\{v_i\}\=\{v_{1}, v_2, v_3\}\,.
\ee
There are three types of symmetry-preserving bit integrals:~$SU(3)$, $SU(2)$ and~$0$-symmetry-preserving bit integrals. These correspond to the choices $\widetilde{\lambda}\,=\,2,\, 1$ and~$0$ respectively.

One of the~$SU(2)$-preserving bits of the~$U(3)$ index is specified by the domain of integration 
\be
\begin{split}
|v_{12}|&\,<\,\delta |\T|\,, \qquad v_{13}\,=\,v_{23}\,\equiv\,\Phi\in\,[-1,1]^\prime\,.
\end{split}
\ee
To keep the discussion close to the presentation given in subsection~\eqref{UNVectorBits}, we restrict the center-of-mass mode to range over the contour~$\mathcal{D}^{(p=0)}_{1,0}$ defined in equation~\eqref{ContourComplexBit}, just as~$v_{12}\,$.

Let~$\rho$ denote the non-vanishing adjoint weights of~$U(3)$ which are
\be
\rho(v)\=\{v_{12},v_{23},v_{31},-v_{12},-v_{23},-v_{31}\}\,.
\ee
As shown before, the integral over the small contour~$\mathcal{D}^{(p=0)}_{(1,0)}\,$ equals the integral over the tilted contour sketched in figure~\ref{fig:ContourGammaMN0}. If we define
\be
v_{12}\,=\,2\xi\,\tau
\ee
then, along the tilted contour,~$\xi$ is real and runs along the small real contour~$\Gamma_{m,n}$ defined in~$\mathcal{F}^{(1,0)}(\rho(v))$ given in~\eqref{ExpansionF0} for~$\rho(v)=\{v_{12}, -v_{12}\}\,$, one obtains
\be
\mathcal{F}^{(1,0)}\Bigl(\rho(v)\,\to\,2\xi\tau\Bigr)\= \mathcal{F}^{(1,0)}(0)\,-\,4 \pi\i \xi^2\,.
\ee
Similarly, for the pieces in the effective action that come from~$v_{23}$ or~$v_{13}$ ranging over the bulk~$[0,1]^\prime\,$, one obtains
\be
\mathcal{F}^{(1,0)}\Bigl(\rho(v)\,\to\,\Phi\Bigr)\= +\,\frac{\, V_2(\Phi)}{\tau^2}\,+\,\frac{ V_1(\Phi)}{\tau}\,+\text{piecewise constant}\,,
\ee
where~$V_{2}$ and~$V_1$ are the ones defined in~\eqref{PotentialBulk} with~$m=1$ and~$n=0\,$. Placing all the pieces together one obtains
\be
\begin{split}
V_{N=3}(v)&\,\equiv\,4\pi\i\,\xi^2\,+\,\frac{\,2 V_2(\Phi)}{\tau^2}\,+\,\frac{2 V_1(\Phi)}{\tau}\,+\text{piecewise constant}\,, \\
      &\,+\, \log \Bigl( 2 \sin{\pi(2 \xi)}\, 2\sin{\pi(-2 \xi)} \Bigr)\,+\,\log(-1)\,.
\end{split}
\ee
The factor of~$2$ in the potentials that depend on~$\Phi$ comes from the two possible choices of~$\rho(v)= \Phi\,$, the ones coming from~$v_{31}$ and~$v_{23}\,$, respectively.
The $SU(2)$-preserving bit integral within the total~$U(3)$ integral is
\be\label{ItermediateBitSymmB}
\propto\,  \frac{\delta}{\tau^{3-1}}\Bigl(\int^{2^\prime}_0 d\Phi\,e^{\frac{\,2 V_2(\Phi)}{\tau^2}\,+\,\frac{2 V_1(\Phi)}{\tau}\,+\,2V_0(\Phi)}\Bigr)\,\times \text{($SU(2)$ CS)}
\ee
The~$SU(2)$ CS stands for~$SU(2)$ Chern-Simons matrix integral, which is just the integral over the variable~$\zeta$. This latter integral factorizes out from the integrals over the variable~$v_{23}=\Phi\,$; there is also a bit integral over the center-of-mass mode which gives the factor of~$\delta$ in the numerator. As we previously showed, reaching the~$SU(2)$ Chern-Simons integral takes one of the factors of~$\tau$ in the denominator, and shifts the total power of~$\tau$ from~$3$ to~$2$.

The bulk integral between parenthesis in~\eqref{ItermediateBitSymmB} is rather similar to the one studied in subsection~\ref{BulkCon}. Indeed, as for the latter
\be
\int^{2^\prime}_0\= 2 \int^{1^\prime}_0 \,\underset{\tau\to0}{\simeq}\, 2 \int^{\frac{1}{3}^\prime}_{-\frac{1}{3}} \=2 \Bigl(\int^{-\delta_\Phi|\tau|}_{-\frac{1}{3}}\,+\,\int^{\frac{1}{3}}_{\delta_\Phi|\tau|}\Bigr)\,.
\ee
where we choose~$\delta_{\Phi}\,\propto\,\delta\,$. Moreover, doing an analogous computation to the one reported in subsection~\ref{BulkCon} (see around equation~\eqref{LimitIL0}), one proves the identity
\be\label{LimitSU2U1}
\frac{1}{\tau^{1}}\Bigl(\int^{\frac{1}{3}^\prime}_{-\frac{1}{3}} d\Phi\,e^{\frac{\,2 V_2(\Phi)}{\tau^2}\,+\,\frac{2 V_1(\Phi)}{\tau}\,+\,2V_0(\Phi)}\Bigr)\,\underset{\tau\,\to\,0}{\simeq}\,\mathcal{O}(1) \,\underset{\delta\,\to\,\infty}{\longrightarrow}\, 0\,.
\ee
In this reduced domain~$[-1/3,1/3]^\prime$, the~$V_0(\Phi)$ is a constant~\footnote{This naive observation, which was explained around~\eqref{PhaseConstantRegions}, is essential. As noted there the function~$e^{V_{0}(x)}$ is constant, in connected regions where the piecewise polynomial contribution coming from the contributions of the piecewise polynomial part of identities~\eqref{IntegrandSoloInitial}, the~$R^{(3)}$, is strictly polynomial i.e. infinitely smooth. The contribution of the integral over any such regions comes with a different leading exponential pre-factor in the expansion near roots of unity. In the case of~$m>1$ the region around a~bit with the leading exponential pre-factor corresponds to the subdomain~$x\,\in\,[p-\frac{1}{3 m}, p+\frac{1}{3 m}]$. }, not just a piecewise constant that could depend on the seed variable $\zeta:=u_{1,2}\,$. This observation implies that the integral over the bulk seed can be factored out the integral over the vector wall (See~\eqref{Factorization} below).

To go from the~$U(3)$ to the~$SU(3)$ integral we use the relation~\eqref{RelationUNSUN}. Doing so, cancels another factor of~$\tau$ in the denominator of~\eqref{ItermediateBitSymmB} and the~$\delta$ in the numerator. Collecting all the pieces, and using~\eqref{LimitSU2U1} one obtains that the~$SU(2)$-preserving bit contributions  ($\widetilde{\lambda}=\lambda=1$) to the~$SU(3)$ index are exponentially suppressed in the double limit in question, i.e., that
\be\label{Factorization}
\frac{1}{\tau^{3-2}}\Bigl(\int^{2^\prime}_0 d\Phi\,e^{\frac{\,2 V_2(\Phi)}{\tau^2}\,+\,\frac{2 V_1(\Phi)}{\tau}\,+\,2V_0(\Phi)}\Bigr)\,\times \text{($SU(2)$ CS)}\,\underset{\tau\,\to\,0}{\simeq}\,\mathcal{O}(1) \,\underset{\delta\,\to\,\infty}{\longrightarrow}\,\, 0\,.
\ee
This is, the $SU(2)$-preserving contribution has a subleading Casimir contribution, as explained in the Introduction.

The same conclusion holds for generic~$N\,>\,3$ and for any symmetry-breaking bit. As explained in the Introduction, in the generic case the integration over bulk seed variables implies a factorization such as~\eqref{Factorization}, i.e., with an exponentially suppressed factor coming from the integration over the seed variables located at the bulk. 

\section{Minimal chiral bit contribution} \label{ChiralWalls}

In this section we present a study of the sector~$\alpha$ for which~$N-2$ seeds are integrated over the bulk region~$\mathcal{M}_{b}(\lambda= N-2)$ and one over a chiral bit $\mathcal{M}_\epsilon(\widetilde{\lambda}=1)$. As said before, this is a subleading sector.

\subsection{Lower dimensional~$A$-models} \label{Amodel}

The exponential of the chiral wall profile~\eqref{ProfileText} can be written in the form
\be\label{CloseForm}
e^{2\pi\i (g-1) \Omega^{(1)}_\rho \,+\,
2\pi\i p \bigl(\mathcal{W}^{(1)}_\rho \,-\,\rho(u) \,\partial_{\rho(u)}\, \mathcal{W}^{(1)}_\rho \,-\, \gamma_{2}\,\partial_{\gamma_2}\, \mathcal{W}^{(1)}_\rho\bigr)\,,}
\ee
where~$\gamma_{2}=\gamma_{2}(m,\ell):=\frac{\Delta_{2}+\ell}{m}$ and~$\ell\in \mathbb{Z}_m$ can depend on~$\rho$ with
\be\label{TwistedPotentialDilaton}
\begin{split}
\Omega^{(1)}_{\rho}&\,:=\, \frac{1}{2\pi\i}\, \Bigl(-{\pi \text{i}}(\rho(u)+\gamma_2)\,+\,\text{Li}_1\bigl(e^{2\pi\i (\rho(u)+\gamma_{2})}\bigr)\,\Bigr)\,.\\
\mathcal{W}^{(1)}_\rho&\,:=\,\frac{1}{4}\,\Bigl(\rho(u)+\gamma_2\Bigr)^2 \,+\,\frac{1}{(2\pi\i)^2}\,\text{Li}_2\bigl(e^{2\pi\i (\rho(u)+\gamma_{2})}\bigr)\,.
\end{split}
\ee

The positive integer parameters~$p$ and~$g$ in~\eqref{CloseForm} are defined as
\be\label{PunctGenus}
p\,:\,=\, m\,\geq\, 1\,, \qquad  g\,-\,1:\= \, \ell^\star \,+\,1\,
\ee
where~$\ell^\star =0,\ldots, m-1$\,, and the bound~$m\,\geq\, 1$ imply that the integer~$g\,\geq\,2\,$. 
We recall that given a set of~$v_i=v_i^{(0)}$'s parametrizing the position of chiral bits, the contribution of a given~$\rho$ to the effective action along such chiral bit takes the form~\eqref{CloseForm} iff~$\rho(v)$ hits a chiral bit. 

Let us briefly explain how~\eqref{ProfileText} can be recast in the form of the exponent of~\eqref{CloseForm}. First, the quadratic term in the twisted superpotential~$2\pi\i m\mathcal{W}^I_\rho$ is the unique choice that gets mapped into a quadratic term of the form~$-\frac{\pi\i}{2}m(\rho(u)+\gamma_2)^2$ after being acted upon by the linear diferential operator~$1-(\rho(u)+\gamma_2) \partial_{\rho(u)}$. Precisely, a quadratic term~$-\frac{\pi\i}{2} m(\rho(u)+\gamma_2)^2$ arises in~\eqref{ProfileText} after applying the identity below-given upon the second Li$_2$ in the second line of~\eqref{ProfileText}
\be
-\text{Li}_2(\frac{1}{z})\=\text{Li}_2(z)\,+\,\frac{\pi^2}{6}\,+\,\frac{1}{2}(\log(-z))^2\,.
\ee
and combining the result with another quadratic contribution coming from the first term in equation~\eqref{ProfileText}. The remainder polylogarithmic contributions can be straightforwardly checked to match in between the two expressions. 

\paragraph{One extra contribution to the twisted superpotential} 
There is an extra quadratic contribution to the twisted superpotential coming from the piecewise polynomial  pre-factor~$R^{(3)}$. For instance, assuming~$n_0=-1$ and $(m+n)\mod3\,\neq\, 0\,$ the contribution coming from the latter pre-factor is
\be\label{CasimirPrefactors}
\begin{split}
-&\frac{2\mathcal{F}^{(m,n)}\bigl(0^{-}+(-\,n\,\frac{\Delta_2\,+\,\ell}{m}\,+\,\Delta_1)+\rho(u)\,\widetilde{\tau} \bigr) \,+\,2\mathcal{F}^{(m,n)}\bigl(0^{+}+(-\,n\,\frac{\Delta_2\,+\,\ell}{m}\,+\,\Delta_1)+\rho(u)\,\widetilde{\tau} \bigr)}{2}\\ &\,=\,\frac{\chi_1(m+n)}{2} \,\pi \text{i} m (\rho(u))^2\,+\,\ldots.
\end{split}
\ee
where the~$\ldots$ denotes $\ell$-dependent imaginary constants that are collected in the phases~$\Phi_\alpha$ in~\eqref{AveragesPartitionF}. We recall that the correct value at the walls is obtained by evaluating the semi-sum of the lateral limits, as illustrated by the limits~$0^{\pm}\,$ of the pre-factor $\mathcal{F}^{(m,n)}$ in~\eqref{CasimirPrefactors}. This polynomial contribution comes from factors~$I(v_{ij})$ that hit chiral bits~$c$. The very same contribution is there for the factors that hit auxiliary chiral bits~$b^{\prime}$.

Note that the quadratic term in the second line of~\eqref{CasimirPrefactors} is $-\frac{1}{2}$ times the analogous result obtained for vector walls,~\eqref{ExpansionF0}~\eqref{KappaMN} (assuming~$n_0=-1$).~ For convenience, and without losing generality, we can assume~$m\,+\,n\,\underset{\text{mod} 3}{=}\,1\,$. The exponential of~\eqref{CasimirPrefactors} can be written in the form 
\be\label{CloseForm2}
e^{2\pi\i p \bigl(\mathcal{W}^{(2)}_\rho \,-\,\rho(u) \,\partial_{\rho(u)}\, \mathcal{W}^{(2)}_\rho \,-\, \gamma_{2}\,\partial_{\gamma_2}\, \mathcal{W}^{(2)}_\rho\bigr)\,,}
\ee
with the choice
\be\label{extraQuadratic}
\widetilde{\mathcal{W}}^{(2)}_\rho\,=\, -\frac{\rho(u)^2}{4}\,.
\ee
Then multiplying the three contributions~\eqref{CloseForm} and~\eqref{CloseForm2} we conclude that the total contribution of a charge vector~$\rho$ along a chiral wall takes the form
\be\label{CloseFormFinal}
\begin{split}
&e^{2\pi\i (g-1) \Omega_\rho \,+\,
2\pi\i p \bigl(\mathcal{W}_\rho \,-\,\rho(u) \,\partial_{\rho(u)}\, \mathcal{W}_\rho \,-\, \gamma_{2}\,\partial_{\gamma_2}\, \mathcal{W}_\rho\bigr)\,,}\\& = \mathcal{C}_{m}[\rho(u),\ell]\times(\text{constant phase})
\end{split}
\ee
with
\be\label{TwistedSuperPotential2}
\begin{split}
\Omega_\rho &\,=\, 3\Omega^{(1)}_\rho\\
\mathcal{W}_{\rho}&\,=\,3\mathcal{W}^{(1)}_{\rho}+\widetilde{\mathcal{W}}^{(1)}_{\rho}\,\\ &=\,\frac{\Bigl(\rho(u)\Bigr)^2}{2} \,+\, \frac{3}{2} \gamma_{(2)} \rho(u) \,+\,\frac{3}{4}\,\gamma_2^2 \,+\,\frac{3}{(2\pi\i)^2}\,\text{Li}_2\bigl(e^{2\pi\i (\rho(u)+\gamma_{2})}\bigr)\,.
\end{split}
\ee
We recall that the chiral building block~$\mathcal{C}_m[x,\ell]$ was defined in~\eqref{HolomorphicBlocks}.

The expression for~$\Omega_\rho$,~\eqref{TwistedPotentialDilaton}, and~$\mathcal{W}_\rho$,~\eqref{TwistedSuperPotential2}, can be interpreted as the dilaton and twisted superpotentials of a 3d~$\mathcal{N}=2$ A-twisted~$SU(2)$ Chern-Simons theory in a 3d manifold~$\mathcal{M}_{g,p}\,$ an oriented circle bundle of degree~$p$ over a closed Riemann surface $\Sigma_g\,$, as defined in~\cite{Closset:2017zgf}.~{The polynomial terms in~\eqref{TwistedPotentialDilaton} are contributions coming from the mixed gauge-R Chern-Simons terms to both the dilaton and twisted superpotential.}

\subsection{Reaching a sum over vacua of the~$A$-model}\label{sec:JKContour}
\begin{figure}\centering
\includegraphics[width=7cm]{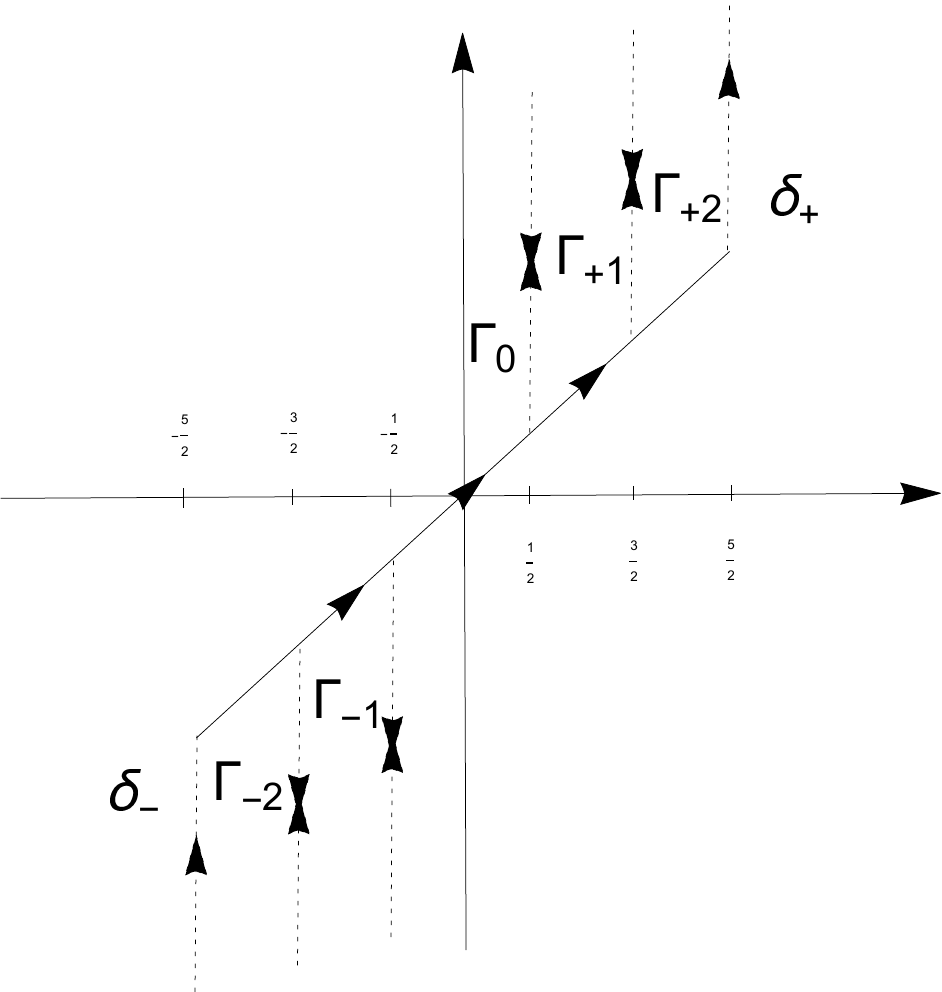} \includegraphics[width=7cm]{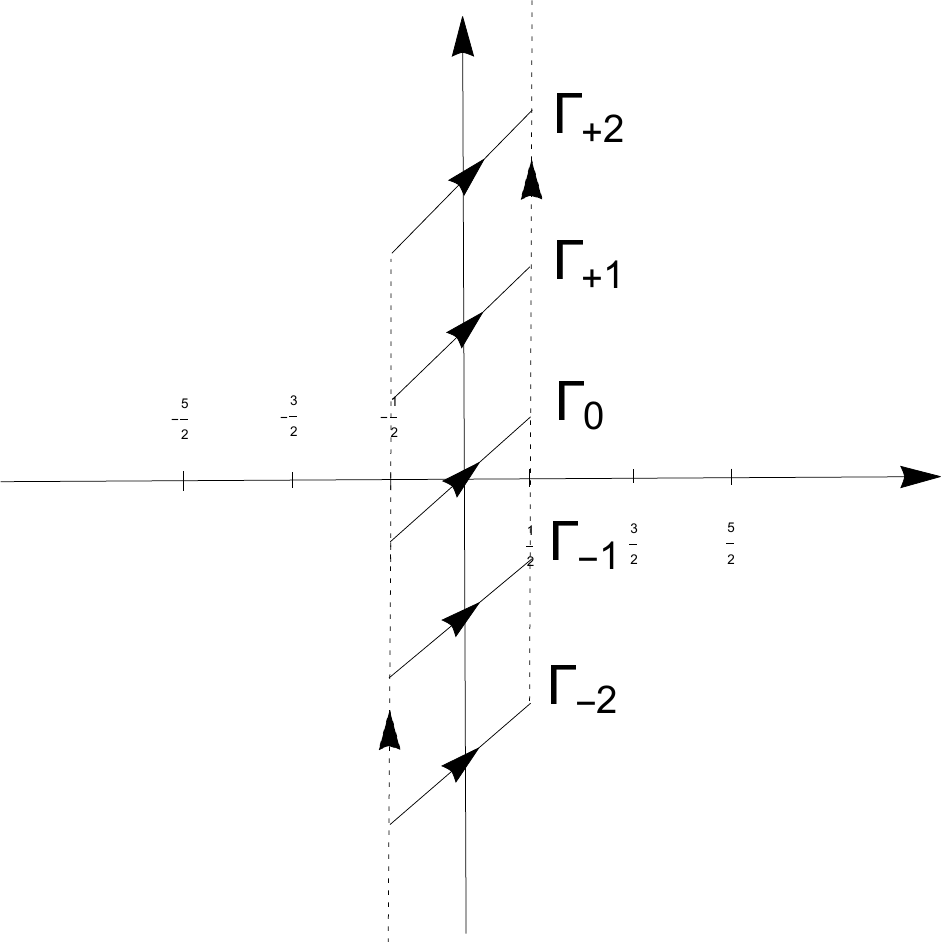} \includegraphics[width=7cm]{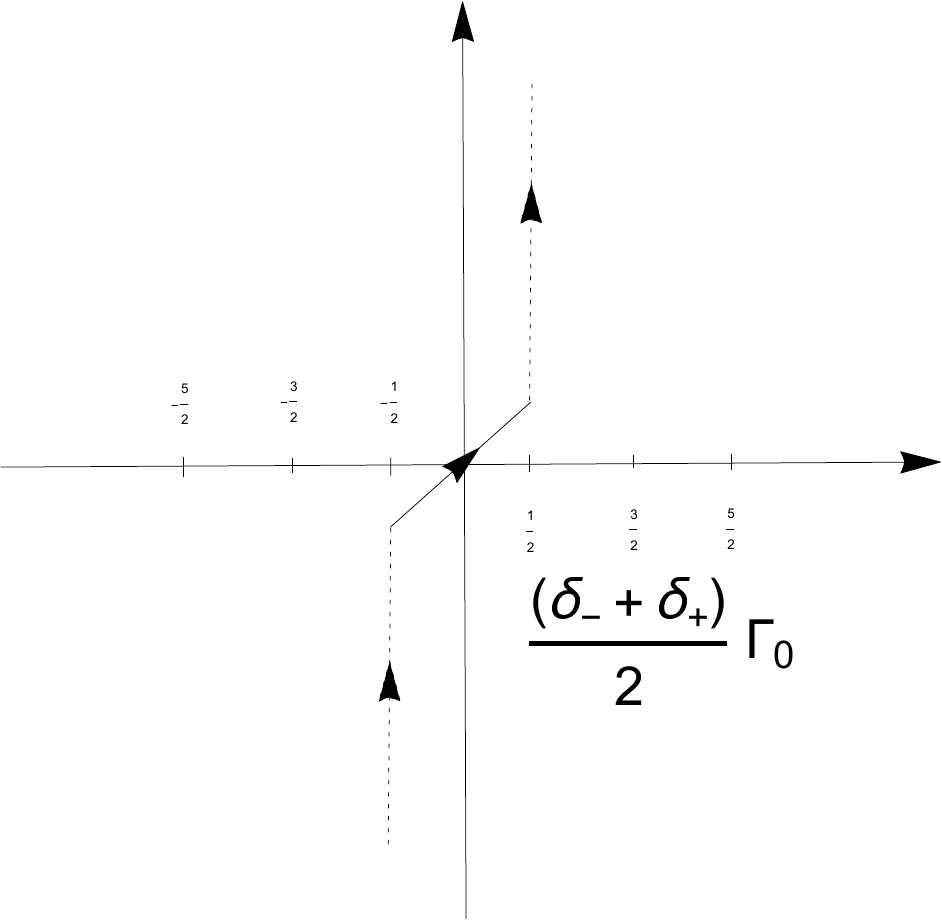} 
\caption{First, one defines a contour that goes to infinity in the quadrants that make integral~\eqref{ChiralBitExample} finite, an infinitesimal deformation of the~$(-\infty,\infty)$. Then the integral is divided in subintegrals~$\Gamma_{p\in \mathbb{Z}}$ that can be pushed to the fundamental domain~Re$u\in [-\frac{1}{2},\frac{1}{2}]$ by using the quasi-periodicity properties~\eqref{Quasiperiodicity}. In particular the integral along the deformed~$\Gamma_{p}$ has extra~$p$-powers of the function~$Q$ inserted. As the deformation is infinitesimal the addition of all the subintegrals can be recast as an integral over the contour $\Gamma_0$ with the insertion of a geometric series in~$Q$ resulting in denominator~$(1-Q)$. The residues that compute the latter integral come from the denominator. There are many details that go into this analysis and they will be revisited elsewhere. }
\label{JKContour}
\end{figure}

At last, we explain how the reduction of the remainning bit integrals
\be\label{ChiralBitExample}
\int_{\Gamma_{\alpha}} du \,\mathcal{C}_{m}[u,\ell]
\ee
to a sum over Bethe vacua occurs. As we are focusing on a single~$\rho$, it is fine to substitute~$\rho(u)\,\to\, 2u$ and drop the subindex~$\rho\,$. The integrand of a chiral bit takes the form
\be
\mathcal{C}_m[u, \ell]\,\propto\,e^{2\pi\i \,(g-1)\,\Omega \,+\,2\pi\i \, p\,\bigl(\mathcal{W} \,-\,u \,\partial_{u}\, \mathcal{W} \,-\, \gamma_2\,\partial_{\gamma_2}\, \mathcal{W}\bigr)}
\ee
This integrand is quasi-periodic under translations
\be\label{Quasiperiodicity}
\mathcal{C}_m(u \,+\,1)\=\,\mathcal{C}_m(u)\, Q(u)
\ee
where the quasiperiodicity factor takes the form
\be
\begin{split}
Q_m(u)&:\=\,\frac{e^{-2\pi\text{i} \,(4m u)}}{(1-e^{2\pi\i (2u+\gamma_2)})^{2}} \\
&\=  e^{-\,2\pi \text{i} \,m\,\partial_{u}\mathcal{W}}
\end{split}
\ee
and it is periodic
\be
Q^k (u+1)\=Q^k (u)\,.
\ee
Just as for the case of pure vector bits, convergence of the chiral bit integral
\be\label{ChiralBitExample}
\int_{\Gamma_{\alpha}} du \,\mathcal{C}_{}(u)
\ee
is determined by the Gaussian factor i.e. by the classical Chern-Simons contributions in~$\mathcal{C}_m[u,\ell]$. The contour $\Gamma_\alpha$ can then be taken to be an infinitesimal deformation of~$(-\infty,\infty)$ that makes the integral~\eqref{ChiralBitExample} convergent. One can then use the quasi-periodicity conditions to express that integral as an integral contour over the contour~$\Gamma_0$ in the figure~\ref{JKContour} which can then be evaluated by computing residues of the solutions to the Bethe-ansatz equation
\be
Q_m(u)\,=\,1
\ee
in the fundamental domain~$-\frac{1}{2}\,<\,\text{Re}(u)\,<\,\frac{1}{2}\,$. The details of this analysis, which are analogous, if not identical, to an approach studied by Closset, Kim and Willet in~\cite{Closset:2017zgf}, will be presented in forthcoming work.

\section{Questions/problems for the future}\label{Sec:Future}

Aside from some technical problems that has been left to address in forthcoming work, following the summary of main results given in the section~\ref{SectionIntroResults}, there are relevant conceptual problems that we find interesting to start thinking about in the near future.

\begin{itemize}
\item \emph{Is there a clear relation to approach 2) and to the observations made in~\cite{PaperPhases}?}  In the particular case of $m=n+1=1$, for instance, there is no average to perform. If we focus on the maximally symmetric bit contribution, then the corresponding partition function is~$SU(N)$ Chern-Simons theory, at quantum corrected level~$k=N$ on~$S_3$, which equals~$1$ (in some convention). So, broadly speaking this individual~$SU(N)$ Chern-Simons integrals are counting single conformal blocks of the universal~$SU(N)$ WZNW at classical level~$k=1$ of~\cite{PaperPhases}. It is possible that a clear relation between these two perspectives exists, in the same way as it is possible that a clear relation in between approaches 1) and 2) exist. It is also interesting to ask~\emph{how are the averages over products of~$\mathbb{Z}_m$'s is interpreted from the perspectives of approach~2) and~\cite{PaperPhases}?}

\item \emph{Can the coarse-grained expansion in sectors~$\alpha$ be interpreted as an expansion of a partition function of a (topological) string theory?} The sum over sectors~$\alpha$ can be organized in a sum over interacting or entangled topologically twisted theories that can be interpreted as two-dimensional theories -- with infinite degrees of freedom -- over Riemann surfaces of genus~$g$ which could be hinting at a string-theory interpretation.~\footnote{The hints of existence of a Gopakumar-Vafa duality~\cite{Gopakumar:1998ki}(as summarized, for instance, in~\cite{Brini:2008ik}) for Lens spaces~$L(p,1)$ are encouraging, given the fact that their open string theory description is related to the sector~$\ell_i=0$ of the maximally symmetric sector~\eqref{ZMaximalBits}.}

\item The so-called coupled sectors~$\alpha$ can involve coupled contributions coming from Riemann surfaces at different genus.~\emph{Is this gluing of contributions from different Riemann surfaces related to a geometrical operation among the underlying Riemann surfaces?}

\end{itemize}

\section*{Acknowledgements}
~It it a pleasure to thank A.A.Ardehali, S. Murthy and L. Pando Zayas for useful comments and/or discussions. This work was supported by the ERC Consolidator Grant N. 681908, “Quantum black holes: A microscopic window into the microstructure of gravity” . I am grateful to the Abdus Salam International Centre for Theoretical Physics (Trieste) for the support it has provided me during the completion of essential part of this work.

\appendix

\section{Definitions and Identities}
\label{app:Conventions}
We use the following relations between the fugacities $\zeta\,$,~$p\,$,~$q$ and the chemical potentials~$z\,$,~$\sigma\,$,~$\tau$
\be
\zeta\,\equiv\,\textbf{e}(z)\,,\, p\=\textbf{e}(\sigma)\,, \, q\,=\textbf{e}(\tau)\,.
\ee
We also define~$\textbf{e}(x)\,\equiv\,e^{2\pi\i x}$\,.
The~$q$-Pochammer symbol~$(\zeta;q)\equiv (\zeta;q)_{\infty}$ has the following product representation
\be
(\zeta;q)\= \prod_{j=0}^{\infty}(1\,-\,q^{j}\,\zeta)\,.
\ee
The quasi-elliptic function~$\th_0(z)=\th_0(z;\t)$ has the following product representation
\be\label{ThetaDef}
\th_{\text{ell}}(\z;q)\=\theta_0(z;\t) \,=\,  (1-\z) \prod_{j=1}^\infty (1-q^j \z) \, (1-q^j \z^{-1}) \,.
\ee
For latter use we note the following relation
\begin{equation}\label{IdentityPochhammer}
(q;q)^2\,=\,\underset{\zeta\to 1}{\text{lim}}\,\frac{\theta_{\text{ell}}(\zeta;q)}{(1-\zeta)}\,.
\end{equation}
The elliptic Gamma functions are defined out of the following product representation
\be\label{GammaeDef}
\G_{\text{ell}}(\z;p,q)\=\Ge(z;\s,\t) \=  \prod_{j,\,k=0}^{\infty}
\frac{1\,-\, \zeta^{-1} p^{j+1} q^{j+1}}{1\,-\,\zeta \,p^j\, q^k} \,.
\ee 
The first three Bernoulli polynomials are
\begin{eqnarray}
 B_1(z) &\,=\,& z-\frac{1}{2}\,,\\
 B_2(z) &\,=\,& z^2-z+\frac{1}{6}\,,\\
 B_3(z) &\,=\,& z^3-\frac{3 \,z^2}{2}+\frac{z}{2}\,.
\end{eqnarray}

\begin{eqnarray}
B_{1per}(z)\,\equiv\, B_1(\{z\}) &\,=\,& \Bigl\{\begin{array}{cc} B_1(z\,-\, \lfloor z \rfloor) &\qquad \text{for}
\qquad z\,\notin\,\mathbb{Z}\, \\ 0 & \qquad \,\text{for}\qquad z\,\in\,\mathbb{Z}\,  \end{array}\,,\\
 B_{2per}(z)\,\equiv\, B_2(\{z\}) &\,\equiv\,& B_2(z-\lfloor z\rfloor)\,,\\
B_{3per}(z)\,\equiv\,  B_3(\{z\}) &\,=\,& B_3(z-\lfloor z\rfloor)\,.
\end{eqnarray}

\begin{eqnarray}\label{Decomp1}
B_{1,1}(z-1\,|\,-1)&\,=\,& -\,B_{1}(z) \,,\\\label{Decomp2}
B_{2,2}(z-1\,| \,\tau,-1) &\,=\,& -\frac{1}{\tau }\, B_2(z)\,+\,B_1(z)\,-\,\frac{\tau }{6}\,,\\\label{Decomp3}
B_{3,3}(z-1\,|\,\sigma,\, \tau,-1) &\,=\,&-\,\frac{1}{\,\tau\,  \sigma }\,B_3(z)\,+\,
\frac{3\,  (\tau +\sigma ) }{2 \,\tau \, \sigma} \, B_2(z) \nonumber \\ 
&&\quad -\,\frac{1}{2} \left(\frac{\tau }{\sigma }\,+\,\frac{\sigma }{\tau}
	+3\right)  B_1(z)\,+\,\frac{\tau +\sigma }{4} \,.
\end{eqnarray}

Next we quote a couple of identities that could be useful for the reader to reproduce some of the results given below. Define
\be
z(y)\= y\,\T+\,\Delta\,,\,\quad \Delta\,\in\,\mathbb{C}\,,
\ee
then
\be\label{B22even}
\begin{split} 
\Delta B_{2,2}(u,\Delta)&\,\equiv\, B_{2,2}(z(y)|\T,-1)\,+\,B_{2,2}(z(-y)|\T,-1)\,-\, 2 B_{2,2}(z(0)|\T,-1)\\&\=-\,2\, y^2\, T \,,
\\
\Delta B_{3,3}(y,\Delta)&\,\equiv\, B_{3,3}(z(y)|\T,\T,-1)\,+\,B_{3,3}(z(-y)|\T,\T,-1)\,-\, 2 B_{3,3}(z(0)|\T,\T,-1)\\&\=6\, y^2\, B_1(T\,-\,\Delta) \,.
\end{split}
\ee
At last, assume
\be\label{B33even}
\Delta\= \widetilde{\Delta}\,+\,\ell \tau\,\equiv\,(\widetilde{\Delta}_2\,+\,\ell)\,\tau+\widetilde{\Delta}_1(\ell)\,,
\ee
then using identities
\be
\sum_{\ell\=0}^{2(m-1)}\,(m\,-\,|\ell\,-\,m\,+\,1|)(m\,-\,\ell)\= \sum_{\ell\=0}^{2(m-1)}\,(m\,-\,|\ell\,-\,m\,+\,1|)\= m^2\,,
\ee
one obtains
\be
\begin{split}
\sum_{\ell\=0}^{2(m-1)}\,(m\,-\,|\ell\,-\,m\,+\,1|)\,\Delta B_{3,3}(y,\widetilde{\Delta}\,+\,\ell \tau)&\= 6 m^2 y^2\, \Bigl((1\,-\,\widetilde{\Delta}_2)\,\frac{\T}{m}\,+\, n \widetilde{\Delta}_2\,-\,\frac{1}{2}\Bigr)  \\
&-\, 6 y^2\, \Bigl(\sum_{\ell\=0}^{2(m-1)}\,(m\,-\,|\ell\,-\,m\,+\,1|)\,\widetilde{\Delta}_1(\ell) \Bigr)\,.
\end{split}
\ee

\section{The contour decomposition at~$N=3$}\label{app:TheCountourDecN3}

Given two co-prime integers~$m$ and~$n\,$ and $n_0=\pm 1$, and the set of bit positions summarized in table~\ref{tab:Tabla} is ($\mathcal{\ell}\in \mathbb{Z}$)
\be
v(\ell)\,:=\,\frac{n\ell}{m}\text{mod}1 \,,\, 
c(\ell)\,:=\,- \frac{(m n_0 -n)}{m}\frac{1}{3}\,+\,\frac{n\ell}{m}\text{mod}1 \,,\,
b^{\prime}(\ell)\,:=\,- \frac{(m n_0 -n)}{m}\frac{2}{3}\,+\,\frac{n\ell}{m}\text{mod}1\,,
\ee
The goal is to derive the particularization of the contour integral~\eqref{Decomposition} for~$N=3$.
Let us define the integration variables in~$\mathcal{I}$ as
\be
\{w_i\}:= \{v_{12},v_{23}\}\,, 
\ee
then the arguments in the factors~$I(v_{ij})$ in the integrand are
\be
\{v_{ij}\}=\{w_1, w_2 , w_1+w_2=v_{13}\}\,.
\ee
Then the integral decomposition
\be
\begin{split}
\int_{-1}^{1} dw_1 \int_{-1}^{1} dw_2 &\,=\,\int_{D^{(2)}} dw_1 dw_2 + \,\\& \,\sum_{\ell_1=0 }^{2(m-1)}\sum_{f_1\in \{v,c,b^\prime\}}\,\Bigl(\int_{D^{(1,1)}} dw_1 dw_2\,+ \,\int_{D^{(1,2)}} dw_1 dw_2 + \,\int_{D^{(1,3)}} dw_1 dw_2 \Bigr)\\&\,+\,\sum_{\ell_1, \ell_2
=0 }^{2(m-1)}\sum_{f_{1}, f_{2
}\in \{v,c,b^\prime\}}\,\int_{D^{(0)}} dw_1 dw_2   
\end{split}
\ee
\be\label{eq:ContourDecompositionAppendix}
\begin{split}
\,\underset{\varepsilon=0^+}{=}\,& \int_{D^{(2)}} dw_1 dw_2 \,+ \, \,3\,\times\,\sum_{\ell_1=0 }^{2(m-1)}\sum_{f_1\in \{v,c,b^\prime\}}\,\int_{D^{(1,1)}} dw_1 dw_2
\\ \qquad&\,+\,\sum_{\ell_1, \ell_2
=0 }^{2(m-1)
}\sum_{f_{1}, f_{2
}\in \{v,c,b^\prime\}}\,\int_{D^{(0)}} dw_1 dw_2
\end{split}
\ee
follows from the definition of the domains
\be
\begin{split}
D^2&:=\{x_1\,\neq_{\varepsilon}\, all \,bits\,,\, x_2\,\neq_{\varepsilon}\, all \, bits\,,\, x_1\,+\,x_2\,\neq_{2\varepsilon}\, all \, bits\}\\
D^{1,1}=D^{1,1}(f_
1,\ell_1)&:=\{x_1\,=_\varepsilon\, f_1(\ell_1)\,,\, x_2\,\neq_{\varepsilon}\, all \, bits\,,\, x_1\,+\,x_2 \,\neq_{2\varepsilon}\, all \, bits\}\\
D^{1,2}=D^{1,2}(f_
1,\ell_1)&:=\{x_1 \,\neq_{\varepsilon}\, all \, bits\,,\, x_2\,=_\varepsilon\, f_1(\ell_1)\,,\, x_1\,+\,x_2\,\neq_{2\varepsilon}\, all \, bits\}\\
D^{1,3}=D^{1,3}(f_
1,\ell_1)&:=\{x_1\,\neq_{\varepsilon}\, all \, bits\,,\, x_2\,\neq_{\varepsilon}\,  all \,bits\,,\, x_1\,+\,x_2\,=_{2\varepsilon}\, f_{1}(\ell_1)\}\\
D^0=D^0(f_{1,2},\ell_{1,2})&:=\{x_1\,=_\varepsilon\, f_1(\ell_1)\,,\, x_2\,=_{\varepsilon}\, f_{2}(\ell_2)\,,\, x_1\,+\,x_2\,=_{2\varepsilon}\, f_{1}(\ell_1)+f_{2}(\ell_2
)\},
\end{split}
\ee
which obey
\be
D^2 \cup D^{1,1} \cup D^{1,2}\cup D^{1,3} \cup D^{0} = [-1,1]^2\,\qquad, \qquad D^{I}\, \cap\, D^{J} \,=\,0 \quad\text{if}\quad \,I\neq J\,,
\ee
~The $x \neq_\varepsilon y$ means~$|x-y|>\underline{\varepsilon =0^+}$ and~$x=_\varepsilon y$ means~$|x-y|\,\leq\,\underline{\varepsilon= 0^+}\,$ ($\varepsilon$ should be thought of as an infinitesimally small number).
The equality~\eqref{eq:ContourDecompositionAppendix} follows from the fact that at $\varepsilon=0^+$ the~$D^{1,1}$,~$D^{1,2}$,~$D^{1,3}$ are isomorphic,~\footnote{At finite~$\varepsilon$ one can almost transform a~$D^{1,3}$ into a~$D^{1,1}$ using the change of variables
\be
x_1=\,-y_2\,,\,x_1+x_2= y_1\,.
\ee
The boundaries of the transformed domain do not remain the ones that we have used in the definition of~$D^{1,1}$ though, but the difference is irrelevant in the limit $\varepsilon=0^+\,$. This is, the integral along the bits is independent on how fast the width of the bit is shrinked to zero.
} and that their corresponding integrals are identified upon a redefinition of variables.
 At last, we note that
\be
\begin{split}
D^2 &= \mathcal{M}_{b}(\lambda=2)\,,\, \\
D^{1,1},D^{1,2},D^{1,3}&\underset{\varepsilon\,=\,0^+}{=} \mathcal{M}_\varepsilon(\widetilde{\lambda}=1)\otimes \mathcal{M}_{b}(\lambda=1)\,,\\
D^{0}&= \mathcal{M}_{\varepsilon}{(\widetilde{\lambda}=2)}\,,
\end{split}
\ee
which implies that
\be\label{app:DecompositionFormula}
\begin{split}
\int_{-1}^{1} dw_1 \int_{-1}^{1} dw_2&\,=\,\int_{\mathcal{M}_b(\lambda=2)} d\underline{y}\,+\,3\,\sum_{\ell_1=0}^{2(m-1)}\sum_{f_1 \in \{v,c,b^\prime\} }\int_{\mathcal{M}_\varepsilon(\widetilde{\lambda}=1)} dx \int_{\mathcal{M}_{b}(\lambda=1)} dy \\ &+ \sum_{\ell_1, \ell_2
=0 }^{2(m-1)
}\sum_{f_{1}, f_{2
}\in \{v,c,b^\prime\}}\,\int_{\mathcal{M}_{\varepsilon}{(\widetilde{\lambda}=2)}} d\underline{x}.
\end{split}
\ee
\eqref{app:DecompositionFormula} is the particularization of the contour decomposition in~\eqref{Decomposition} for~$N=3$.

\section{Chern-Simons partition function on~$S_3$}\label{app:CSW}
In this appendix, we compute the integral
\begin{equation}\label{CSMIntegral}
\ZGUN\,\equiv\,
\frac{1}{N!}\,\int_{\Gamma^N} \frac{d^N\sigma}{(2\pi)^N}\, e^{-S_{\text{CS},\, k}(\sigma)}  \=\frac{1}{N!} \,\int_{\Gamma^N} \frac{d^N\sigma}{(2\pi)^N}\, \Bigl(\prod_{s=1}^{N} f_{k,\xi}(\sigma_s)\Bigr)\, \Delta_N(\sigma)
\end{equation}
where
\begin{equation}
f_{k,\xi}(\sigma)\,\equiv\,e^{-\,\frac{\text{i} \,k}{4\pi}\,\sigma^2\,-\,\text{i}\, \frac{\xi}{2}\, \sigma}\,, \qquad \Delta_N (\sigma)\,\equiv\, \prod_{i\neq j}2 \sinh{\frac{\sigma_i-\sigma_j}{2}}\,.
\end{equation}
The contour~$\Gamma^N$ will be defined below. Starting from the Weyl denominator formula
\be
\prod_{i< j}2 \sinh{\frac{\sigma_i-\sigma_j}{2}}\,=\,\det_{i,j}\Bigl(e^{\sigma_i (j-\frac{N+1}{2})}\Bigr)
\ee
we obtain
\begin{equation}\label{FormulaTwo}
\Delta_N (\sigma)\,=\,\det_{i,j}\Bigl(e^{\sigma_{i} j}\Bigr)\,\times\,\det_{k,\ell}\Bigl(e^{-\sigma_k \ell}\Bigr)\,.
\end{equation}

From~\eqref{FormulaTwo} and identity
\begin{equation}\label{IdentityInte}
\begin{split}
\int dx_i \,\Bigl(\prod_{p=1}^N h(x_p)\Bigr)\,\Bigl(\det_{i,j}{g_{1i}(x_j)}\Bigr)\,\Bigl(\det_{i,j}{g_{2i}(x_j)}\Bigr)\,=\, N!\,\det_{i,j}\Bigl(\int dx\, h(x)\,g_{1i}(x) g_{2j}(x)\Bigr)\,,
\end{split}
\end{equation}
with~$i$ and~$j=1,\ldots, N\,$, and~$g_{1 i}(\sigma)\to e^{\sigma i}$ and~$g_{2\text{j}}(\sigma)\to e^{-\sigma j}\,$, we obtain
\begin{equation}\label{DeterminantalFormula}
\begin{split}
   \ZGUN&\,=\,\det_{ij}{\Bigl(\int_\Gamma \frac{\diff\sigma}{2\pi}\,f_{k,\xi}\, e^{\sigma\, (i-j)}\Bigr)}\,.
    \end{split}
\end{equation}
If the boundary~$\partial \Gamma$ is composed by two points in the complex $u$-plane at polar coordinates~$\pm\, r \,e^{\text{i} \phi}\,$, with~$r\gg1$ and~$0\,<\,\phi\,<\,\pi$ then the single-variable integral in the determinant~\eqref{DeterminantalFormula} is convergent iff
\begin{align}\label{ContourApp}
\left\{\begin{array}{cc}
0\,<\,\phi\,<\,\frac{\pi}{2}& \text{for} \qquad k\,<\,0\,, \\
\frac{\pi}{2}\,<\,\phi\,<\,\pi&\text{for} \qquad k\,>\,0\,.
\end{array}\right.
\end{align}
In those cases, one obtains
\begin{equation}\label{FormulaCS}
\int_\Gamma \frac{d\sigma}{2\pi}\,f_{k,\xi}(\sigma)\, e^{\sigma\, (i-j)}\,=\,\signo\,\frac{e^{-\frac{\text{i} \,\pi\,(-i+j+\text{i}\xi)^2}{k}}}{\sqrt{\text{i}\,k}}\,+\,\ldots\,,
\end{equation}
assuming~$\Gamma$ runs from~$-r e^{\i \phi}$  to~$r e^{\i \phi}\,$, namely, from the lower half-plane to the upper half-plane. 
The~$\ldots$ in~\eqref{FormulaCS} stand for non-pertubative corrections in the~$\frac{1}{r}$ expansion, that can be exactly computed. Finally, combining~\eqref{DeterminantalFormula} with~\eqref{FormulaCS}, and after algebraic manipulations we obtain
\be\label{UNCSIntegralApp}
\ZGUN\underset{r\to \infty}{\longrightarrow}\,{\signo^{N}}\,\frac{e^{-\frac{\pi\i}{6\,k} \,N\,(N+1)\,(N-1)}}{(\i\,k)^{\frac{N}{2}}}\,\prod_{L\=1}^{N-1}\bigl(2\,\i \sin \frac{\pi L}{k}\bigr)^{N-L}\,.
\ee
This result is up to a normalization the~$U(N)$ Chern-Simons partition function on~$S_3\,$.

\section{The~$\theta_0$ and~$\Gamma_e$ along the~$0$-th wall}
\label{app:MoreOnMNReps}

In reproducing many of the results below given, the reader may come across quantities that hit a discontinuity coming from the use of~\emph{floor} and~\emph{ceiling} functions. Below, we will deal with these technical difficulties by introducing \emph{ad hoc} deformations and then taking limits. There is a simpler way though to recover the correct answer:

\emph{Rule:} Each time a discontinuity is encountered in computing a function~$f$, the value of~$f$ at the discontinuity matches the semi-sum of its lateral limits. 

Recall that for~$\theta_0$ the~$(m,n)$-representations are
\begin{align}\label{IdentityTheta0}
\log \theta_0(z)\,\equiv\, \pi\, \text{i}\,\sum_{\ell=0}^{m-1} B_{2,2}(\xi_{\ell}|\T,-1)\,+\,\text{i}\sum_{j=1}^{\infty} \frac{1}{j \sin \frac{\pi j}{\T}}\sum_{\ell=0}^{m-1} \,\cos\Bigl(\pi j\,\frac{2\xi_{\ell}+1}{\T} \Bigr)\,,
\end{align}
where
\be
\widetilde{\tau}\,\equiv\,m \,\tau\,+\,n\,,
\ee
and
\begin{equation}\label{DefinitionZetaLTau}
\xi_{\ell}\,=\, \xi_{\ell}(z;\tau)\,=\, \xi_{\ell}(z)\,\equiv\,z\,-\,\text{Integer}\,+\,\ell\, \tau\,.
\end{equation}
The Integer is selected by the condition
\begin{equation}\label{ConditionNoLog}
-1<\xi_{\ell \perp}<0
\end{equation}
where the real numbers~$\xi_{\ell \perp}\,$, and its~\emph{dual}~$\xi_{\ell ||}\,$, are components of the complex number~$\xi_{\ell}$ in the basis of the complex plane given by~$1$ and~$\T\,$, i.e.~$\xi_{\ell}=\xi_{\ell\perp}+\T \xi_{\ell ||}\,$.
For later convenience we define
\be\label{DefVU}
v\,=\, u \T\,\equiv\,z\={z}(u)
\ee
and assume, for the moment,~
\be\label{RegionUInitial}
u\,=\,\frac{z(u)}{\T}\,\in\, \mathbb{R}_{\pm}\,.
\ee

Generically, in an expansion around~$\tau=-\frac{n}{m}$ the second term in the right-hand side of~\eqref{IdentityTheta0} is exponentially suppressed. Only for values of~$\ell$ for which~$\xi_{\ell}$ is infinitesimally close to, either~$0$ (from below) or~$-1$ (from above), the latter term becomes relevant. Let us assume that only~$\zeta_{\ell=0\perp }$ is infinitesimally close to~$0$ or~$-1\,$, then
\begin{align}
\log \theta_0(z)&\,\underset{\tau\to -\frac{n}{m}}{\longrightarrow}\, \pi\, \text{i}\,\sum_{\ell=0}^{m-1} B_{2,2}(\xi_{\ell}(z)|\T,-1)\,\nonumber\\ &+\,\text{i}\sum_{j=1}^{\infty} \frac{1}{j \sin \frac{\pi j}{\T}} \,\cos\Bigl(\pi j\,\frac{2\zeta_{\ell=0}(z)+1}{\T} \Bigr)\,+\,\ldots\,,\nonumber\\&\,=\, \pi\, \text{i}\,\sum_{\ell=0}^{m-1}B_{2,2}(\xi_{\ell}(z)|\T,-1)\,-\,\sum_{j=1}^{\infty} \frac{\cos\Bigl(2\pi j u \Bigr)}{j} \,+\,\ldots\,. \label{SeriesCos}
\end{align}
A computation shows that
\begin{equation}\label{Limitnm}
\log \theta_0(z)\,\underset{\tau\to -\frac{n}{m}}{\longrightarrow}\, \pi\, \text{i}\,\sum_{\ell=0}^{m-1}B_{2,2}(\xi_{\ell}(z)|\T,-1)\,+\,\log{\Bigl(2\,|\sin \pi {u}|\Bigr)}\,+\,\ldots\,.
\end{equation}
The~$\ldots$ denote corrections that are odd under~$u\to-u\,$. Note that the logarithm term does not depend on~$m$ and~$n\,$. To resum the series in~\eqref{SeriesCos} to a logarithm, we have also assumed~$u\notin \mathbb{Z}\,$.
We have also assumed
\begin{equation}\label{Conditionzeta}
\zeta_{\ell=0}(\tau)   \,=\, \left\{\begin{array}{cc}
      u \,\T+0^- &\,\equiv\, u^-\,,  \\
     \,u \,\T-1^+ \,. & 
    \end{array}\right.
\end{equation}
and moreover, that every other~$\xi_{\ell} (\tau)\,$, with~$\ell\neq0$ is such that~$\xi_{\ell}$ is not infinitesimally close to either~$0^-$ or~$-1^+\,$. Later on we will relax the latter assumption, and find there can be contributions coming from other values of~$\ell\,\neq\, 0\,$.

From the identity
\be
\sin (\pi |u|)^2 \= (-1)\times \sin (\pi u) \sin (\pi (-u))\,
\ee
it follows that one can drop the absolute value in the argument of the logarithm in~\eqref{Limitnm}, and use instead
\be\label{IdentityTheta02}
\begin{split}
\log \theta_0(z)\theta_0(-z)&\underset{\tau\to -\frac{n}{m}}{\longrightarrow}\, \pi\, \text{i}\,\sum_{\ell=0}^{m-1}B_{2,2}(\xi_{\ell}(z)|\T,-1)\,+\,(z\,\to\,-z)\,\\&\qquad\,+\,\log{\Bigl(2 \,\sin \pi {u}\Bigr)\Bigl(2 \,\sin \pi {(-u)}\Bigr)}\,+\,\log(-1)\,.
\end{split}
\ee

As we will further explain in subsection~\ref{app:GeneralLogs}, the equation~\eqref{IdentityTheta02} can be used in a thin ribbon of the complex plane parallel to the rays~$u\in\mathbb{R}\,$. More precisely, it can be extended to~
\be
u \= v\=v_{||}\,+\,\frac{v_\perp}{\T}\,\in\, \mathbb{C}\,,
\ee
where~$v_{||}$ is an arbitrary real number and
\be
-|\T|\,\delta\,<\,v_{\perp}\,<\,|\T|\,\delta.
\ee
The~$\delta\in\mathbb{R}$ is to be identified with the splitting parameter of the contour of integration.~$\delta$ is independent of~$\tau\,$.

\paragraph{Expansion of the~$q$-Pochhammer symbol} Assuming~$z$ lies in the domain defined by equations~\eqref{DefVU} and~\eqref{RegionUInitial}, and using the expansion for~$\log \theta_0$ given in~\eqref{Limitnm}, together with identity~\eqref{IdentityPochhammer} we obtain
\be\nonumber
\begin{split}
\log\,(q;q)&\,=\,\frac{1}{2}\,\,\underset{z\to0}{\lim}\,\log\,\frac{\theta_0(z)}{(2\pi\text{i}z)} \\\label{ExpPoch} &\underset{\tau\to -\frac{n}{m}}{\longrightarrow}\, \frac{\pi\, \text{i}}{2}\,\sum_{\ell=0}^{m-1}B_{2,2}(\xi_{\ell}(z\to 0)|\T,-1)\,-\,\frac{1}{2}\,\log{\Bigl(\text{i}\, \T\Bigr)}\,.
\end{split}
\ee
The derivation of the relation in the first line above can be found, for instance, in equation (1.11) of~\cite{Dolan:2008qi}.
We have left the limit~$z\to 0$ in the second line because the result obtained by naively substituting~$z$ by~$0$ does not match the limits. Due to condition~\eqref{Conditionzeta}, we are interested in the result of the limit~$z\to 0^+$ which up to exponentially suppressed contributions is
\be\label{QPochIdentity}
\begin{split}
\log{(q;q)}&\,\underset{\tau\to -\frac{n}{m}}{\longrightarrow}\,-\,\frac{ \pi\i }{12 m \T}\,-\,\frac{\pi\i  \T}{12 m}
\\
&+\pi\i\,\sum_{\ell\=0}^{m-1} \frac{ \, (m-2 \ell) B_1\bigl(\{\ell\frac{ n}{m}+0^{+}\}\bigr)}{2 m}\,-\,\frac{1}{2}\,\log{\Bigl(\text{i}\, \T\Bigr)}\,.
\end{split}
\ee

\paragraph{Contribution from vector multiplets}~\footnote{Our treatment of the vector multiplet contribution will not require the use of the regulator~$\epsilon$ used in~\cite{Cabo-Bizet:2019eaf}. The difference one obtains by using the $\epsilon$-regularization of~\cite{Cabo-Bizet:2019eaf} arises at order~$\mathcal{O}(1)$ in the large-$N$ expansion, more precisely, in the pure imaginary constant~$\varphi$ that was left undetermined in such a reference, and later on fixed in~\cite{Cabo-Bizet:2020ewf}. For the same reason, should one used the~$\epsilon$-regularization, for instance for~$(m,n)\,=\,(1,0)$ one obtains~$k_v\,+\,3k_I\,=\,0$ as one can check from the data reported in Tables 3, 4, and equation~$(C.41)$ of~\cite{Cabo-Bizet:2019eaf}. For the approach here taken, we obtain the exact result, which is different from zero i.e.~$k_v\,+\,3k_I\,\neq\,0\,$. The approach here presented fixes this error, which was introduced by the use of the~$\epsilon$ regularization in~\cite{Cabo-Bizet:2019eaf}.} This contribution comes from using the following definition
\be \label{FcdVector}
- \,2\,\mathcal{F}_{V}^{(m,n)}(x) \={\pi\,\i }  \, \sum_{\ell=0}^{(m-1)}  \,
\Big( B_{2,2} \bigl(\xi_{\ell}(z(x))| \T,-1 \bigr) + \left(x \rightarrow -x\right)\Big)\,.
\ee
and~$x\in\mathbb{C}\,$. In this appendix we are interested in the specialization~$x=u^-$. 
A computation shows that~\footnote{This result also follows from the naive use of~\eqref{B22even}.}
\be\label{auxeq2}
- 2\, \,\mathcal{F}^{(m,n)}_V (u) \,+\,2\,\mathcal{F}^{(m,n)}_V (0)\,+\,   2\pi \i m^2\, \tau \, u^2\,= \kappa_{V} \,\pi \i \,u^2\,.
\ee
In this equation~$\kappa_V \,=\,-\, 2 m n\in \mathbb{Z}$ and
\be\label{DefinitionF}
-\,\mathcal{F}^{(m,n)}_V (0)\= -\frac{ \pi \,\i}{6 m}
\,\frac{1}{\T}\, +\,{2\,\pi\,\i} \,  \sum _{l=0}^{m-1} \frac{\ell}{m} \,B_1\bigl(\Bigl\{\frac{\ell n}{m}\Bigr\}\bigr)\,-\,\frac{\pi\,\i}{6 m}\, \T\,.
\ee

\subsection{The expansions of~$\Gamma_e$ along the~$0$-th wall}

To compute chiral multiplet contributions we use the representations
\begin{equation} \label{repmnG}
\begin{split}
\log \Gamma_e(z;\, \tau,\,\tau) & \,=\, \sum_{\ell=0}^{2 (m-1)}  \frac{\pi i}{3}\,(m-|\ell-m+1|)\, B_{3,3}(\xi_{\ell}|\T,\T,-1)  \\
& \; \,+\,  \sum_{\ell=0}^{2 (m-1)} {\text{i}}\,(m-|\ell-m+1|)\,\Bigl(\,\sum_{j=1}^{\infty}  \frac{-\,2\, \T +2 \xi_{\ell}+1 }{2 \,j\, \T\, \sin \left(\frac{\pi  j}{\T }\right) }\,\times\\&\quad
	 \cos  \frac{\pi j (2\xi_{\ell}+1)}{\T }\,- \, \frac{\pi  j \cot \left(\frac{\pi  j}{\T }\right)+\T }{2 \pi\,  j^2\, \T \, \sin \left(\frac{\pi  j}{\T }\right)} \,
	\sin{ \frac{\pi  j (2\xi_{\ell} +1)}{\T}}\,\Bigr) \,.
\end{split}
\end{equation}
For coprimes~$m\,$ and~$n\,$, complex~$z= u \T$ and~$u$ real, the~$\xi_{\ell}$ was defined in~\eqref{DefinitionZetaLTau}.
\begin{itemize}
\item[a)] For~$m$ and~$n$ such that for every~$0\leq\ell\leq 2(m-1)$ none of the~$\xi_{\ell}$ is of the form~\eqref{Conditionzeta}, thus
\end{itemize}
\begin{equation}
\log \Gamma_e(z;\, \tau,\,\tau) \,\liMN\, \sum_{\ell=0}^{2 (m-1)}  \frac{\pi i}{3}\,(m-|\ell-m+1|)\,\, B_{3,3}(\xi_{\ell}|\T,\T,-1)\,,
\end{equation}
up to exponentially suppressed contributions. In this last expression~$z$ and~$u$ can be generic complex numbers. For the cases we will study, condition~a) will be always satisfied.

For convenience we define, for~$x\in\mathbb{C}\,$, 
\be \label{Fcd}
- \,2\,\mathcal{F}_{I}^{(m,n)}(x) \,\equiv\,\frac{\pi\,\i }{3}  \, \sum_{\ell=0}^{2(m-1)} \, (m-|\ell -m+1|) \,
\Big( B_{3,3} \bigl(\xi_{\ell}(z_I(x);\tau)| \T,\T,-1 \bigr) + \left(x \rightarrow -x\right)\Big)\,.
\ee
The label~$I$ refers to the~$I$-th~$\mathcal{N}=1$ chiral multiplet with~$U(1)$ $R$-charge~$r_I$ and~
\be
z_I(v)\,\equiv\, v\,+\,\frac{r_I}{2} (2\tau-n_0)\,=\, u \T+\frac{r_I}{2} (2\tau-n_0)\,.
\ee
As noted in~\cite{Cabo-Bizet:2019eaf}~\footnote{This can be shown using~\eqref{B33even} with~$y=\rho(u)$,~$\widetilde{\Delta}_{2}=r_I$ and~$\widetilde{\Delta}_{1}=-\,\Bigl(\frac{n_0 r_I}{2}\,-\,\lfloor\frac{n (\ell\,+\,r_I)}{m}\,+\,\frac{n_0 r_I}{2}\rfloor\Bigr)\,$.}
\be\label{auxeq2}
- 2\, \,\mathcal{F}^{(m,n)}_I (z_I(v)) \,+\,2\,\mathcal{F}^{(m,n)}_I (z_I(0))\,+\,   2\pi \i m^2\, \tau \, u^2 \, (r_I-1)\,= \kappa_{I} \,\pi \i \,u^2\,,
\ee
where the real number~$\kappa_I$ is
\be
\kappa_I\,\equiv\,\sum _{\ell=0}^{2 (m-1)} \frac{(m-\left| \ell-m+1\right| ) \left(2 m\, B_1\Bigl(\Br{\frac{n(\ell\,+\, r_I)}{m}+\frac{n_0 r_I}{2}}\Bigr)\,-\,2 n (\ell+r_I-m)\right)}{m}\,.
\ee

\section{The integrand of~$\mathcal{I}(q)$ along the~$0$-th wall}\label{EffectiveActionComputation}
For the $R$-charge of a chiral multiplet in~$\mathcal{N}=4$ SYM~$r_I=\frac{2}{3}\,$, the~$k_I$ is an integer multiple of~$\frac{1}{3}\,$.~In $\mathcal{N}=1$ language the theory is built out of three chiral multiplets of $R$-charge~$r_I=\frac{2}{3}$ labeled as~$I=1,2,3$, and a vector multiplet labeled as~$V$. If one defines
\be\label{DefF}
\mathcal{F}^{(m,n)}(\rho(v)) \,\equiv\,\sum_{\text{a}\,\in\,\{\,V,\,I\,=\,1,\,2,\,3\}}\mathcal{F}_{\text{a}}^{(m,n)}(z(\rho(v)))\,,
\ee
and uses definitions~\eqref{FcdVector} and~\eqref{Fcd}, one obtains
\be\label{ExpansionF0}
-\,2\, \mathcal{F}^{(m,n)}(\rho(v)) \=-\,2\, \mathcal{F}^{(m,n)}(0)\,+\, \kappa_{m,n} \,\pi \i \,\rho(u)^2\,,
\ee
where
\be\label{KappaMN}
\kappa_{m,n}\,\equiv\, \kappa_V\,+\, 3\, \kappa_{I}\= m\,\chi_1(m\,-\,n_0\,n)\,,
\ee
and~$\chi_1(x)\,\equiv\,\{-1,0,1\}$ if~$x\,\text{mod}\,3\=\{2,0,1\}\,$, respectively.
Using~\eqref{Fcd} a computation shows that~\cite{Cabo-Bizet:2019eaf}
\begin{eqnarray}\label{ConstantPhase0}
 \mathcal{F}^{(m,n)}(0)&\=&\mathcal{S}_{(m,n)}\,+\,\pi\i O( \tau^0)\,,
\end{eqnarray}
where
\be
\begin{split}
\mathcal{S}_{(m,n)}\,\equiv\, \frac{\pi\i}{27 m}\frac{(2 \T-n_0 \chi_1(m-n_0n))^3}{\T^2}\,.
\end{split}
\ee
The constant phase in~\eqref{ConstantPhase0} can be explicitly evaluated from~\eqref{FcdVector} and~\eqref{Fcd}.

\subsection{The effective action along a vector wall} \label{app:ExpansionAction}
Let us write down the profile of the effective action along the contour~$\Gamma^{p=0}_{m,n}$ i.e. for real~$u\=\frac{v}{\T}\,$. This is the profile that will be relevant to compute the contribution of the~$p=0$ bit and its other~$N-1$ replica images. 

\begin{figure}\centering
\includegraphics[width=9cm]{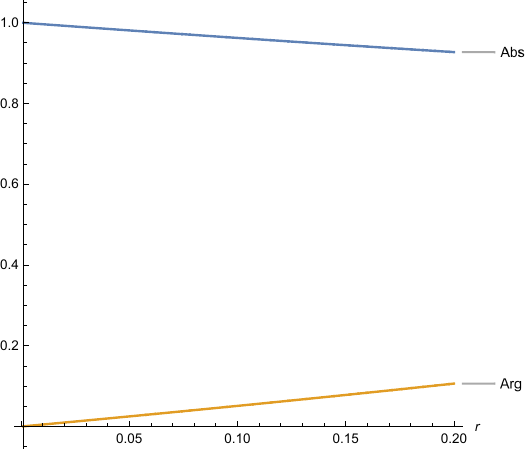} 
\caption{Numerical plot of the absolute value (Abs) and argument(Arg) of the quotient between the left and right-hand sides of equation~\eqref{IdentityHalfNEq} in the case~$N=k=2$ and~$\frac{1}{2\delta}= r e^{\frac{2\pi \i}{3}}$ with~$r=|\frac{1}{2\delta}|\,$. Note that for~$\frac{1}{2\delta}\longrightarrow0$ the result tends to one, as predicted by~\eqref{IdentityHalfNEq}.}
\label{IdentityhalfN}
\end{figure}

Collecting previous results we obtain
\be\label{EffectiveActionSmn}
\begin{split}
S_{\text{eff}\,p=0}(u)&\,\liMN\,\sum_{\rho\,\neq\,0}\,\Bigl( \mathcal{F}^{(m,n)}(\rho(v))\,-\, \log 2\sin \rho(u)\Bigr)\,+\,\mathcal{O}(1)\,.
\end{split}
\ee
Plugging the adjoint weights, and using the relations~\cite{GonzalezLezcano:2020yeb}
\be
\begin{split}
\frac{1}{2}\,\sum_{\rho} (\rho(u))^2 \=\frac{1}{2}\,\sum_{i,j=1\atop i\neq j}^{N} (u_i\,-\,u_j)^2&\= N\, \sum_{i\=1}^N \Bigl(u_i\,-\,\frac{1}{N} \sum_{j=1}^N u_j\Bigr)^2
\end{split}
\ee
we obtain
\be
\begin{split}
\int_{\Gamma_{m,n}} \prod_{i=1}^N \,du_i\,e^{-S_{\text{eff}\,p=0}(u)}\,
\liMN e^{-(N^2-N)\,\mathcal{S}_{(m,n)}\,+\,\pi\i \mathcal{O}(1) }\,\int_{\Gamma_{m,n}} \prod_{i=1}^N \,\diff u_i \,e^{ V(u)}\,,
\end{split}
\ee
where the constant phase is under control, although we do not report it's analytic expression, and
\be\label{appCSWAction}
\begin{split}
\,V(u)&\=\pi\i\, k \sum_{i\,=\,1}^N \Bigl(u_i\,-\,\frac{1}{N}\sum_{j\,=\,1}^N u_j\Bigr)^2\, +
\,\sum_{i\,\neq\, j}\log 2 \sin{ \pi u_{ij}}\,,
\end{split}
\ee
with
\be
k\,\equiv\, n_0 N m\,\chi_1(m\,-\,n_0 n)\,.
\ee
Note that for limits~$(m,n)$ such that~$\chi_1(m\,-\,n_0 m)=0$ there is no exponential growth of the integrand of the index, as $|e^{-\mathcal{S}_{(m,n)}}|$ is of order one, and moreover there is no polynomial contribution in~$u\,$. 
 At last, we prove the identity
\be\label{IdentityHalfNEq}
\begin{split}
\int_{\Gamma_{m,n}} \prod_{i=1}^N \,du_i e^{V(u)}&\underset{\delta\,\to\,\infty}{\simeq}\, 2\delta\,\times\, {N}\,\int_{\Gamma_{m,n}} \prod_{i=1}^{N-1} \,du_i e^{V(u)}\,,
\end{split}
\ee
where in the right-hand side~$u_N\=-\sum_{i=1}^{N-1} u_i\,$. 
\noindent\emph{Proof:} 
\be\label{IntegralOriginalApp}
\begin{split}
\int_{\Gamma_{m,n}} \prod_{i=1}^N \,\diff u_i e^{V(u)}&\=\int_{\Gamma_{m,n}} \prod_{i=1}^N \, \diff u_i e^{\pi\i\, k \sum_{i\,=\,1}^N \Bigl(u_i\,-\,\frac{1}{N}\sum_{j=1}^N u_j\Bigr)^2} \prod_{i\neq j} 2\sin \pi u_{ij}\,\\ &\=\int_{\Gamma_{m,n}} \prod_{i=1}^N \, \diff u_i e^{\pi\i\, k \sum_{i\,=\,1}^N \Bigl(u_i\,-\,\frac{1}{N}\sum_{j=1}^N u_j\Bigr)^2} \prod_{i\neq j} 2\sin \pi u_{ij}\, \\ &\qquad\qquad\times\,\Bigl(1\,\longrightarrow\,\int_{\Gamma_{m,n}} \diff C\, \delta\Bigl(C-\frac{1}{N}\sum^{N}_{j=1} u_j\Bigr)\Bigr) \\&\=\int_{\Gamma_{m,n}}\diff C \int_{\Gamma_{m,n}} \prod_{i=1}^N \, \diff u_i e^{\pi\i\, k \sum_{i\,=\,1}^N \Bigl(u_i\,-\,C\Bigr)^2} \prod_{i\neq j} 2\sin \pi u_{ij}\,\\&\qquad\qquad\times \delta\Bigl(C-\frac{1}{N}\sum^{N}_{j=1} u_j\Bigr)\,\\
&\= \int_{\Gamma_{m,n}}\diff C \int_{\Gamma_{m,n}\,+\,C} \prod_{i=1}^N \, \diff \widetilde{u}_i\, e^{\pi\i\, k \sum_{i\,=\,1}^N \widetilde{u}^{i2}} \prod_{i\neq j} 2\sin \pi \widetilde{u}_{ij}\,\\&\qquad\qquad\times \delta\Bigl(\frac{1}{N}\sum^{N}_{j=1}\widetilde{u}_j\Bigr)\,\\&
\={N}\,\int_{\Gamma_{m,n}}\diff C \int_{\Gamma_{m,n}\,+\,C} \prod_{i=1}^{N-1} \, \diff \widetilde{u}_i\, e^{\pi\i\, k \sum_{i\,=\,1}^N \widetilde{u}^{i2}} \prod_{i\neq j} 2\sin \pi \widetilde{u}_{ij}\, \\
&\,\underset{\delta\,\to\,\infty}{\simeq}  {N}\,\int_{\Gamma_{m,n}}\diff C \int_{\Gamma_{m,n}} \prod_{i=1}^{N-1} \, \diff \widetilde{u}_i\, e^{\pi\i\, k \sum_{i\,=\,1}^N \widetilde{u}^{i2}} \prod_{i\neq j} 2\sin \pi \widetilde{u}_{ij}\, 
\end{split}
\ee
\be
\begin{split}
&\underset{\delta\,\to\,\infty}{\simeq}2\delta\times{N}\, \int_{\Gamma_{m,n}} \prod_{i=1}^{N-1} \, \diff \widetilde{u}_i\, e^{\pi\i\, k \sum_{i\,=\,1}^N \widetilde{u}^{i2}} \prod_{i\neq j} 2\sin \pi \widetilde{u}_{ij}\\
&\underset{\delta\,\to\,\infty}{\simeq}2\delta\times{N}\,\int_{\Gamma_{m,n}} \prod_{i=1}^{N-1} \,\diff u_i e^{V(u)}\,.
\end{split}
\ee
In the second line we have used that for every~$u_j\in \Gamma_{m,n}\,$, and from the fact~$\Gamma_{m,n}$ is a straight segment, it follows that
\be
\int_{\Gamma_{m,n}} \diff C\, \delta\Bigl(C-\frac{1}{N}\sum^{N}_{j=1} u_j\Bigr)\=1\,.
\ee
In the sixth step we have used that the integrals over~$\Gamma_{m,n}$ and~$\Gamma_{m,n}+h\,$, with~$|h|$ arbitrary and finite, are equal in the limit~$\delta\to \infty\,$. Thus in the limit~$\delta\to \infty\,$, one can safely deform~$\Gamma_{m,n}+C$ into~$\Gamma_{m,n}\,$, and the integral of~$C$ becomes~$2\delta\,$. See plot~\ref{IdentityhalfN}.

\section{The integrand of~$\mathcal{I}(q)$ along vector walls}

\label{OtherBits}

So far we have mainly focused on the rays~$z=v\=u\, \T$ with~$u\in \mathbb{R}\,$. However, the original integration contour lies along~$v\,\in\,\mathbb{R}\,$. Before starting, let us recall that, for fixed~$m$ and~$n$, we associate the two real numbers~$x_\perp$ and~$x_{||}$ defined by the relation
\be
x\=  x_{||}\, \T\,+\, x_{\perp}\,,
\ee
to the complex number complex~$x$.

\subsection{Other walls in complex~$z$-plane}\label{app:GeneralLogs}

We have already shown that the analytic extension of the vector multiplet contribution to the integrand of the superconformal index gives, in Cardy-like limits, a logarithmic contribution along the~$0$-th wall~$\frac{v}{\T}\=u\in\mathbb{R}$ (See around~\eqref{IdentityTheta02}). Can there be other such non-analyticities? This section classifies all such possible walls of non-analyticities,~\footnote{Chiral multiplets can also give logarithmic contributions. These contributions are of a different nature though, as they come from poles of the elliptic Gamma functions associated to chiral multiplets. Those contributions will be analyzed in deeper detail elsewhere.} namely, those coming from
\be
\log \theta_0(z)\,\theta_0(-z)\,,
\ee
specifically from the series in the right-hand side of the~$(m,n)$ representations in~\eqref{IdentityTheta0} (See equation~\eqref{PortionSeries} below). In the following subsection we will complete the analysis for the piecewise polynomial contributions.

Recall the definition 
 \be
 \xi_{\ell}(\tau)\= z\,+\, \ell \tau \,-\,k_0\=\Bigl(z_{||}\,+\,\frac{1}{m}\,\ell\Bigr)\,\T\,+\,z_{\perp}-\frac{n}{m}\ell\,-\, k_0\,,
 \ee
 where~$k_0=\lfloor z_{\perp \T}\,-\,\frac{n}{m}\ell\rfloor-1\,$. Before, we were assuming~$z_{||}=u\in\mathbb{R}$ and~$z_{\perp}=0\,$, now there are two cases we want to explore 
 \begin{itemize}
 \item[] Case 1):
 \be\label{AsumptionRealContour}
 z_{||}\=0\,,\qquad ~z_{\perp}\,-\,p\,\ll\, 1\,,
 \ee
 \item[] Case 2): \be\label{AsumptionRealContourCase2}
 \begin{split}
 z_{||}\=u\,\in\,\mathbb{R}\,, \qquad ~z_{\perp}\,-\,p\,\ll\, 1\,,
 \end{split}
 \ee
 \end{itemize}
 for some real number~$p\,$ that will be fixed by requiring convergence.
 
 Let us start with Case 1). Specifically, we assume that
 \be
 \begin{split}
 \frac{z_{\perp}\,-\,p}{|\T|}\,\liMN\, y\,,
 \end{split}
 \ee
 where~$y$ is an arbitrary and finite real number i.e.~$|y|<\delta\,$. $\delta$ is to be identified with the splitting parameter used to divide the original contour of integration.

 For small enough~$z_{\perp}-p$
 \be
 k_0\=\lfloor p -\frac{n}{m}\ell\rfloor\,-\,1\,,
 \ee
and
\be
\xi_{\ell}(\tau)\= \Bigl(y\,e^{\i\Phi_{m,n}}\,+\,\frac{1}{m}\ell\Bigr)\,\T\,+\,\{p-\,\frac{n}{m} \,\ell\}\,-\,1\,,
\ee
where~$\{x\}\equiv x-\lfloor x\rfloor$ and
\be
e^{\i \Phi_{m,n}}\,\equiv\,\underset{\tau\,\to\,-\frac{n}{m}}{\lim}\,\frac{|\T|}{\T}\,.
\ee
The condition for the series 
\be\label{PortionSeries}
\begin{split}
L^{(\ell)}(z)\,\equiv\, \Lz
\end{split}
\ee
to be absolutely convergent is
\be
\begin{split}
|\text{Im} \Bigl(\frac{2\xi_{\ell}(z)+1}{\T}\Bigr)| &\,<\, |\text{Im} \Bigl(\frac{1}{\T}\Bigr)| \\
|y \sin \Phi_{m,n} \,+\,\text{Im} \Bigl(\frac{2\,\{p-\,\frac{n}{m} \,\ell\}\,-\,1}{\T}\Bigr)|&\,<\, |\text{Im} \Bigl(\frac{1}{\T}\Bigr)| \,.
\end{split}
\ee
In the asymptotic Cardy-like limit~$\tau\,\to\,-\frac{n}{m}$ the conditions reduce to
\be
|2\,\{p-\,\frac{n}{m} \,\ell\}\,-\,1|\,<\,1\,.
\ee
This condition is satisfied for generic real~$p\,$, but not for those values that make the left-hand side equal~$1\,$. Those are also the solutions to
\be
\{p-\,\frac{n}{m} \,\ell\}\= 0 \quad\text{or} \quad 1
\ee
which are
\be\label{Operation}
p(\ell)\,\text{mod}\,1\= \frac{n}{m} \ell\,.
\ee
In these cases
\be\label{IdentityTheta0RealContour}
\begin{split}
L^{(\ell)}(z)&\,\liMN\,\sum_{p(\ell)} \,\textbf{P}(z_{\perp}-p)\,\log 2\,|\sin \pi\Bigl(y\, \cos{\Phi_{m,n}}\,+\,\frac{\ell}{m}\Bigr)|\,+\,\ldots \\
&\=\sum_{p(\ell)} \,\textbf{P}(z_{\perp}-p)\,\log 2\,|\sin\pi \Bigl(\frac{z_{\perp}\,-\,p}{|\T|}\,\cos \Phi_{m,n}\,+\,\frac{\ell}{m}\Bigr)|\,+\,\ldots\\
&\= \sum_{ \widetilde{j} \,\in \,\mathbb{Z}}\,\textbf{P}(z_{\perp}-\frac{\ell^*}{m}+\widetilde{j})\,\log 2\,|\sin\pi \Bigl(\text{Re}\Bigl(\frac{z_{\perp}\,-\,\frac{\ell^*}{m}\,+\,\widetilde{j}}{\T}\Bigr)\,+\,\frac{\ell}{m}\Bigr)|\,+\,\ldots\,,
\end{split}
\ee
The meaning of~$\ldots$ will be given below.

In these equations the~\emph{projectors}~$\textbf{P}(x)$ is defined as
\be
\textbf{P}(x)\= 1\,
\ee
in the infinitesimal domain~$|x|<|\T|\,\delta\liMN0\,$, and vanishes exponentially fast outside that region. Thus, multiplying~$\textbf{P}(z_\perp -p)$ by a function~$f(z)$ projects the function to zero outside the infinitesimal ribbon located at~$z_{\perp}\=p\,$, and leaves it unchanged within the ribbon. These ribbons are the walls.

The integer~$\ell^*=0,\ldots, m-1\,$, which is a function of~$\ell\,$, is determined by the condition
\be\label{DefLStar}
\frac{\ell^*(\ell)}{m}\=p(\ell)\,\mod{\mathbb{Z}}\,.
\ee
The~$\ldots$ in~\eqref{IdentityTheta0RealContour} denote plus terms that cancel after adding the contribution with~$z_{\perp}\to-z_{\perp}\,$, and summing over~$\ell$ and~$\widetilde{j}$; this conclusion can be understood to follow from the following observation. For each~$\ell=0,1,\ldots,m-1$ there exists the inverse~$m\,-\,\ell \mod{m}=0,1,\ldots,m-1\,$. For these pairs
\be
p(\ell) \= -p(m\,-\,\ell)\mod{\mathbb{Z}}\,.
\ee
Thus, two mutually inverse values of~$\ell$'s are mapped to mutually inverse values of~$\ell^*$'s by the map~$p=p(\ell)\,$. That implies that the sum in the second line of the right-hand side of~\eqref{IdentityTheta0RealContour} is invariant under the simultaneous substitution of~$p\to-p$ and~$\ell\to-\ell\,$, and thus (because of the presence of the absolute value) it is also invariant under~$z_{\perp}\to-z_{\perp}\,$. 

At last, the previous invariances imply that the sum in the first line in the right-hand side of~\eqref{IdentityTheta0RealContour} is invariant under the transformations~$y\to -y\,$, $\ell\to-\ell\,$, and consequently, odd terms under such transformation cancel out, after summing over~$\ell\,$.

In summary, definning
\be
L_{+}(z)\,\equiv\,\sum_{\ell=0}^{m-1}\,\Bigl(L^{(\ell)}(z)\,+\,L^{(\ell)}(-z)\Bigr)\,,
\ee
one obtains
\be
\begin{split}
L_{+}(z)\,\liMN\,
\sum_{\ell=0}^{m-1}\sum_{ \widetilde{j} \,\in \,\mathbb{Z}}\,\textbf{P}(z_{\perp}-\frac{\ell^*}{m}+\widetilde{j})\,\Bigl(\log\Bigl( 2\,\sin\pi& \Bigl(y\,\cos{\Phi_{m,n}}\,+\,\frac{\ell}{m}\Bigr)\,\times\\ \times\,& 2\,\sin\pi \Bigl(-y\,\cos{\Phi_{m,n}}\,-\,\frac{\ell}{m}\Bigr)\Bigr)\,+\,\log{(-1)}\Bigr)\,.
\end{split}
\ee
In this equation
\be
y\=\frac{z_{\perp}\,-p}{|\T|}\=\frac{z_{\perp}\,-\,\frac{\ell^*}{m}\,+\,\widetilde{j}}{|\T|}\,.
\ee
Due to the multiplication by the projector~$\textbf{P}(z_{\perp}-\frac{\ell^*}{m}+\widetilde{j})$, a given value of~$p$ picks up a unique~$\ell\,\in\,\mathbb{Z}$ mod $m\,$, and vice versa. That implies that at a vicinity of~$z\approx p$ there is a single logarithmic term in the potential, the one associated to the value of~$\ell$ corresponding to~$p$ via the inverse map~$p^{-1}\,$. By~$p^{-1}$ we mean the inverse map of the operation~$p:\ell \mapsto p(\ell)$ where~$p(\ell)$ was defined in equation~\eqref{Operation}.

Repeating the same steps one can reach a similar formula for the Case 2. If in this case we assume~$y\=0$~\footnote{As explained before this introduces an error that is exponentially suppressed in the Cardy-like expansion.} the result is
\be\label{LPlusU}
\begin{split}
L_+(z)\,\liMN\,
\sum_{\ell=0}^{m-1}\,\sum_{ \widetilde{j} \,\in \,\mathbb{Z}}\,\textbf{P}(z_{\perp}-\frac{\ell^*}{m}+\widetilde{j})\,\Bigl(\log 2\,\sin\pi \Bigl(u\,+\,\frac{\ell}{m}\Bigr)\,& 2\,\sin\pi \Bigl(-u\,-\,\frac{\ell}{m}\Bigr)\,+\,\log(-1)\Bigr)\,.
\end{split}
\ee
This formula will be used to compute the bits of integral for~$N=2$ (these results can be used at any value of~$N$ though): it is the profile of the (limit of the) series along the corresponding wall.

\subsection{The polynomial part along a generic vector wall}
\label{PhasesLens}
For~$z\=z_{||}\,\T\,+\,z_{\perp}$ and
\be
\frac{z_{\perp}\,-\,p}{\T}\= y\,<\,\infty\,, \quad \text{with} \,\quad\, p\,\text{mod}\,1\,\equiv \frac{\ell^*}{m}\,,
\ee
a computation shows that
\be\label{ExpansionF0PthWall}
-\,2\,\mathcal{F}^{(m,n)}(z) \=-\,2\, \mathcal{F}^{(m,n)}(z=p)\,+\, \kappa_{m,n} \,\pi \i \,z_{||}^2\,,
\ee
where~$\kappa_{m,n}$ was defined in~\eqref{KappaMN}.

In the main body of the paper we will use the following definition
\be\label{PhaseP}
-2 \pi \i \varphi^{(m,n)}_p\,\equiv\,-\,2\, \mathcal{F}^{(m,n)}(z = p)\,+\,\,2\, \mathcal{F}^{(m,n)}(z=0)\,.
\ee
~$\varphi^{(m,n)}_p$ is a real constant that depends only on the choice of the wall. The piece-wise polynomial function~$\mathcal{F}^{(m,n)}(z)$ was defined in~\eqref{KappaMN}.

\subsection{From the horizontal to the diagonal contour }\label{OnCourtourDiagonal} To justify the use of the diagonal contour in~figure~\ref{fig:ContourGammaMN0} instead of the horizontal one, we build upon an observation about~$L_{+}(z)$ when~$y\neq 0$ and~$u\neq0\,$.

Consider
\be\label{Zform}
z\= \Bigl(y\,e^{\i\Phi_{m,n}}\,+\, u\Bigr)\,\T\,,
\ee
with~$y\in\mathbb{R}$ and~$u\in \mathbb{R}$.  With~\eqref{Zform}, the new version of~\eqref{LPlusU} can be obtained from the latter after substituting
\be
u \longrightarrow \,  \Bigl(y\,e^{\i\Phi_{m,n}}\,+\, u\Bigr)\,.
\ee
 Everything else remains the same e.g. the positions of the walls, etc. Now, from property~$L_+(z)\=L_{+}(-z)\,$, it follows that
\be\label{PropertyLp}
L_{+}(z= p+\delta p)\=L_{+}(z=-p-\delta p)
\ee
where again,~$p=\frac{\ell^*}{m}+\text{integers}\in\mathbb{R}$ and~$\delta_p\in\mathbb{C}$ is a complex number with small enough absolute value. As the integrand is even in the integration variable~$z\,$, the same property~\eqref{PropertyLp} applies to the full integrand of the index, not just for the contributions coming from the exponential of~$L_+\,$. 

In Cardy-like limit the previous statement implies that the integral along the dashed vertical line to the right (resp. left) in figure~\ref{fig:ContourGammaMN0}, in a vicinity of the bit~$z=p\,\text{mod}\,1\,$, cancels the integral along the dashed vertical line to the left (resp. right), but this time in a vicinity of the bit~$z=\,-\,p\,\text{mod}\,1\,$. As one must sum over all bits intersected by the original contour of integration, and these always include both, the~$p$ and~$-p $ bits, it follows that to compute the localized contributions we can use the diagonal contour in figure~\ref{fig:ContourGammaMN0} instead of the horizontal one.

\section*{Some comments about~~$\mathcal{K}$ in~\eqref{LeadingBehaviour}}

{For generic $m$ and~$n$ the approach~$2)$ predicts that
\be\label{Kappa02}
\mathcal{K}\,=\, \sum_{I\,\in\,\text{fixed points leading at~$\tau\to-\frac{n}{m}$}} e^{2\pi\i \phi_{I}}\,=\,{N} \,+\,\sum_{I\,\in\, \text{Other possible fixed points}\atop
\text{leading in the limit~$\tau\to -\frac{n}{m}$}} e^{2\pi\i \phi_{I}}\,,
\ee
where the factor of~$N$ counts certain configurations that we call fixed points or Bethe roots, indistinctly. For the Cardy-like limits that we study, and for~$N=2$~\cite{Benini:2021ano,Lezcano:2021qbj} it has been already shown that~$
\sum_{I\,\in\, \text{Other possible fixed points}\atop
\text{leading in the limit~$\tau\to -\frac{n}{m}$}} e^{2\pi\i \phi_{I}}\,=\,0\,
$. {This was also argued to hold in the limit~$\tau\to0\,$, for generic~$N$~\cite{GonzalezLezcano:2020yeb,Amariti:2021ubd,ArabiArdehali:2021nsx,Cassani:2021fyv}, and our conclusions confirm so (See also the discussion in~\cite{PaperPhases}).}
}~
{In the gravitational side of the duality these~$N$ should correspond to a subset of the Euclidean configurations considered in~\cite{Aharony:2021zkr}. }

~{These solutions are conjectured to be related to massive vacua of~$\mathcal{N}=1^*$ theory on~$\mathbb{R}^{1,3}$~\cite{Benini:2021ano,Hong:2018viz}\cite{Donagi:1995cf}. The set of solutions corresponding to the second factor in~\eqref{Kappa02} should correspond to continuum sets of Bethe roots~\cite{ArabiArdehali:2019orz} and are expected to correspond to vacua of the~$\mathcal{N}=1^*$ theory containing massless photons~\cite{ArabiArdehali:2019orz,Benini:2021ano}\cite{Donagi:1995cf,Dorey:1999sj}.} It would be interesting to explore whether the symmetry-breaking classification introduced in section~\ref{SymmetryBreakingBits} relates to the classification of vacua of~$\mathcal{N}=1^*$ on $\mathbb{R}^{3,1}$ of~\cite{Donagi:1995cf}\cite{Dorey:1999sj}. The latter have been conjectured to correspond to Bethe roots of the~$SU(N)$~$\mathcal{N}=4$ index~\cite{Benini:2021ano,ArabiArdehali:2019tdm}.

\providecommand{\href}[2]{#2}\begingroup\raggedright\endgroup

\bibliographystyle{JHEP}

\end{document}